\documentclass[a4paper,12pt]{article}
\usepackage[a4paper,left = 2cm, right= 2cm,top = 2cm, bottom = 2cm]{geometry}
\usepackage{cite}
\usepackage{amsmath,amssymb,amsfonts}
\usepackage{multirow}
\usepackage{algorithmic}
\usepackage{graphicx}
\usepackage{textcomp}
\usepackage{amssymb,
amsthm,
amsfonts,
amsmath,
enumerate,
caption,
subfigure,diagbox,booktabs,bm,array,tikz,cuted
}
\allowdisplaybreaks
\usepackage[ruled,vlined,commentsnumbered]{algorithm2e}
\newcommand{\me}{\mathrm{e}}

\newtheorem{defi}{Definition} 
\newtheorem{thm}{Theorem}
\newtheorem{corollary}{Corollary}
\newtheorem{lem}{Lemma}
\newtheorem{pf}{Proof}

\usepackage{authblk}

\begin{document}
\bibliographystyle{ieeetr}
\title{On the Convergence of Sigmoid and tanh Fuzzy General Grey Cognitive Maps}
\date{}
\author[a]{Xudong Gao} 
\author[a]{Xiaoguang Gao} 
\author[b]{Jia Rong} 
\author[c]{Ni Li}
\author[d]{Yifeng Niu}
\author[a,e]{Jun Chen\thanks{Corresponding Author, e-mail:junchen@nwpu.edu.cn}}

\affil[a]{School of Electronics and Information, Northwestern Polytechnical University, Xi'an 710072, Shaanxi, China}
\affil[b]{Department of Data Science and AI, Monash University, Clayton, Melbourne,VIC3800, Victoria, Australia}
\affil[c]{School of Aeronautics, Northwestern Polytechnical University, Xi'an 710072, Shaanxi, China}
\affil[d]{College of Intelligence Science and Technology, National University of Defense Technology, Changsha, 410073, Hunan, China}
\affil[e]{Chongqing Institute for Brain and Intelligence, Guangyang Bay Laboratory, Nan'an District, 400064, Chongqing, China}

\newcommand\blfootnote[1]{
\begingroup
\renewcommand\thefootnote{}\footnote{#1}%
\addtocounter{footnote}{-1}
\endgroup
}

\maketitle

\begin{abstract}
  Fuzzy General Grey Cognitive Map (FGGCM) and Fuzzy Grey Cognitive Map (FGCM) are extensions of Fuzzy Cognitive Map (FCM) in terms of uncertainty. FGGCM allows for the processing of general grey number with multiple intervals, enabling FCM to better address uncertain situations. Although the convergence of FCM and FGCM has been discussed in many literature, the convergence of FGGCM has not been thoroughly explored. This paper aims to fill this research gap. 
 First, metrics for the general grey number space and its vector space is given and proved using the Minkowski inequality. By utilizing the characteristic that Cauchy sequences are convergent sequences, the completeness of these two space is demonstrated. On this premise, utilizing Banach fixed point theorem and Browder-Gohde-Kirk fixed point theorem, combined with Lagrange's mean value theorem and Cauchy's inequality, deduces the sufficient conditions for FGGCM to converge to a unique fixed point when using tanh and sigmoid functions as activation functions. The sufficient conditions for the kernels and greyness of FGGCM to converge to a unique fixed point are also provided separately. Finally, based on Web Experience and Civil engineering FCM, designed corresponding FGGCM with sigmoid and tanh as activation functions by modifying the weights to general grey numbers. By comparing with the convergence theorems of FCM and FGCM, the effectiveness of the theorems proposed in this paper was verified. It was also demonstrated that the convergence theorems of FCM are special cases of the theorems proposed in this paper. The study for convergence of FGGCM is of great significance for guiding the learning algorithm of FGGCM, which is needed for designing FGGCM with specific fixed points, lays a solid theoretical foundation for the application of FGGCM in fields such as control, prediction, and decision support systems.

\end{abstract}

\section{Introduction}  \label{sec:introduction}

The fuzzy cognitive map (FCM) was proposed by Kosko in 1986, consisting of nodes and weights between nodes. Its structure resembles that of a recurrent neural network, and it utilizes activation functions such as sigmoid or tanh for reasoning. Compared to neural networks, FCM exhibits excellent interpretability \cite{Napoles2023, Li2024}. Due to the tangible meaning of its nodes and the absence of hidden layers, FCM emulates the cognitive processing of the human brain. Over the past $40$ years, FCM have been successfully applied in various domains \cite{Zhe2020}. The FCM model can be defined as follow:

\begin{defi}
  Fuzzy Cognitive Map (FCM) is defined as a 4-tuple $M = \{\bm C, \mathbf{W}, \bm{A}, f\}$ where $\bm C$ denotes the collection of all map neurons, $\mathbf{W}: (C_i, C_j) \rightarrow w_{ij}$ defines the causal weight matrix $\mathbf{W}$, $\bm{A}: C_i \rightarrow A^{(t)}_i $ represents the function that determines the activation level of each neuron $C_i$ at a specific discrete time step $t$ (where $t = 1, 2, \cdots, T$) and $f(\cdot)$ is the activation function. 
\end{defi}

Eq. \eqref{fcm_iter} illustrates the interaction of above elements, demonstrating how the vector $\bm{A^{(t)}} = [A^{(t)}_1, A^{(t)}_2, \cdots , A^{(t)}_M]^{\mathsf{T}}$ is iteratively calculated from the initial state vector $\bm{A^{(0)}} = [A^{(0)}_1, A^{(0)}_2,\cdots,$ $ A^{(0)}_M]^{\mathsf{T}}$. 
\begin{equation}
  \label{fcm_iter}
  \bm{A^{( t+1 )}}= f \left(\mathbf{W} \bm{A^{(t )}} \right) 
\end{equation}

Eq. (\ref{fcm_iter}) serves as the general updating rule, as it employs the current neuron activation value $A^{( t )}_i$ to determine the subsequent state $A^{( t+1 )}_i$. However, in certain applications, this recursive feature is prohibited, as it implicitly includes self-reinforcing causal links, and there are situations where a variable should not be influenced by its own state, in such cases, $w_{ii}=0$.

The elements of the weight matrix are typically constrained within the range of $[-1, 1]$, while the activation values of nodes, denoted as $A$, are usually set within the intervals $[0, 1]$ or $[-1, 1]$. In discrete activation states, the activation functions commonly employed is binary function or trivalent function; whereas in continuous activation states, sigmoid function (Eq.\ref{sigmoid}) or hyperbolic tangent (Eq.\ref{tanh}) function is typically used. When the activation function is a sigmoid function, the range of node activations is $[0, 1]$; conversely, when the activation function is a tanh function, the range of node activations is $[-1, 1]$.
\begin{equation}
  \label{sigmoid}
  S(x) = \frac{1}{1 + \me^{-\lambda x}}
\end{equation}
\begin{equation}
  \label{tanh}
  \tanh(x) = \frac{\me^{\lambda x} - \me^{-\lambda x}}{\me^{\lambda x} + \me^{-\lambda x}}
\end{equation}

In Eq. (\ref{sigmoid}) and Eq. (\ref{tanh}), the parameter $\lambda$ is used to control the slope of the activation function, taking a value greater than $0$, and its magnitude is closely related to the convergence behavior of the FCM.

Numerous scholars have proposed several extensions to enhance the uncertainty representation and dynamic reasoning capabilities of FCMs. Among them, Fuzzy General Grey Cognitive Map (FGGCM) is an extension of FCMs in terms of uncertainty representation. FGGCM builds upon fuzzy grey cognitive maps (FGCM) and incorporates the concept of general grey numbers (GGN) from grey system theory. By employing the kernel and the degree of greyness of GGN for iterative reasoning, FGGCM acquires the ability to simultaneously handle multiple interval data (represented using GGN).

The convergence of FCMs holds significant guidance for various aspects such as large-scale system simulation, learning algorithm, and pattern recognition. Unclear convergence can pose significant challenges to the design of FCMs, in such cases, designed FCMs may converge to undesired states or chaotic states, rendering them ineffective. Thus, the convergence of FCMs has gradually attracted the attention of many scholars. As the iterative reasoning of FCMs progresses, their nodes can converge to fixed points, limit cycles, or chaotic states. Their standard mathematical definition are as follows:

\begin{defi}
  Fixed point: $\exists t_\alpha \in \{1, 2, \ldots, T - 1\}, \forall t \geq t_\alpha, s.t.\bm A^{(t+1)} = \bm A^{(t)}$. The system enters a state of equilibrium after $t_\alpha$ , resulting in a consistent output where $\bm A^{(t_\alpha) }= \bm A^{(t_\alpha+1)} = \bm A^{(t_\alpha+2)} = \ldots = \bm A^{(T)}$.
\end{defi}

\begin{defi}
  Limit cycle: $\exists t_\alpha, P \in \{1, 2, \cdots, T - 1\}, \forall t \geq t_\alpha, s.t. \bm A^{(t+P)} = A^{(t)}$. The system enters a periodic state after $t_\alpha$ step, resulting in a repeated output pattern such that $\bm A^{(t_\alpha)} = \bm A^{(t_\alpha+P)} = \bm A^{(t_\alpha+2P) }= \ldots = \bm A^{(t_\alpha+jP)}$, where $j \in \mathbb{N}$.
\end{defi}

\begin{defi}
  Chaos: The system persistently generates distinct state vectors over consecutive steps. In these instances, the FCM fails to reach a stable state, resulting in unpredictable and ambiguous system responses.
\end{defi}

For the sake of convenience, subsequent references to the convergence, kernel, and greyness of FCM, FGCM, and FGGCM refer to the convergence of node values, the kernel of node values, and the greyness of node values of models.

\begin{figure}[htbp]
  \centering
  \begin{tikzpicture}[every node/.style={circle,draw,inner sep=5pt,font=\footnotesize}]
    \def\side{2}
  \foreach \i in {1,2,...,7}  {        \pgfmathsetmacro{\angle}{\i * 51.42857142857143} 
  \pgfmathsetmacro{\x}{\side * cos(\angle)}
  \pgfmathsetmacro{\y}{\side * sin(\angle)}
  \node[fill=blue!20] (node\i)at (\x,\y){$C_{\i}$};};
  \foreach \i in{1,2,...,7} {\foreach \j in{1,2,...,7}  \ifnum \i = \j
  \else \draw[-latex] (node\i)--(node\j)
\fi;} 
  \draw[-latex] (node3.north) .. controls (-2,2) and (-2.2,1.5 ) .. (node3.north west); 
  \draw[-latex] (node4.west) .. controls (-3,-2) and (-2,-1.5 ) .. (node4.south); 
  \draw[-latex] (node5.south west) .. controls (-1,-3) and (-0.5,-3 ) .. (node5.south);
  \draw[-latex] (node6.south) .. controls (1,-3) and (2,-2.5 ) .. (node6.south east);
  \draw[-latex] (node7.south east) .. controls (2.5,-1) and (3.5,0 ) .. (node7.east);
  \end{tikzpicture}
  \caption{The user's Web Experience FCM}
  \label{figfcmexample}
\end{figure}
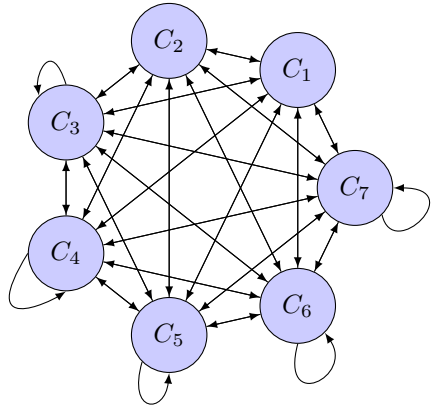

For example, Fig. \ref{figfcmexample} employs FCMs to model a user's web experience on the World Wide Web \cite{Meghabghab2001,Meghabghab2003}, with the meanings of the nodes as Table \ref{nodes_web}. For the convenience, this FCM will be called the Web Experience FCM in the following text. 

\begin{table}[htbp]
  \centering
  \caption{The nodes' meanings of Web Experience FCM.}\label{nodes_web}
  \begin{tabular*}{.35\linewidth}{cc}
    \toprule    
    Nodes & Meanings \\ 
    \midrule 
    $C_1 $& Exploration  \\ 
    $C_2 $& Scanning   \\ 
    $C_3 $& Temporal Restrictions  \\ 
    $C_4 $& Data Restrictions  \\ 
    $C_5 $& Achievement  \\ 
    $C_6 $& Pertinence \\ 
    $C_7 $& Unsuccess  \\ 
    \bottomrule    
    \end{tabular*}
  \end{table}

This FCM employs the sigmoid function (Eq. \ref{sigmoid}) as its activation function, and its weight matrix is as follows:
\begin{equation}
  \label{wweb}
\begin{split}
    \mathbf{W}_{web}=
  \begin{pmatrix}
    0.0 & -0.9 & -0.88 & 1.0 & -0.85 & -0.83 & 1.0 \\
    1.0 & 0.0 & -0.93 & -0.89 & -0.9 & -0.94 & 1.0 \\
    -0.98 & -0.93 & -1.0 & -1.0 & 1.0 & 1.0 & 1.0 \\
    -0.99 & -0.89 & -1.0 & -0.39 & 0.73 & 0.58 & 0.7 \\
    1.0 & 1.0 & 1.0 & 1.0 & -0.8 & 0.51 & 1.0 \\
    1.0 & 1.0 & 0.83 & 1.0 & 0.51 & -0.39 & 1.0 \\
    1.0 & 1.0 & 1.0 & 1.0 & -0.71 & -0.49 & -0.67
    \end{pmatrix}
\end{split}
\end{equation}

\begin{figure}[htbp]
  \centering
    \includegraphics[width=.75\linewidth]{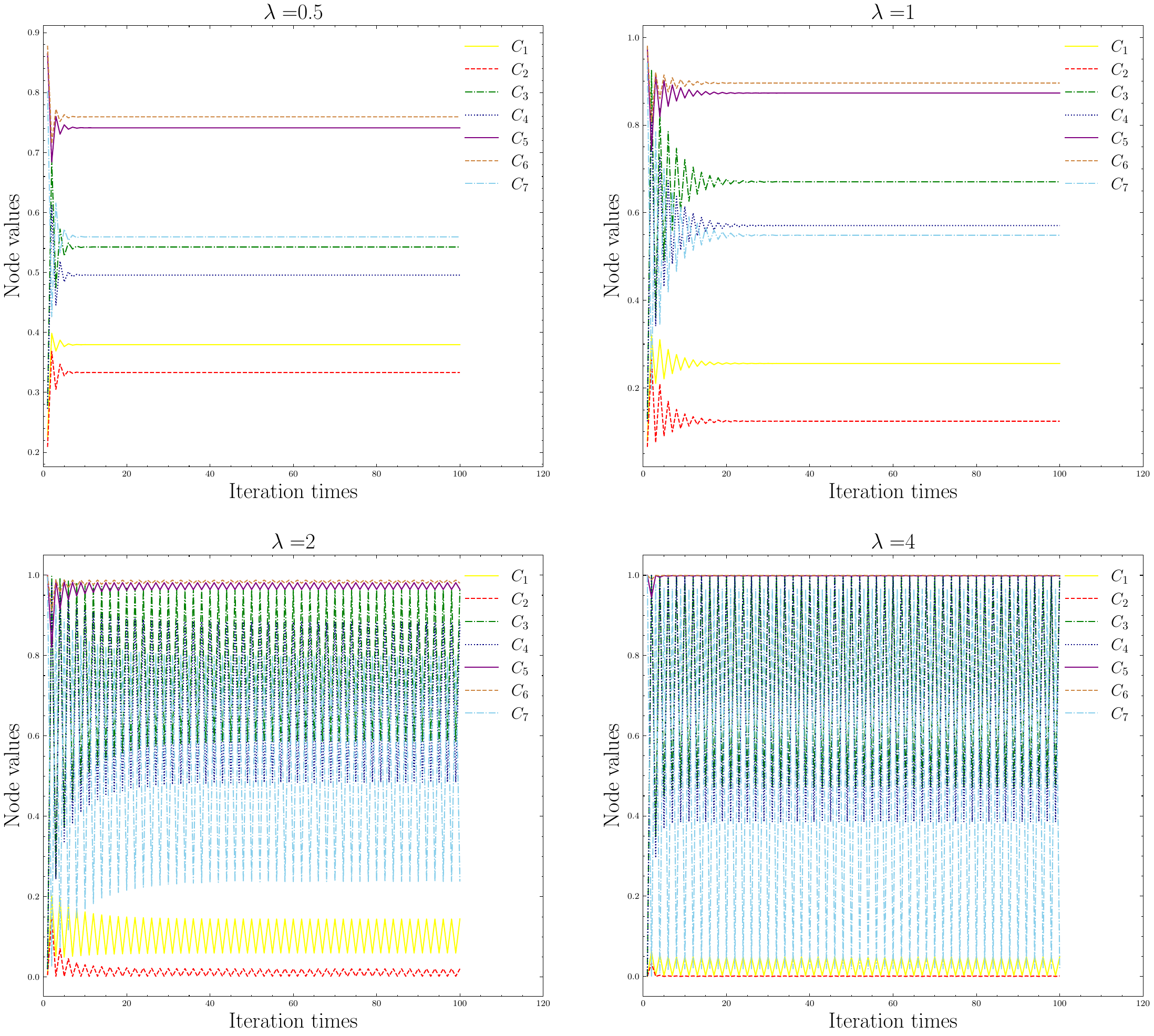}
  \caption{The output of the Web Experience FCM under different $\lambda$s.}
  \label{figfcmweb}
\end{figure}

Utilizing an initial vector of $\setlength{\arraycolsep}{2pt}
\begin{pmatrix}
  1 & 1 & 1 & 1 & 1 & 1 & 0
  \end{pmatrix}$. Fig. \ref{figfcmweb} illustrates the simulation outcomes for each node when \(\lambda\) is set to $0.5$, $1$, $2$, and $4$.

From Fig. \ref{figfcmweb}, it is observed that the FCM converges to a fixed point when \(\lambda\) is equal to $0.5$ and $1$, and forms a limit cycle when \(\lambda\) is set to $2$ and $4$.

In 2009, Boutalis proposed the sufficient conditions for FCMs to converge to a fixed point \cite{Boutalis2009}, subsequently leading to a plethora of research endeavors that explored the convergence properties of FCMs and introduced novel extensions of FCMs based on these studies. In 2018, Harmati provided the sufficient conditions for FGCMs to converge to fixed points \cite{Harmati2018, Harmati2020}. However, since the introduction of FGGCMs in 2020, there has been a lack of literature investigating the convergence of FGGCMs. This study aims to fill this gap by employing the Banach fixed point theorem and Browder-Gohde-Kirk fixed point theorem to establish a sufficient condition for FGGCMs to converge to fixed points when the activation functions are sigmoid and tanh, respectively.

The primary contributions and of this paper can be summarized as follows:

\begin{itemize}
  \item This paper introduces metrics for general grey number spaces and general grey number vector spaces, and demonstrates that these spaces are complete.
  \item Building upon the Banach fixed point theorem and Browder-Gohde-Kirk fixed point theorem, this study establishes sufficient conditions for FGGCMs to converge to fixed points when the activation functions are sigmoid and tanh, respectively. 
  \item The effectiveness of the proposed theorems has been verified through multiple examples, which provides a theoretical foundation for the design, analysis, and development of learning algorithms for FGGCM.
\end{itemize}

The structure of this paper is as follows: Section \ref{sec:related_work} reviews previous studies by scholars on the convergence properties of FCMs and FGCMs. Section \ref{sec:basicsFGGCM} revisits the operational rules of general grey numbers and the reasoning rules of FGGCMs when the activation functions are sigmoid and tanh, respectively. Section \ref{sec:completeness} introduces matrices for general grey number spaces and general grey number vector spaces, along with the proof of the completeness of these spaces. Section \ref{sec:convergence} presents sufficient conditions for FGGCMs to converge to fixed points when the activation functions are sigmoid and tanh, respectively. Section \ref{sec:case_study} designed examples for two activation functions to illustrate the convergence theorems proposed in this paper, Section \ref{sec:result} presented the convergence results for the corresponding examples, and Section \ref{sec:discussion} discussed and analyzed the convergence results. Section \ref{sec:conclusion} summarizes the entire paper and attempts to outline future research directions regarding the convergence properties of FGGCMs.

\section{Related Work} \label{sec:related_work}

According to the literature available to the author, the study of the convergence properties of FCMs originated with B. Kosko, who introduced the FCM model. Specifically, Kosko devised an analytical technique that utilized Liapounov functions to achieve stable outcomes for Feedback Standard Additive Models (SAM), which share numerous similarities with FCMs. However, Kosko concluded that these robust mathematical conditions and theorems are not readily applicable to FCMs due to the extensive network of feedback connections inherent in FCM-based architectures.

Subsequently, Athanasios K. Tsadiras's team analyzed the dynamic behavior of certainty neuron FCMs (CNFCMs). They were the first to describe in the literature the three convergence scenarios of FCMs and identified how two parameters influence the system's dynamic behavior, transitioning it from a fixed point to a limit cycle \cite{Tsadiras1999}. Additionally, Taber et al. demonstrated that with small data, the construction of the weight matrix is more likely to lead to FCMs exhibiting a limit cycle state \cite{Taber2001, Taber2007}.

By 2009, Boutalis et al. utilized the Banach fixed point theorem to analyze the sufficient conditions for FCMs with sigmoid activation functions to converge to a unique fixed point, as well as the conditions for the existence of fixed points that may not be unique \cite{Boutalis2009}. Subsequently, based on the convergence properties of FCMs, Boutalis' team proposed the fuzzy cognitive network (FCN) model and applied it to various domains such as indirect adaptive inverse control of power plants, DC motor control, motor damage and diagnosis, time series prediction, and fault detection of induction generators \cite{Karatzinis2018, Karatzinis2018a, Karatzinis2018b, Karatzinis2021, Karatzinis2021a}.

In 2014, Nápoles et al. found that the number of sigmoidal FCM's fixed point is dependent on the magnitude of $\lambda$. When $\lambda$ is tiny and positive, the system yields a solitary solution, signifying a fixed point of stability in the sigmoid FCM. With $\lambda$ substantially positive, the system may present multiple solutions, some being stable fixed points. These findings are especially pertinent in control systems, where steady outputs are regardless of initial conditions, such as in system identification and control strategy development \cite{Napoles2014}. Christopher Knight established a condition with a clear parameter boundary to ensure that a sigmoid FCM has a unique and stable fixed point \cite{Knight2014}, which was applied in a case study on the transition to a bio-based economy in the Humber region of the United Kingdom. Miklos F. Hatwagner and colleagues investigated several organization models defined by real experts and conducted simulations to study their asymptotic behavior, particularly in seeking all fixed point attractors or identifying chaotic behavior \cite{Hatwagner2017}.

Harmati, through the analysis of a rescaling algorithm, delved into the behavior of FCMs, offering new insights into their convergence \cite{Harmati2020c}. In 2021, the Harmati team proposed a new condition for FCMs to globally asymptotically stabilize, comparing it with previous convergence criteria \cite{Harmati2023a}. Koutsellis identified a specific range of parameters for the common transfer functions in FCMs to address output uncertainty \cite{Koutsellis2022}. In 2023, Maximov provided convergence conditions for multi-valued FCMs \cite{Maximov2023}.
The Nápoles team discovered that the state space of any FCM with a activation function contracts to infinity, and it cannot be guaranteed to converge to a fixed point, but rather to its limit state space \cite{Concepcion2021}. They also introduced a model for quantifying implicit bias in structured datasets and studied its convergence, deriving analysis conditions for the existence and uniqueness of fixed point attractors \cite{Napoles2022}.
Luo Chao, based on algebraic dynamics, analyzed and quantitatively studied FCMs and their stabilization, proposing a necessary and sufficient condition to determine attractors and implementing an algorithm to calculate all attractors. Additionally,  using the semi-tensor product (STP) of matrices, derived algebraic expressions for FCMs and proposed a condition for global stability, which was achieved through the design of a state feedback controller \cite{Xiaojie2022}. From these literatures, the conditions for the existence of fixed points of FCM can be summarized as:

\begin{thm}
  \label{thmfcm}
  In an FCM utilizing the sigmoid function $f = 1/(1 + \me ^{-\lambda x})$ as activation function, there exists a unique fixed point for every concept value $A_i$, provided that the condition 
  \begin{equation}
    \label{eqfcmsig}
    \left\lVert \mathbf{W} \right\rVert _F  < \frac{4}{\lambda}
  \end{equation} 
  holds true.
 If $\left\lVert \mathbf{W} \right\rVert _F = 4/\lambda,$ 
  then for each concept value $A_i$ in any FCM, there is at least one fixed point, although uniqueness cannot be guaranteed.

  Furthermore, in an FCM utilizing the tanh function $f(x) = \tanh(x) = \frac{\me^{\lambda x} - \me^{-\lambda x}}{\me^ {\lambda x} + \me^{-\lambda x}}$, there exists a unique fixed point for every concept value $A_i$, provided that the condition 
   \begin{equation}
    \label{eqfcmtanh}
    \left\lVert \mathbf{W} \right\rVert _F  < \frac{1}{\lambda}
   \end{equation} 
  holds true,
 If $\left\lVert \mathbf{W} \right\rVert _F = 1/\lambda,$ 
  then for each concept value $A_i$ in any FCM, there is at least one fixed point, although uniqueness cannot be guaranteed.
\end{thm}

$ \left\lVert \mathbf{W} \right\rVert _F $ denotes the Frobenius norm of $ \mathbf{W} $, which can be calculated as:
$$
\| \mathbf{W} \|_{F}=\left( \sum_{i=1}^{n} \sum_{j=1}^{n} w_{i j}^{2} \right)^\frac{1}{2}
$$

The convergence conditions of FGCMs have been a subject of recent research. Harmati analyzed the convergence conditions of FCMs with tanh as the activation function \cite{Harmati2018, Harmati2020}. In 2019, the team further analyzed the fixed point convergence conditions of FGCMs under the sigmoid and tanh functions \cite{Harmati2019a, Harmati2019, Harmati2020a}. The Nápoles team revealed the behavior of FGCM models with sigmoid functions and proposed sufficient conditions for the existence and uniqueness of fixed point attractors \cite{Concepcion2020}. To summarized, the conditions for the existence of fixed points of FGCM are:

\begin{thm}
  \label{thmfgcm}
  If  $\otimes \mathbf{W} $ is the extended weight matrix of a FGCM, which includes possible feedback, where the weights $\otimes w_{ij} = \left[\underline{w_{ij}},\overline{w_{ij}}\right]$ are either nonnegative or nonpositive interval grey numbers, and define $w_{ij}^{*}$ as the elements of matrix $\mathbf{W}^*$ as follow:
  \begin{equation}
    \label{wstar}
    w_{i j}^{*}=\left\{\begin{array} {c c} {{ \left  \vert \underline{w_{ij}} \right \vert}} & \mathrm{if} \,\, \underline{w_{ij}}\leqslant \overline{w_{ij}} \leqslant 0 \\ 
      &\\
      \overline{w_{ij}}& \mathrm{if} \,\, 0 \leqslant \underline{w_{ij}}\leqslant \overline{w_{ij}}   \\ \end{array} \right.
  \end{equation}
  
    Let $\lambda > 0$ be the parameter of the sigmoid function used in the iterative process. If the condition 
$$\left\lVert \mathbf{W}^* \right\rVert _F < \frac{4}{\lambda}$$
is met, then the FGCM will have a unique grey fixed point, regardless of the initial concept values.

Furthermore,   Let $\lambda > 0$ be the parameter of the tanh function used in the iterative process. If the condition 
$$\left\lVert \mathbf{W}^* \right\rVert _F < \frac{1}{\lambda}$$
is met, then the FGCM will have a unique grey fixed point, regardless of the initial concept values.
\end{thm}

The convergence characteristics of FCMs can be utilized to guide the design of FCMs. The convergence of an FCM to a fixed point often corresponds to the system gradually converging to a fixed state. For instance, Peng utilized FCMs based on steady-state rules to model and analyze the Sanjiangyuan ecosystem, providing a reference for the sustainable evolution of the ecosystem \cite{Peng2016}. Additionally, Behrooz designed an air conditioning system controller that reduces energy consumption while meeting system requirements by leveraging the simple and easily convergent nature of FCMs \cite{Behrooz2019}. More recently, the convergence characteristics of FCMs have been applied to the intelligentization of tennis teaching \cite{Song2024}. For FCMs, a convergence to a limit cycle typically signifies that the entire system undergoes periodic state transitions without external intervention. For example, Biloslavo successfully simulated the periodic nature of climate change awareness and environmental impact using an FCM with 28 nodes \cite{Biloslavo2012}. However, it is crucial to avoid FCMs that iterate into chaotic states.

The convergence characteristics of FCMs have also been applied to the design of FCM learning algorithms. To avoid FCMs converging to unwanted steady states, Papageorgiou et al. developed an unsupervised learning algorithm based on nonlinear Hebbian rules, known as Active Hebbian Learning \cite{Papageorgiou2005}. In 2014, they introduced a learning algorithm for FCMs based on cultural algorithms, which constructs the weight matrix to allow the FCM algorithm to reach its final steady state \cite{Ahmadi2014}. Additionally, they presented a method that includes supervised and unsupervised learning algorithms to calculate causal weights, aiming to optimize the network topology in large FCMs and enhance the global convergence of continuous FCMs. Altundoğan developed a FCM PSO method that considers the initial vector state of the system, the weights between vector nodes, and the desired steady state vector \cite{Altundogan2018}. In 2012, Motlagh et al. developed a system for a simple hexapod walking expert model that does not rely on weight training to avoid FCMs converging to unwanted steady states. Nápoles et al. considered setting a separate activation function for each node of the FCM to improve its convergence characteristics in 2014, and continued to refine this method in subsequent studies, enhancing the efficiency and convergence speed of the learning algorithm \cite{Napoles2016, Napoles2017, Napoles2018}.

The above studies present the convergence conditions of FCM and FCGM, the specific applications of FCM under different convergent state, and the learning algorithms developed based on the convergence of FCM. However, up to now, there are no literatures conducting relevant research on the convergence of FGGCM, and there are also no related research results on the applications of the convergence of FGCM and FGGCM and the learning algorithms inspired thereby. In this paper, the completeness of the general grey number space is first proved through rigorous demonstration. On this basis, the convergence conditions of FGGCM are clearly given, and a detailed comparative analysis is conducted with the convergence conditions of FCM and FCGM. 

\section{Basics of FGGCM} \label{sec:basicsFGGCM}

FGGCM serves as an extension of FGCM, its primary objective is to enhance the uncertainty modeling capacity of FGCM. This is achieved by leveraging the broader range of uncertainty processing capabilities inherent in Grey System Theory, which were not fully utilized by FGCM. FGGCM adopts the general grey number (GGN) as its fundamental building block, as opposed to the interval grey number (IGN) used in FGCM.

\subsection{The General Grey Number}
The GGN is an amalgamation of IGN, defined as $g^\pm \in \bigcup\limits^n_{i=1}[\underline{a}_i,\overline{a}_i]$. For instance, $[0,1.2]\cup [1.5,2] \cup [3,5]\cup 6$ constitutes a GGN, while $[0,1]$ not only serves as an IGN but also represents a special case of a GGN. The mathematical operations between GGNs and IGNs is different. IGNs utilize their upper and lower bounds for mathematical operations \cite{Qiao2024}, whereas GGNs rely on their kernel and greyness. Both GGNs and IGNs possess a kernel that signifies the most probable crisp value within the grey number's range. The kernel of an IGN can be computed as the expectation of the crisp value, while the kernel of a GGN is determined using Eqs. \eqref{uniong} and \eqref{unionp} \cite{Jiang2021}.
\begin{equation}
  \hat{g}=\frac{1}{n}(\sum \limits^{n}_{i=1}\hat{a_i}).
  \label{uniong}
\end{equation}
\begin{equation}
  \hat{g}=\sum \limits^{n}_{i=1}p_i\hat{a_i}.
  \label{unionp}
\end{equation}
In Eqs. \eqref{uniong} and \eqref{unionp}, $\hat{a_i}$ represents the kernel of the interval $[\underline{a}_i,\overline{a}_i]$. It is calculated as the expectation of the crisp value. When the distribution of the GGN is unknown, Eq. \eqref{uniong} can be employed to calculate the kernel of the GGN. Conversely, if the distribution of a GGN is known, and the probability that $g^\pm$ falls within the interval $[\underline{a}_i,\overline{a}_i]$ is $p_i$ for $i=1,2,\cdots,n$, with $\sum^{n}_{i=1}p_i=1$ and $p_i>0$ for $i=1,2,\cdots,n$, then Eq. \eqref{unionp} can be used to compute the kernel of the GGN.

Given the domain $\Omega$ and the measure $\mu(\cdot)$, the greyness degree of a GGN $g^\pm$ is determined by the following expression:
\begin{equation}
  g^\circ (g^\pm)=\frac{1}{\hat{|g|}}\sum \limits^{n}_{i=1}|\hat{a_i}|\mu(\otimes_i)/\mu(\Omega).
  \label{union}
\end{equation}
This greyness value quantifies the degree of uncertainty associated with the GGN $g^\pm$. For convergence, the greyness degree will be called greyness in the following text.

The representation of a GGN using its kernel and greyness is given by the simplified form $\hat{g}_{g^\circ}$, where $\hat{g}$ represents the kernel and $g^\circ$ signifies the greyness. A GGN is also can be represent as $g^\pm = \hat{g}_{g^\circ}$ or $\otimes g = \hat{g}_{g^\circ}$, while a IGN is often represent as $\otimes g = [\underline{g},\overline{g} ]$ The operating rules for GGNs are as follows:

\begin{itemize}
  \item Equality: Two GGNs are equal if and only if their kernels are equal and their greyness values are the same:
  $$
  \hat{g_1}_{(g_1^\circ)} = \hat{g_2}_{(g_2^\circ)} \Longleftrightarrow \hat{g_1} = \hat{g_2} \quad \text{and} \quad g_1^\circ = g_2^\circ
  $$
  \item Addition: The addition of two GGNs is calculated as the sum of their kernels, with the greyness values weighted according to the magnitudes of the kernels:
  $$
  \hat{g_1}_{(g_1^\circ)} + \hat{g_2}_{(g_2^\circ)} = (\hat{g_1} + \hat{g_2})_{(w_1g_1^\circ + w_2g_2^\circ)}
  $$
  where:
  $$
  w_1 = \frac{|\hat{g_1}|}{|\hat{g_1}| + |\hat{g_2}|}, \quad w_2 = \frac{|\hat{g_2}|}{|\hat{g_1}| + |\hat{g_2}|}
  $$
\item Subtraction: The subtraction of one GGN from another is performed in a similar manner:
$$
\hat{g_1}_{(g_1^\circ)} - \hat{g_2}_{(g_2^\circ)} = (\hat{g_1} - \hat{g_2})_{(w_1g_1^\circ + w_2g_2^\circ)}
$$
where:
$$
w_1 = \frac{|\hat{g_1}|}{|\hat{g_1}| + |\hat{g_2}|}, \quad w_2 = \frac{|\hat{g_2}|}{|\hat{g_1}| + |\hat{g_2}|}
$$
\item Scalar Multiplication: Multiplying a GGN by a scalar preserves the greyness and adjusts the kernel:
$$
k \cdot \hat{g}_{(g^\circ)} = (k \cdot \hat{g})_{(g^\circ)}, \quad k \in \mathbb{R}
$$
\item Multiplication: The multiplication of two GGNs involves the product of their kernels, with the greyness values taken from the maximum of the two greyness values:
$$
\hat{g_1}_{(g_1^\circ)} \times \hat{g_2}_{(g_2^\circ)} = (\hat{g_1} \times \hat{g_2})_{(\max(g_1^\circ, g_2^\circ))}
$$
\item Division: Dividing one GGN by another is similar to multiplication, with the greyness value taken from the maximum of the two greyness values:
$$
\frac{\hat{g_1}_{(g_1^\circ)}}{\hat{g_2}_{(g_2^\circ)}} = \left(\frac{\hat{g_1}}{\hat{g_2}}\right)_{(\max(g_1^\circ, g_2^\circ))}, \quad (\hat{g_2} \neq 0)
$$
\item Inverse: The inverse of a non-zero GGN is obtained by taking the reciprocal of its kernel:
$$
\frac{1}{\hat{g_1}_{(g_1^\circ)}} = \left(\frac{1}{\hat{g_1}}\right)_{(g_1^\circ)}, \quad (\hat{g_1} \neq 0)
$$
\item Power: Raising a GGN to a power preserves the greyness and adjusts the kernel:
$$
(\hat{g}_{(g^\circ)})^k = (\hat{g})^k_{(g^\circ)}, \quad k \in \mathbb{R}
$$
\end{itemize}

These rules provide a framework for performing arithmetic operations on GGNs, which are essential for modeling and analyzing uncertainty in grey system theory.
It is noteworthy that the set $\mathbb{G}$ of all general grey numbers $g^\pm = \hat{g}_{(g^\circ)}$ does not constitute a linear space. For any $g^\pm \in \mathbb{G}$, it is not necessarily the case that there exists a $-g^\pm$ such that $g^\pm + (-g^\pm) = 0$. This is because the inequality $\left(g^\pm + (-g^\pm)\right)^\circ \geqslant g^\circ$ necessarily holds. Only when $g^\circ = 0$ does $g^\pm + (-g^\pm) = 0$ hold. In all other cases, $g^\pm + (-g^\pm) = 0$ does not hold, hence $\mathbb{G}$ is not a linear space.

\subsection{The Activation Functions of FGGCM}

The FGGCM adopts the GGN as its fundamental element, replacing the IGN used in the original FGCM. This substitution extends the FGGCM's ability to model uncertainty, as GGNs are capable of representing a broader range of uncertainty than IGNs. The sigmoid and tanh activation functions, which have been validated in \cite{Chen2020}, are employed in the FGGCM, as shown in Eqs. \eqref{ggsigmoid} and \eqref{ggtanh}.

The sigmoid function, as referenced in \cite{Chen2020}, is defined as follows:
\begin{equation}
  \label{ggsigmoid}
  S(g^\pm)=\frac{1}{1+\me^{-\lambda g^\pm}}.
\end{equation}

 $S(g^\pm)$ is a new GGN, the kernel is
\begin{equation}
  \hat {S}(g^\pm)=\frac{1}{1+\me^{-\lambda \hat {g}}},
  \label{sk}
\end{equation}
the greyness is
\begin{equation}
  S^\circ(g^\pm)= \frac{1}{1+\me^{-\lambda \hat{g}}}g^\circ.
  \label{gsigmoid}
\end{equation}

The hyperbolic tangent function in the context of GGNs is defined as:
\begin{equation}
  \label{ggtanh}
  \tanh(g^\pm) = \frac{\me^{\lambda g^\pm}-\me^{-\lambda g^\pm}}{\me^{\lambda g^\pm}+\me^{-\lambda g^\pm}}.
\end{equation}
The kernel is 
\begin{equation}
  \tanh(\hat g) = \frac{\me^{\lambda \hat g}-\me^{-\lambda \hat g}}{\me^{\lambda \hat g}+\me^{-\lambda \hat g}},
  \label{tk}
\end{equation}
the greyness is
\begin{equation}
  (\tanh(g^\pm))^\circ=g^\circ 
  \label{gtanh}
\end{equation}

Like Eq. \eqref{sigmoid} and Eq. \eqref{tanh}, in Eq. \eqref{ggsigmoid} to Eq. \eqref{ggtanh}, the $\lambda$ is a positive parameter to set the steepness of the curve or surface. Using these two activation functions, FGGCM can carry out reasoning within the range of node values between $[0, 1]$ or $[-1, 1]$. 

\section{The Completeness of Related Spaces} \label{sec:completeness}
This paper employs the Banach fixed point theorem to deduce the convergence conditions of FGGCM.

\begin{lem}
  \label{Banach}
  (Banach Fixed Point Theorem)
Consider a complete metric space $X$ and a mapping $T: X \rightarrow X$ that satisfies the contraction property:
\[ d(Tx, Ty) \leqslant k \cdot d(x, y) \]
for all $x, y \in X$, where $0 \leqslant k < 1$, $T$ is referred to as a contraction mapping.
 Then, there exists a unique point $x^* \in X$ such that:
\[ T(x^*) = x^* \]
This point $x^*$ is known as the fixed point of $T$.
\end{lem}

It can be observed that, in order to explore the convergence conditions of FGGCM, it is first necessary to ensure that the space under discussion is complete. This section first provides a metric on $\mathbb{G}$ and proves its completeness. On this basis, a metric on $\mathbb{G}^n$ is given, and its completeness is also proved.

\subsection{The Completeness of $\mathbb{G}$ }

To begin,  define a metric $d_2$ on the set $\mathbb{G}$.

\begin{thm}
  Given elements $g_1^\pm = \hat{g_1}_{(g_1^\circ)}$ and $g_2^\pm = \hat{g_2}_{(g_2^\circ)}$ in $\mathbb{G}$, let
  $$d_2(g_1^\pm,g_2^\pm ) = \left( \left|\hat{g_1} - \hat{g_2} \right|^2 + \left| g_1^\circ - g_2^\circ \right|^2 \right)^{\frac{1}{2}}. $$
Then $d_2$ is a metric on $\mathbb{G}$, and the pair $(\mathbb{G},d_2)$ is referred to as the general grey number metric space.
\end{thm}

\begin{pf}
  \begin{enumerate}[(1)]
    \item Non-negativity: $\forall g_1^\pm, g_2^\pm \in \mathbb{G},$ 
    $$
     d_2(g_1^\pm, g_2^\pm) = \left( |\hat{g_1} - \hat{g_2}|^2 + |g_1^\circ - g_2^\circ|^2 \right)^{\frac{1}{2}} \geqslant 0 
    $$
    This property is evidently satisfied, as the square root of a sum of squares is always non-negative.
    \item Non-degeneracy:
    If $g_1^\pm, g_2^\pm \in \mathbb{G}$, then 
    $$
    d_2(g_1^\pm, g_2^\pm) = \left( |\hat{g_1} - \hat{g_2}|^2 + |g_1^\circ - g_2^\circ|^2 \right)^{\frac{1}{2}} = 0 
    $$
    only when $ \hat{g_1} = \hat{g_2} $ and $ g_1^\circ = g_2^\circ $, that is, when $g_1^\pm = g_2^\pm$.
    \item Symmetry:
    $$
    \forall g_1^\pm, g_2^\pm \in \mathbb{G}, \quad d_2(g_1^\pm, g_2^\pm) = d_2(g_2^\pm, g_1^\pm) 
    $$
    This property is clearly satisfied.

    \item Triangle Inequality:

    For all $g_1^\pm, g_2^\pm, g_3^\pm \in G$, by the Minkowski inequality:
    \begin{equation}
    \begin{split}
      \left(\sum_{n=1}^{\infty}\mid x_{n}+y_{n}\mid^{p}\right)^{\frac{1}{p}}
      \leqslant \left(\sum_{n=1}^{\infty}\mid x_{n}\mid^{p}\right)^{\frac{1}{p}}+\left(\sum_{n=1}^{\infty}\mid y_{n}\mid^{p}\right)^{\frac{1}{p}},
    \end{split}
    \end{equation}
    where $1<p<\infty$, $x_n,y_n \in \mathbb{R}$, it can be derived:
    \begin{equation*}
      \begin{split}
        d_{2}(g_1^\pm ,g_2^\pm) 
        = &\,{\left(\left|\hat{g_1} - \hat{g_2} \right|^2 + \left| g_1^\circ - g_2^\circ \right|^2\right)}^{1/2}\\
        = & \left(\left| (\hat{g_1} -\hat{g_3}) + (\hat{g_3}- \hat{g_2} )\right|^2 \right. \\
        & + \left. \left| (g_1^\circ - g_3^\circ)+ (g_3^\circ - g_2^\circ)\right|^2\right)^{1/2} \\
         \leqslant &  \;\left(\,\left|\hat{g_1} - \hat{g_3} \right|^2 + \left| g_1^\circ - g_3^\circ \right|^2\,\right)^{\frac{1}{2}}\\
         &+\left(\,\left|\hat{g_3} - \hat{g_2} \right|^2 + \left| g_3^\circ - g_2^\circ \right|^2\,\right)^{\frac{1}{2}}\,\\
        =&\,d_{2}(g_1^\pm ,g_3^\pm)\,+\,d_{2}(g_3^\pm ,g_2^\pm) 
      \end{split}
    \end{equation*}
    This demonstrates that the triangle inequality is satisfied, as the sum of the distances between $g_1^\pm$ and $g_2^\pm$, is less than or equal to the distance between $g_1^\pm$ and $g_3^\pm$, plus the distance between $g_3^\pm$ and $g_2^\pm$.
  \end{enumerate}

Therefore, $d_2$ is a metric on $\mathbb{G}$, and $(\mathbb{G},d_2)$ constitutes a metric space.
\end{pf}

The following illustrates the completeness of the general grey number space $(\mathbb{G},d_2)$.

\begin{defi}
  Let $(X,d)$ be a metric space, and let $\{x_n\}$ be a sequence in $X$. We call $\{x_n\}$ a Cauchy sequence if, for any $\epsilon > 0$, there exists an integer $N \geq 1$ such that for all $m, n \geq N$, the following inequality holds: $d(x_m, x_n) < \epsilon$. A metric space $(X,d)$ is called complete if every Cauchy sequence in $X$ is convergent.
\end{defi}

\begin{lem}
 The real number space $(\mathbb{R},d)$ is complete, where $d = \left\lvert x - y\right\rvert$ for all $x, y \in \mathbb{R}$.
\end{lem}

\begin{thm}
 The general grey number metric space $(\mathbb{G},d_2)$ is complete.
\end{thm}

\begin{pf}
  Let $\{\bm{g_n^\pm}\} = g_1^\pm, g_2^\pm, \cdots, g_n^\pm$ be a Cauchy sequence in the general grey number metric space $(\mathbb{G},d_2)$. Then, for any $\epsilon > 0$, there exists an integer $N \geq 1$ such that for all $m, k \geq N$, the following inequalities hold:
$$ \left( \left|\hat{g_m} - \hat{g_k} \right|^2 + \left| g_m^\circ - g_k^\circ \right|^2 \right)^{\frac{1}{2}} < \frac{\epsilon}{2} $$
This implies:
\begin{equation}
  \label{kernel_cauchy}
  \left|\hat{g_m} - \hat{g_k} \right| < \frac{\epsilon}{2}
\end{equation}
and
\begin{equation}
  \label{greyness_cauchy}
  \left| g_m^\circ - g_k^\circ \right| < \frac{\epsilon}{2}
\end{equation}
These inequalities indicate that the sequences $\{\bm{\hat{g_n}}\} = \hat{g_1}, \hat{g_2}, \cdots, \hat{g_n}$ and $\{\bm{g_n^\circ}\} = g_1^\circ, g_2^\circ, \cdots, g_n^\circ$ are Cauchy sequences in the real number space $(\mathbb{R},d)$. Since $(\mathbb{R},d)$ is complete, there exist real numbers $\hat{g} \in \mathbb{R}$ and $g^\circ \in \mathbb{R}$, respectively, such that $\hat{g_n} \rightarrow \hat{g}$ and $g_n^\circ \rightarrow g^\circ$. Let
$$ g^\pm = \hat{g}_{g^\circ} \in \mathbb{G} $$
In the inequalities (\ref{kernel_cauchy}) and (\ref{greyness_cauchy}), fix $m \geq N$ and let $k \rightarrow \infty$. There are
$$ \left|\hat{g_m} - \hat{g} \right| < \frac{\epsilon}{2} < \epsilon $$
and
$$ \left| g_m^\circ - g^\circ \right| < \frac{\epsilon}{2} < \epsilon. $$
This is equivalent to saying that for $m \geq N$,
$$ d_2(g_m^\pm, g^\pm) = \left( \left|\hat{g_m} - \hat{g} \right|^2 + \left| g_m^\circ - g^\circ \right|^2 \right)^{\frac{1}{2}} < \epsilon $$
Thus, $g_m^\pm \rightarrow g^\pm$. This demonstrates that the general grey number metric space $(\mathbb{G},d_2)$ is complete.

\end{pf}

\subsection{The Completeness of $\mathbb{G}^n$ }

Similar to the proof of the completeness of $\mathbb{G}$, it can be defined a metric on $\mathbb{G}^n$ and then demonstrate its completeness.

\begin{thm}
  Given elements $\bm{x^\pm} = (x_1^\pm, x_2^\pm, \cdots, x_n^\pm) =$ $ (\hat{x_1}_{x_1^\circ}\hat{x_2}_{x_2^\circ}, \cdots, \hat{x_n}_{x_n^\circ})$ and
  $\bm{y^\pm} = (y_1^\pm,y_2^\pm,$ $\cdots, y_n^\pm) = (\hat{y_1}_{y_1^\circ },\hat{y_2}_{y_2^\circ },\cdots, \hat{y_n}_{y_n^\circ }) $ in $\mathbb{G}^n$, define the metric $d$ on $\mathbb{G}^n$ as follows:
  \begin{equation}
    \begin{split}
      d(\bm{x^\pm, y^\pm}) = \left(\sum_{i=1}^{n} d_2^2(x_i^\pm, y_i^\pm) \right)^{\frac{1}{2}} = \left(\sum_{i=1}^{n} \left( \left|\hat{x_i} - \hat{y_i} \right|^2 + \left| x_i^\circ - y_i^\circ \right|^2 \right) \right)^{\frac{1}{2}}
    \end{split} 
  \end{equation}
  where $d_2$ is the metric defined on the individual components of the vectors in $\mathbb{G}$. Therefore, $d$ is a metric on $\mathbb{G}^n$, and the pair $(\mathbb{G}^n, d)$ is referred to the GGN vector metric space.
\end{thm}

\begin{pf}
  The non-negativity, non-degeneracy, and symmetry of the metric $d$ on $\mathbb{G}^n$ are easily verified. Thus, it is necessarily to demonstrate that $d$ satisfies the triangle inequality.
For any $\bm{x^\pm}, \bm{y^\pm}, \bm{z^\pm} \in \mathbb{G}^n$, since $d_2$ satisfies the triangle inequality, and from the Minkowski inequality, there is
\begin{equation*}
  \begin{split}
    d(\bm{x^\pm,y^\pm}) &= \left(\sum_{i=1}^{n}d^2_2(x_i^\pm,y_i^\pm)\right)^{\frac{1}{2}} \\
    & \leqslant \left(\sum_{i=1}^{n}\left(d_2(x_i^\pm,z_i^\pm) + d_2(z_i^\pm,y_i^\pm) \right)^2 \right)^{\frac{1}{2}} \\
    & \leqslant \left(\sum_{i=1}^{n}d^2_2(x_i^\pm,z_i^\pm)\right)^{\frac{1}{2}} + \left(\sum_{i=1}^{n}d^2_2(y_i^\pm,z_i^\pm)\right)^{\frac{1}{2}} \\
    & = d(\bm{x^\pm,z^\pm}) + d(\bm{z^\pm,y^\pm})
  \end{split}
\end{equation*}
Therefore, $d$ is a metric on $\mathbb{G}^n$, and the pair $(\mathbb{G}^n, d)$ constitutes a metric space.
\end{pf}

\begin{thm}
The general grey number metric space $(\mathbb{G}^n,d)$ is complete.
\end{thm}

\begin{pf}
  Let $\bm{g}^{\pm(m)} = (g_1^{\pm(m)}, g_2^{\pm(m)}, \cdots, g_n^{\pm(m)})$ be a sequence in $\mathbb{G}^n$, and let $\{\mathbf{g^{\pm(m)}}\}$ be a Cauchy sequence in the metric space $(\mathbb{G}^n, d)$. Then, for any $\epsilon > 0$, there exists an integer $N \geq 1$ such that for all $m, k \geq N$, the following inequality holds:
$$ d(\bm{g}^{\pm(m)},\bm{g}^{\pm(k)}) < \frac{\epsilon}{2\sqrt{n}} $$
This implies:
$$ \left(\sum_{i=1}^{n}d_2^2(g_i^{\pm(m)},g_i^{\pm(k)})\right)^{\frac{1}{2}} < \frac{\epsilon}{2\sqrt{n}} $$
Thus, for any $1 \leq i \leq n$, we have:
\begin{equation}
  \label{vector_e}
  d_2(g_i^{\pm(m)},g_i^{\pm(k)} ) < \frac{\epsilon}{2\sqrt{n}} 
\end{equation}
This indicates that $\{g_i^{\pm(m)}\}$ is a Cauchy sequence in the metric space $(\mathbb{G}, d_2)$. Utilizing the completeness of $(\mathbb{G}, d_2)$, there exists a sequence $\{g_i^{\pm(m)}\}$ that converges to $g_i^{\pm} \in \mathbb{G}$. Let
$$ \bm{g^\pm} = (g_1^\pm, g_2^\pm, \cdots, g_n^\pm) \in \mathbb{G}^n $$
In the inequality (\ref{vector_e}), fix $m \geq N$ and let $k \rightarrow \infty$. We have:
$$ d_2(g_i^{\pm(m)}, g_i^{\pm}) < \frac{\epsilon}{2\sqrt{n}} < \epsilon $$
This is equivalent to saying that for $m \geq N$,
$$ d(\bm{g}^{\pm(m)},\bm{g}^{\pm}) = \left(\sum_{i=1}^{n}d_2^2(g_i^{\pm(m)},g_i^{\pm})\right)^{\frac{1}{2}} < \frac{\epsilon}{2} < \epsilon $$
Thus, $\bm{g_m^\pm} \rightarrow \bm{g^\pm}$. This demonstrates that the general grey number metric space $(\mathbb{G}^n, d)$ is complete.
\end{pf}

\section{Convergence of FGGCM} \label{sec:convergence}

The common method for finding fixed points is iterative methods. Starting from an arbitrary fixed initial point $ x_0 \in X $, define $ x_1 = T x_0, x_2 = T x_1 = T^2 x_0, \cdots, x_n = T x_{n-1} = T^n x_0 $, $T$ is a contraction mapping. It can be observed that this iterative process for the initial point is similar to the reasoning process of FCM and the FGGCM. Therefore, examining the convergence conditions of FGGCM is equivalent to examining under what conditions FGGCM is a contraction mapping. This section will separately illustrate the sufficient conditions for FGGCM to converge to fixed points when the activation function is either the tanh or sigmoid function.

\subsection{The Convergence of tanh FGGCM}
First, present a lemma to illustrate a property of the tanh function.
\begin{lem}
  \label{tanh_bound}
  Given the function $f(x) = \tanh(x) = \frac{\me^{\lambda x} - \me^{-\lambda x}}{\me^{\lambda x} + \me^{-\lambda x}}$, where $x \in \mathbb{R}$ and $\lambda > 0$, it follows that for any $a, b \in \mathbb{R}$, there is necessarily:
\begin{equation*}
  |f(b) - f(a)| \leqslant \lambda|(b - a)|
\end{equation*}
\end{lem}

\begin{pf}
  By differentiating $f(x)$, it obtains:
\begin{equation}
  \label{derivative}
  f'(x) = \lambda (1 - (f(x))^2) \leqslant \lambda
\end{equation}
The equality holds if and only if $f(x) = 0$.
According to the Lagrange Mean Value Theorem, if a function $f(x)$ is continuous on the closed interval $[a, b]$ and differentiable on the open interval $(a, b)$, then there exists at least one point $\xi$ in the open interval $(a, b)$ such that:
$$
f(b) - f(a) = f'(\xi)(b - a)
$$
Substituting Eq. \eqref{derivative} into the Lagrange Mean Value Theorem, we obtain:
$$
|f(b) - f(a)| \leqslant \lambda|(b - a)|
$$

\end{pf}

\begin{thm}
  \label{thmFGGCMtanh}
  Given that $\mathbf{W^\pm} = \mathbf{\hat{W}_{W^\circ}}$ is the weight matrix of FGGCM, $w^\pm_{ij} = \hat{w}{_{ij}}_{w_{ij}^\circ} \in \mathbb{G}$ are the elements of $\mathbf{W^\pm}$, and $\lambda > 0, \lambda \in \mathbb{R}$ is the parameter of the tanh activation function of the FGGCM model with $n$ nodes,
$$
\tanh(g^\pm) = \frac{\me^{\lambda g^\pm} - \me^{-\lambda g^\pm}}{\me^{\lambda g^\pm} + \me^{-\lambda g^\pm}}
$$
the FGGCM will converge to a unique fixed point if:
\begin{equation}
  \label{tanhkernelconvergence}
  \left(\sum_{i=1}^n\sum_{j=1}^n \hat{w}_{ij}^2 \right)^\frac{1}{2} < \frac{1}{\lambda}
\end{equation}
and
\begin{equation}
  \label{tanhgreynessconvergence}
    \left(\sum_{i=1}^{n} \sum_{j=1}^{n}\left(\frac{\left|{\hat{A}_{j} \hat{w}_{ij}}\right|\theta\left(C_j^\circ - \hat{w}_{ij}^\circ\right) }{\sum_{j=1}^{n}\left|\hat{w}_{ij}\hat{A}_j\right|}\right)^2\right)^{\frac{1}{2}} < 1.
\end{equation}
where, $C_j^\circ$ is the greyness of the node $C_j$ in FGGCM at any iteration step. $\theta(x)$ is Heaviside function, defined as: 
\[
\theta(x) = \begin{cases}
0 & \text{if } x < 0 \\
1 & \text{if } x \geq 0
\end{cases}
\]

Specifically, if \begin{equation}
  \label{tanheqkernel}
  \left(\sum_{i=1}^n\sum_{j=1}^n \hat{w}_{ij}^2 \right)^\frac{1}{2} = \frac{1}{\lambda} 
\end{equation}
and  \begin{equation}
  \label{tanheqgreyness}
  \left(\sum_{i=1}^{n} \sum_{j=1}^{n}\left(\frac{\left|{\hat{A}_{j} \hat{w}_{ij}}\right|\theta\left(C_j^\circ - \hat{w}_{ij}^\circ\right) }{\sum_{j=1}^{n}\left|\hat{w}_{ij}\hat{A}_j\right|}\right)^2\right)^{\frac{1}{2}} = 1
\end{equation} hold, then FGGCM must have at least one fixed point.

\end{thm}
\begin{pf}
  $\forall \bm{A,A'} \in \mathbb{G}^n$, calculate $d^2(f(\mathbf{W^\pm} \bm A),f(\mathbf{W^\pm} \bm A'))$ as Eq. \eqref{dTAA},
    \begin{equation}
     \label{dTAA}
    %  \footnotesize
   \begin{split}
     &d^2(f(\mathbf{W^\pm} \bm A),f(\mathbf{W^\pm} \bm A'))\\ = &\sum_{i=1}^n d_2^2(f(\bm{W_iA}),f(\bm{W_iA'}))\\
     =&\sum_{i=1}^n d_2^2 \left(\left(\frac{\me^{\lambda \bm{\hat{W_i} \hat{A}}}-\me^{-\lambda \bm{\hat{W_i} \hat{A}}}}{\me^{\lambda \bm{\hat{W_i} \hat{A}}}+\me^{-\lambda \bm{\hat{W_i} \hat{A}}}}\right)_{\bm{(W_iA)^\circ}}, \left(\frac{\me^{\lambda \bm{\hat{W_i} \hat{A'}}}-\me^{-\lambda \bm{\hat{W_i} \hat{A'}}}}{\me^{\lambda \bm{\hat{W_i} \hat{A'}}}+\me^{-\lambda \bm{\hat{W_i} \hat{A'}}}}\right)_{\bm{(W_iA')^\circ}}\right)\\
     =& \sum_{i=1}^n \left(\left(\frac{\me^{\lambda \bm{\hat{W_i} \hat{A}}}-\me^{-\lambda \bm{\hat{W_i} \hat{A}}}}{\me^{\lambda \bm{\hat{W_i} \hat{A}}}+\me^{-\lambda \bm{\hat{W_i} \hat{A}}}} - \frac{\me^{\lambda \bm{\hat{W_i} \hat{A'}}}-\me^{-\lambda \bm{\hat{W_i} \hat{A'}}}}{\me^{\lambda \bm{\hat{W_i} \hat{A'}}}+\me^{-\lambda \bm{\hat{W_i} \hat{A'}}}} \right)^2 + \left(\bm{(W_iA)^\circ}-\bm{(W_iA')^\circ}\right)^2\right) \\
     = & \sum_{i=1}^n \left(\frac{\me^{\lambda \bm{\hat{W_i} \hat{A}}}-\me^{-\lambda \bm{\hat{W_i} \hat{A}}}}{\me^{\lambda \bm{\hat{W_i} \hat{A}}}+\me^{-\lambda \bm{\hat{W_i} \hat{A}}}} - \frac{\me^{\lambda \bm{\hat{W_i} \hat{A'}}}-\me^{-\lambda \bm{\hat{W_i} \hat{A'}}}}{\me^{\lambda \bm{\hat{W_i} \hat{A'}}}+\me^{-\lambda \bm{\hat{W_i} \hat{A'}}}} \right)^2 + \sum_{i=1}^n \left(\bm{(W_iA)^\circ}-\bm{(W_iA')^\circ}\right)^2\\
     \leqslant &  \sum_{i=1}^n \lambda^2 \left(\bm{\hat{W_i} \hat{A}} - \bm{\hat{W_i} \hat{A'}} \right)^2 + \sum_{i=1}^n \left(\left(\sum_{j=1}^{n} w_{ij}A_j\right)^\circ - \left(\sum_{j=1}^{n} w_{ij}A_j'\right)^\circ\right)^2\\
     = &\lambda^2 \sum_{i=1}^n \left(\sum_{j=1}^n \hat{w}_{ij}\hat{A}_j - \sum_{j=1}^n \hat{w}_{ij}\hat{A'}_j \right)^2   +GR \\
     = & \lambda^2 \sum_{i=1}^n \left(\sum_{j=1}^n \hat{w}_{ij} \left(\hat{A}_j - \hat{A'}_j\right) \right)^2 +GR \\
     \leqslant & \lambda^2 \sum_{i=1}^n \left(\sum_{j=1}^n \hat{w}_{ij}^2 \right) \left(\sum_{j=1}^n \left(\hat{A}_j - \hat{A'}_j\right)^2 \right) +GR 
   \end{split}
   \end{equation}

where
\begin{equation*}
  \begin{split}
    GR=&\sum_{i=1}^n \left(\left(\sum_{j=1}^{n} w_{ij}A_j\right)^\circ - \left(\sum_{j=1}^{n} w_{ij}A_j'\right)^\circ\right)^2\\
    =&\sum_{i=1}^n \left(\sum_{j=1}^{n}\frac{\max(w_{ij}^\circ,A_j^\circ)\left|\hat{w}_{ij}\hat{A}_j\right|}{\sum_{j=1}^{n}\left|\hat{w}_{ij}\hat{A}_j\right|}- \sum_{j=1}^{n}\frac{\max(w_{ij}^\circ,{A'}_j^\circ)\left|\hat{w}_{ij}\hat{A'}_j\right|}{\sum_{j=1}^{n}\left|\hat{w}_{ij}\hat{A'}_j\right|}\right)^2 \\
  \end{split}
\end{equation*}

In Eq. (\ref{dTAA}), the first inequality is derived from Lemma \ref{tanh_bound}, while the second inequality is a consequence of the Cauchy-Schwarz inequality.
On the other hand,
\begin{equation}
  \label{dAA}
  \begin{split}
    d^2(\bm{A, A'}) = &\sum_{j=1}^n d_2^2(A_j, {A'}_j) \\
    =&\sum_{j=1}^n \left(\hat{A}_j - \hat{A'}_j\right)^2 + \sum_{j=1}^n \left(A_j^\circ - {A'}_j^\circ\right)^2
  \end{split}
\end{equation}

Observing Eqs. (\ref{dTAA}) and (\ref{dAA}), it can be found that the distance are divided into two parts: the kernel and the greyness. The convergence of the greyness does not affect the convergence of the kernel. Focusing on the kernel part, if
\begin{equation*}
\begin{split}
    \lambda^2 \sum_{i=1}^n \left(\sum_{j=1}^n \hat{w}_{ij}^2 \right) \left(\sum_{j=1}^n \left(\hat{A}_j - \hat{A'}_j\right)^2 \right)
     < \sum_{j=1}^n \left(\hat{A}_j - \hat{A'}_j\right)^2
\end{split}
\end{equation*}
holds, which is equivalent to the condition
\begin{equation*}
  \left(\sum_{i=1}^n\sum_{j=1}^n \hat{w}_{ij}^2 \right)^{\frac{1}{2}} < \frac{1}{\lambda}
\end{equation*}
from Lemma \ref{Banach}, then the kernel of FGGCM must converge to a unique fixed point.

Observing the greyness part, according to the reasoning process of the FGGCM when the activation function is tanh, the calculation method of the greyness at a certain iterative step is as follows: 
\begin{equation}
  \label{itertanh}
  \begin{split}
    {A^\circ_j}^{t + 1}=&f({A_1^\circ}^t,{A_2^\circ}^t,\cdots,{A_n^\circ}^t,\hat{A}_1^t,\hat{A}_2^t\cdots \hat{A}_n^t)\\
     = &\sum_{j=1}^{n}\frac{\max (w_{ij}^\circ,{A_j^\circ}^t)\left|\hat{w}_{ij}\hat{A}_j^t\right|}{\sum_{j=1}^{n}\left|\hat{w}_{ij}\hat{A}_j^t\right|}\\
  \end{split}
\end{equation}

According to the mean value theorem of multivariate functions, there must be: 
  \begin{equation}
    \label{multivariate}
    \begin{split}
      &f({A_1^\circ},{A_2^\circ},\cdots,{A_n^\circ},\hat{A}_1,\hat{A}_2\cdots \hat{A}_n) -  f({{A'}_1^\circ},{{A'}_2^\circ},\cdots,{{A'}_n^\circ},\hat{{A'}}_1,\hat{{A'}}_2\cdots \hat{{A'}}_n)\\
      =& \bigtriangledown f \cdot \bigtriangleup  \bm{A}\\
      =& \begin{pmatrix}
        \frac{\partial f}{\partial C_1^\circ} &\frac{\partial f}{\partial C_2^\circ} & \cdots&\frac{\partial f}{\partial C_n^\circ} &\frac{\partial f}{\partial \hat{C}_1}&\frac{\partial f}{\partial \hat{C}_2}&\cdots &\frac{\partial f}{\partial \hat{C}_n}\\
      \end{pmatrix} \cdot \\
      &\begin{matrix}
        ({A_1^\circ} - {{A'}_1^\circ} &{A_2^\circ} - {{A'}_2^\circ}& \cdots&{A_3^\circ} - {{A'}_3^\circ} &
        \hat{A}_1 - \hat{{A'}}_1&\hat{A}_2 - \hat{{A'}}_2&\cdots &\hat{A}_n - \hat{{A'}}_n ) ^\top
      \end{matrix}\\
      =& \sum_{j=1}^{n} \frac{\partial f}{\partial C_j^\circ} ({A_j^\circ} - {{A'}_j^\circ}) + \sum_{j=1}^{n} \frac{\partial f}{\partial \hat{C}_j} (\hat{A}_j - \hat{{A'}}_j)
    \end{split}
  \end{equation}
where, $\hat{C}_j = \hat{A}_j +t(\hat{{A'}}_j-\hat{A}_j)$, $C_j^\circ = {{A}_j^\circ}+t({{A'}_j^\circ} - {{A}_j^\circ})$,$0<t<1$.
It can be found that in the distance measurement between greyness, the kernel contributes a part of the distance. The partial derivative with respect to the greyness is calculated as: 
\begin{equation}
  \frac{\partial f}{\partial C_j^\circ}= \frac{\left|{\hat{C}_{j} w_{ij}}\right| \theta\left(C_j^\circ - \hat{w}_{ij}^\circ\right)}{\sum_{j=1}^{n}\left|\hat{w}_{ij}\hat{C}_j\right|}
\end{equation}

Suppose the kernel of the FGGCM has converged to a fixed point. Both $A$ and $A'$ are taken from the set after converging to the fixed point, then it is inevitable that $\hat{A}_j = \hat{{A'}}_j$ holds, thus, there is Eq. \eqref{greynesstanh} holds.

  \begin{figure*}[htbp]
    \begin{equation}
      \label{greynesstanh}
      \begin{split} 
        &\sum_{i=1}^n \left(\sum_{j=1}^{n}\frac{\max(w_{ij}^\circ,A_j^\circ)\left|\hat{w}_{ij}\hat{A}_j\right|}{\sum_{j=1}^{n}\left|\hat{w}_{ij}\hat{A}_j\right|}- \sum_{j=1}^{n}\frac{\max(w_{ij}^\circ,{A'}_j^\circ)\left|\hat{w}_{ij}\hat{A'}_j\right|}{\sum_{j=1}^{n}\left|\hat{w}_{ij}\hat{A'}_j\right|}\right)^2\\ 
        =&\sum_{i=1}^{n}(f({A_1^\circ},{A_2^\circ},\cdots,{A_n^\circ},\hat{A}_1,\hat{A}_2\cdots \hat{A}_n) - f({{A'}_1^\circ},{{A'}_2^\circ},\cdots,{{A'}_n^\circ},\hat{{A'}}_1,\hat{{A'}}_2\cdots \hat{{A'}}_n))\\
        =& \sum_{i=1}^{n} \left(\sum_{j=1}^{n}\frac{\partial f}{\partial C_j^\circ} ({A_j^\circ} - {{A'}_j^\circ})\right)^2\\
        =& \sum_{i=1}^{n} \left(\sum_{j=1}^{n}\frac{\left|{\hat{C}_{j} w_{ij}}\right| \theta\left(C_j^\circ - \hat{w}_{ij}^\circ\right)}{\sum_{j=1}^{n}\left|\hat{w}_{ij}\hat{C}_j\right|} ({A_j^\circ} - {{A'}_j^\circ})\right)^2\\
        \leqslant & \sum_{i=1}^{n} \sum_{j=1}^{n}\left(\frac{\left|{\hat{C}_{j} \hat{w}_{ij}}\right|\theta\left(C_j^\circ - \hat{w}_{ij}^\circ\right) }{\sum_{j=1}^{n}\left|\hat{w}_{ij}\hat{C}_j\right|}\right)^2\sum_{j=1}^{n} \left({A_j^\circ} - {{A'}_j^\circ}\right)^2 \\
      \end{split}
    \end{equation}
  \end{figure*}

Obviously, if
\begin{equation*}
  \sum_{i=1}^{n} \sum_{j=1}^{n}\left(\frac{\left|{\hat{C}_{j} \hat{w}_{ij}}\right|\theta\left(C_j^\circ - \hat{w}_{ij}^\circ\right) }{\sum_{j=1}^{n}\left|\hat{w}_{ij}\hat{C}_j\right|}\right)^2 < 1
\end{equation*}
holds, then the greyness of FGGCM will converge to a fixed point.

And because $\hat{A}_j = \hat{{A'}}_j$, that is, $\hat{C}_j = \hat{A}_j + t(\hat{{A'}}_j - \hat{A}_j) = \hat{A}_j$, then the above formula can be reduced to
\begin{equation*}
  \left(\sum_{i=1}^{n} \sum_{j=1}^{n}\left(\frac{\left|{\hat{A}_{j} \hat{w}_{ij}}\right|\theta\left(C_j^\circ - \hat{w}_{ij}^\circ\right) }{\sum_{j=1}^{n}\left|\hat{w}_{ij}\hat{A}_j\right|}\right)^2\right)^{\frac{1}{2}} < 1
\end{equation*}

If the inequalities \eqref{tanhkernelconvergence} and \eqref{tanhgreynessconvergence} were to transform into equalities, the convergence of FGGCM would be affected. This is established by the following Lemma \ref{Browder}.

\begin{lem}
  \label{Browder}
  (Browder-Gohde-Kirk Fixed Point Theorem)
  Let $X $ be a Banach space, and $C $ a closed, bounded, and convex subset of  $X $. If $U$ is a mapping of $C$  into itself, then $U$  is said to be nonexpansive maping, if for every pair of elements $ x, y$  in  $C$ , the following inequality holds:
  \[ \| Ux - Uy \| \leqslant \| x - y \| \]
  where $\left\lVert \cdot \right\rVert $ is a norm of $C$.

  Let $X $ be a uniformly convex Banach space, and  $U$ a nonexpansive mapping of the bounded, closed, convex subset  $C$  of  $X $ into  $C$ . Then  $U$ has at least one fixed point in  $C$ .
\end{lem}

According to this lemma, it can be concluded that under the condition where both $$\left(\sum_{i=1}^n\sum_{j=1}^n \hat{w}_{ij}^2 \right)^{1/2} = \frac{1}{\lambda}$$ and  $$\left(\sum_{i=1}^{n} \sum_{j=1}^{n}\left(\frac{\left|{\hat{A}_{j} \hat{w}_{ij}}\right|\theta\left(C_j^\circ - \hat{w}_{ij}^\circ\right) }{\sum_{j=1}^{n}\left|\hat{w}_{ij}\hat{A}_j\right|}\right)^2\right)^{\frac{1}{2}} = 1$$ 
 hold, the FGGCM at least has one fixed point. Furthermore, if Eq. \eqref{tanheqkernel} holds, then the kernel of FGGCM at least has one fixed point.

\end{pf}

From the above proof process, the following corollaries can be easily drawn:

\begin{corollary}
  \label{corollary_tanh1}
  If Inequality \eqref{tanhkernelconvergence} holds, then the kernel of the FGGCM node must converge to a unique fixed point; if Eq. \eqref{tanheqkernel} holds, then the kernel of FGGCM has at least one fixed point.
\end{corollary}

\begin{corollary}
  \label{corollary_tanh2}
  If the kernel of the FGGCM converges and inequality \eqref{tanhgreynessconvergence} holds, then the greyness of the FGGCM node must converge to a unique fixed point. On the other hand, if the kernel of the FGGCM converges and Eq. \eqref{tanheqgreyness} holds, then the greyness of the FGGCM node must have at least one fixed point.
\end{corollary}

For the tanh function, it is clear that $\tanh (\bm{0}) = \bm{0}$. Therefore, it can be easily deduced that:
   
\begin{corollary}
  \label{corollary_tanh3}
  If Inequality \eqref{tanhkernelconvergence} holds, then the kernels of the FGGCM must converge to a unique fixed point, and this fixed point is $\bm{0}$.
\end{corollary}

It is important to note that the premises of Corollary \ref{corollary_tanh1}, \ref{corollary_tanh2}, and \ref{corollary_tanh3} are that tanh is the activation function of the FGGCM.

Based on an in-depth analysis of the greyness iteration process in the FGGCM, the Theorem \ref{thm_tanh_grey_convergence} can be proved. Given that the Eq. \eqref{tanh_grey_convergence_eq} is essentially a special case of Eq. \eqref{tanhgreynessconvergence} in Theorem \ref{thmFGGCMtanh}, this discovery provides further evidence for the discriminant of greyness convergence in Eq. \eqref{tanhgreynessconvergence} of Theorem \ref{thmFGGCMtanh}.

\begin{thm}
  \label{thm_tanh_grey_convergence}
  If a FGGCM uses the hyperbolic tangent as the activation function, and for any iteration step, it holds that $\max (w_{ij}^\circ, A_j^\circ) = A_j^\circ$, the greyness of the FGGCM will have at least one fixed point. 

  Particularly, if \begin{equation} \label{tanh_grey_convergence_eq}
    \sum_{i=1}^{n} \sum_{j=1}^{n}\left(\frac{\left|\hat{A}_{j} \hat{w}_{ij}\right|}{\sum_{j=1}^{n}\left|\hat{w}_{ij}\hat{A}_j\right|}\right)^2 = {\left\lVert \mathbf{M}\right\rVert }^2_F <1
  \end{equation}
   is satisfied, then the greyness of the FGGCM will necessarily converge to a unique fixed point. This fixed point is a characteristic vector corresponding to the eigenvalue $1$ of the $\mathbf{M}$ or zero vector, where
   \begin{equation*}
     \mathbf{M} = 
     \begin{pmatrix}
       \frac{\left|\hat{w}_{11}\hat{A}_1^t\right|}{\sum_{j=1}^{n}\left|\hat{w}_{1j}\hat{A}_j^t\right|} & \frac{\left|\hat{w}_{12}\hat{A}_2^t\right|}{\sum_{j=1}^{n}\left|\hat{w}_{1j}\hat{A}_j^t\right|} & \cdots &  \frac{\left|\hat{w}_{1n}\hat{A}_n^t\right|}{\sum_{j=1}^{n}\left|\hat{w}_{1j}\hat{A}_j^t\right|}\\
       \frac{\left|\hat{w}_{21}\hat{A}_1^t\right|}{\sum_{j=1}^{n}\left|\hat{w}_{2j}\hat{A}_j^t\right|} & \frac{\left|\hat{w}_{22}\hat{A}_2^t\right|}{\sum_{j=1}^{n}\left|\hat{w}_{2j}\hat{A}_j^t\right|} & \cdots &  \frac{\left|\hat{w}_{2n}\hat{A}_n^t\right|}{\sum_{j=1}^{n}\left|\hat{w}_{2j}\hat{A}_j^t\right|}\\
       \vdots &\vdots & \ddots & \vdots \\
       \frac{\left|\hat{w}_{n1}\hat{A}_1^t\right|}{\sum_{j=1}^{n}\left|\hat{w}_{nj}\hat{A}_j^t\right|} & \frac{\left|\hat{w}_{n2}\hat{A}_2^t\right|}{\sum_{j=1}^{n}\left|\hat{w}_{nj}\hat{A}_j^t\right|} & \cdots &  \frac{\left|\hat{w}_{nn}\hat{A}_n^t\right|}{\sum_{j=1}^{n}\left|\hat{w}_{nj}\hat{A}_j^t\right|}\\
     \end{pmatrix}.
   \end{equation*}
\end{thm}

\begin{pf}

Suppose the greyness of the FGGCM converges, then there must be: 
\begin{equation*}
  {A^\circ_j}^{t + 1} = {A^\circ_j}^{t}
\end{equation*}
For the tanh function, by substituting it into Eq. \eqref{itertanh}, the equation can be obtained:
\begin{equation*}
  {A^\circ_j}=
   \sum_{j=1}^{n}\frac{\max (w_{ij}^\circ,{A_j^\circ})\left|\hat{w}_{ij}\hat{A}_j^t\right|}{\sum_{j=1}^{n}\left|\hat{w}_{ij}\hat{A}_j^t\right|}
\end{equation*}

Written as a matrix equation as: 
\begin{equation}
  \label{matrixeq}
  \bm{A^\circ} =\mathrm{diag}  (\mathbf{M} \cdot \max_{1\leqslant i\leqslant n} (\mathbf{w_{i}^\circ,{A^\circ}})^{^\top})
\end{equation}
where,
\begin{equation*}
  \begin{split}
    \max_{1\leqslant i\leqslant n} (\mathbf{w_{i}^\circ,{A^\circ}}) = \begin{pmatrix}
      \max({w}_{11}^\circ,{A}_1^{\circ t}) & \max({w}_{12}^\circ,{A}_2^{\circ t}) & \cdots &  \max({w}_{1n}^\circ,{A}_n^{\circ t})\\
      \max({w}_{21}^\circ,{A}_1^{\circ t}) & \max({w}_{22}^\circ,{A}_2^{\circ t}) & \cdots &  \max({w}_{2n}^\circ,{A}_n^{\circ t})\\
      \vdots &\vdots & \ddots & \vdots \\
      \max({w}_{n1}^\circ,{A}_1^{\circ t}) & \max({w}_{n2}^\circ,{A}_2^{\circ t}) & \cdots &  \max({w}_{nn}^\circ,{A}_n^{\circ t})\\
    \end{pmatrix}
  \end{split}
\end{equation*}
$\mathrm{diag}(\cdot)$ indicates taking the elements on the main diagonal.
If for any $i$ and $j$, there is $\max (w_{ij}^\circ, {A_j^\circ} )= {A_j^\circ}$, then the equation can be simplified to: 
\begin{equation}
  \label{simplified_eq}
  \bm{A^\circ} =\mathbf{M} \cdot  \bm{A}{^\circ}^{\top}
\end{equation}
It can be found that the sum of each row of the matrix $\mathbf{M}$ is $1$. Examine its characteristic equation and eigenvalue. Its characteristic equation is: 
\begin{equation}
  \label{eq_tanh}
  \det(\mathbf{M}-\mu\mathbf{I}) = 0
\end{equation}
In the formula, $\mathbf{I}$ represents the identity matrix, and $\mu$ represents the eigenvalue of the matrix. Considering that the row sum of the matrix $\mathbf{M}$ is $1$, then according to the properties of the determinant, the following transformation can be made: 
  \begin{equation*}
    \begin{split}
      \det(\mathbf{M}-\mu\mathbf{I})
      =& \begin{vmatrix}
        \frac{\left|\hat{w}_{11}\hat{A}_1\right|}{\sum_{j=1}^{n}\left|\hat{w}_{1j}\hat{A}_j\right|} -\mu & \frac{\left|\hat{w}_{12}\hat{A}_2\right|}{\sum_{j=1}^{n}\left|\hat{w}_{1j}\hat{A}_j\right|} & \cdots &  \frac{\left|\hat{w}_{1n}\hat{A}_n\right|}{\sum_{j=1}^{n}\left|\hat{w}_{1j}\hat{A}_j\right|}\\
      \frac{\left|\hat{w}_{21}\hat{A}_1\right|}{\sum_{j=1}^{n}\left|\hat{w}_{2j}\hat{A}_j\right|} & \frac{\left|\hat{w}_{22}\hat{A}_2\right|}{\sum_{j=1}^{n}\left|\hat{w}_{2j}\hat{A}_j\right|}-\mu & \cdots &  \frac{\left|\hat{w}_{2n}\hat{A}_n\right|}{\sum_{j=1}^{n}\left|\hat{w}_{2j}\hat{A}_j\right|}\\
      \vdots &\vdots & \ddots & \vdots \\
      \frac{\left|\hat{w}_{n1}\hat{A}_1\right|}{\sum_{j=1}^{n}\left|\hat{w}_{nj}\hat{A}_j\right|} & \frac{\left|\hat{w}_{n2}\hat{A}_2\right|}{\sum_{j=1}^{n}\left|\hat{w}_{nj}\hat{A}_j\right|} & \cdots &  \frac{\left|\hat{w}_{nn}\hat{A}_n\right|}{\sum_{j=1}^{n}\left|\hat{w}_{nj}\hat{A}_j\right|}-\mu\\
      \end{vmatrix}\\
  = & \begin{vmatrix}
    \sum_{j=1}^{n}\frac{\left|\hat{w}_{1j}\hat{A}_j\right|}{\sum_{j=1}^{n}\left|\hat{w}_{1j}\hat{A}_j\right|} -\mu & \frac{\left|\hat{w}_{12}\hat{A}_2\right|}{\sum_{j=1}^{n}\left|\hat{w}_{1j}\hat{A}_j\right|} & \cdots &  \frac{\left|\hat{w}_{1n}\hat{A}_n\right|}{\sum_{j=1}^{n}\left|\hat{w}_{1j}\hat{A}_j\right|}\\
    \sum_{j=1}^{n}\frac{\left|\hat{w}_{2j}\hat{A}_j\right|}{\sum_{j=1}^{n}\left|\hat{w}_{2j}\hat{A}_j\right|} -\mu & \frac{\left|\hat{w}_{22}\hat{A}_2\right|}{\sum_{j=1}^{n}\left|\hat{w}_{2j}\hat{A}_j\right|}-\mu & \cdots &  \frac{\left|\hat{w}_{2n}\hat{A}_n\right|}{\sum_{j=1}^{n}\left|\hat{w}_{2j}\hat{A}_j\right|}\\
  \vdots &\vdots & \ddots & \vdots \\
  \sum_{j=1}^{n}\frac{\left|\hat{w}_{nj}\hat{A}_j\right|}{\sum_{j=1}^{n}\left|\hat{w}_{nj}\hat{A}_j\right|} -\mu & \frac{\left|\hat{w}_{n2}\hat{A}_2\right|}{\sum_{j=1}^{n}\left|\hat{w}_{nj}\hat{A}_j\right|} & \cdots &  \frac{\left|\hat{w}_{nn}\hat{A}_n\right|}{\sum_{j=1}^{n}\left|\hat{w}_{nj}\hat{A}_j\right|}-\mu\\
  \end{vmatrix}\\
  = & (1-\mu)\begin{vmatrix}
    1 & \frac{\left|\hat{w}_{12}\hat{A}_2\right|}{\sum_{j=1}^{n}\left|\hat{w}_{1j}\hat{A}_j\right|} & \cdots &  \frac{\left|\hat{w}_{1n}\hat{A}_n\right|}{\sum_{j=1}^{n}\left|\hat{w}_{1j}\hat{A}_j\right|}\\
   1 & \frac{\left|\hat{w}_{22}\hat{A}_2\right|}{\sum_{j=1}^{n}\left|\hat{w}_{2j}\hat{A}_j\right|}-\mu & \cdots &  \frac{\left|\hat{w}_{2n}\hat{A}_n\right|}{\sum_{j=1}^{n}\left|\hat{w}_{2j}\hat{A}_j\right|}\\
  \vdots &\vdots & \ddots & \vdots \\
  1 & \frac{\left|\hat{w}_{n2}\hat{A}_2\right|}{\sum_{j=1}^{n}\left|\hat{w}_{nj}\hat{A}_j\right|} & \cdots &  \frac{\left|\hat{w}_{nn}\hat{A}_n\right|}{\sum_{j=1}^{n}\left|\hat{w}_{nj}\hat{A}_j\right|}-\mu\\
  \end{vmatrix} = 0\\
    \end{split}
  \end{equation*}

Apparently, $\mu = 1$ must be an eigenvalue of the matrix $\mathbf{M}$, which means that the equation $\bm{A^\circ} = \mathbf{M} \cdot {A^\circ}$ must have a solution. That is, when $\max (\bm{w_{i}^\circ,{A^\circ}}) ={\bm{A}^\circ}$, the greyness of the FGGCM using the tanh function must have a fixed point, which is a certain eigenvector corresponding to the eigenvalue 1 of the $\mathbf{M}$ matrix or the $0$ vector.

Furthermore, when $\max (\bm{w_{i}^\circ,{A^\circ}}) ={\bm{A}^\circ}$ holds, there is
\begin{equation*}
  \begin{split}
    \sum_{i=1}^{n} \sum_{j=1}^{n}\left(\frac{\left|\hat{A}_{j} \hat{w}_{ij}\right|\theta(C_j^\circ - \hat{w}_{ij}^\circ)}{\sum_{j=1}^{n}\left|\hat{w}_{ij}\hat{A}_j\right|}\right)^2 
    = \sum_{i=1}^{n} \sum_{j=1}^{n}\left(\frac{\left|\hat{A}_{j} \hat{w}_{ij}\right|}{\sum_{j=1}^{n}\left|\hat{w}_{ij}\hat{A}_j\right|}\right)^2 = {\left\lVert \mathbf{M}\right\rVert }^2_F
  \end{split}
\end{equation*}

Then, it can be obtained that when tanh is the activation function and when $\max (\bm{w_{i}^\circ,{A^\circ}}) ={\bm{A}^\circ}$ and when the kernel of the FGGCM converges to a fixed point, if ${\left\lVert \mathbf{M}\right\rVert }_F < 1$, then the greyness of the FGGCM must converge to a unique fixed point, which is a certain eigenvector or 0 vector corresponding to the eigenvalue $1$ of the $\mathbf{M}$ matrix. 
\end{pf}

\subsection{The Convergence of Sigmoid FGGCM}
This part is presented to illustrate the convergence of FGGCM when the activation function is the sigmoid function. Similar to the derivation of the convergence of FGGCM when the activation function is tanh, a lemma is presented firstly.
\begin{lem}
  \label{sigmoid_lemma}
  Given the function $f(x) = \mathrm{sigmoid}(x) = \frac{1}{1 + \me^{-\lambda x}}$, where $x \in \mathbb{R}$ and $\lambda > 0, \lambda \in \mathbb{R}$, it follows that for any $a, b \in \mathbb{R}$, there is necessarily:
\begin{equation*}
  |f(b) - f(a)| \leqslant \frac{\lambda}{4}|(b - a)|
\end{equation*}
\end{lem}
\begin{pf}
  By differentiating $f(x)$, it can be obtained:
\begin{equation}
  \label{sigmoidderivative}
  f'(x) = \frac{\lambda}{2 + \me^{-\lambda x} + \frac{1}{\me^{-\lambda x}}} \leqslant \frac{\lambda}{4}
\end{equation}
The equality holds if and only if $x = 0$. Thus,
\begin{equation*}
  |f(b) - f(a)| \leqslant \frac{\lambda}{4}|(b - a)|
\end{equation*}
is derived by substituting Eq. \eqref{sigmoidderivative} into the Lagrange Mean Value Theorem.
\end{pf}

Therefore, it can be deduced that:
\begin{thm}
  \label{thmFGGCMsig}
  Given that $\mathbf{W^\pm} = \mathbf{\hat{W}_{W^\circ}}$ is the weight matrix of FGGCM, $w^\pm_{ij} = \hat{w}{_{ij}}_{w_{ij}^\circ} \in \mathbb{G}$ are the elements of $\mathbf{W^\pm}$, and $\lambda > 0, \lambda \in \mathbb{R}$ is the parameter of the sigmoid activation function of the FGGCM model with $n$ nodes, the sigmoid function is
  $$\mathrm{sigmoid}(g^\pm) = \frac{1}{1+\me^{-\lambda g^\pm}},$$
the FGGCM will converge to a unique fixed point if:
\begin{equation}
  \label{sigkernelconvergence}
  \left(\sum_{i=1}^n\sum_{j=1}^n \hat{w}_{ij}^2 \right)^\frac{1}{2} < \frac{4}{\lambda}
\end{equation}
and
\begin{equation}
  \label{siggreynessconvergence}
  \left(\sum_{i=1}^{n} \sum_{j=1}^{n}\left(\frac{\hat{A}_{i}\left|{\hat{A}_{j} \hat{w}_{ij}}\right|\theta\left(C_j^\circ - \hat{w}_{ij}^\circ\right) }{\sum_{j=1}^{n}\left|\hat{w}_{ij}\hat{A}_j\right|}\right)^2\right)^\frac{1}{2} < 1
\end{equation}

Specifically, if \begin{equation}
  \label{sigeqkernel}
  \left(\sum_{i=1}^n\sum_{j=1}^n \hat{w}_{ij}^2 \right)^\frac{1}{2} = \frac{4}{\lambda} 
\end{equation}
and  \begin{equation}
  \label{sigeqgreyness}
  \left(\sum_{i=1}^{n} \sum_{j=1}^{n}\left(\frac{\hat{A}_{i}\left|{\hat{A}_{j} \hat{w}_{ij}}\right|\theta\left(C_j^\circ - \hat{w}_{ij}^\circ\right) }{\sum_{j=1}^{n}\left|\hat{w}_{ij}\hat{A}_j\right|}\right)^2\right)^{\frac{1}{2}} = 1
\end{equation} hold, then FGGCM must have at least one fixed point.
\end{thm}

\begin{pf}
  $\forall \bm{A,A'} \in \mathbb{G}^n$,calculate $d^2(f(\mathbf{W^\pm }\bm A),f(\mathbf{W^\pm} \bm A'))$ as Eq. \eqref{dSAA},

  \begin{equation}
    \label{dSAA}
    % \footnotesize
  \begin{split}
    &d^2(f(\mathbf{W^\pm }\bm A),f(\mathbf{W^\pm} \bm A'))\\ 
    = &\sum_{i=1}^n d_2^2(f(\mathbf{W_iA}),f(\mathbf{W_iA'}))\\
    =&\sum_{i=1}^n d_2^2 \left(\left(\frac{1}{1+\me^{-\lambda \bm{\hat{W_i} \hat{A}}}}\right)_{\left(\frac{1}{1+\me^{-\lambda \bm{\hat{W_i} \hat{A}}}}\right)^\intercal\bm{(W_iA)^\circ}},\left(\frac{1}{1+\me^{-\lambda \bm{\hat{W_i} \hat{A'}}}}\right)_{\left(\frac{1}{1+\me^{-\lambda \bm{\hat{W_i} \hat{A'}}}}\right)^\intercal\bm{(W_iA')^\circ}}\right)\\
    =& \sum_{i=1}^n \left(\left(\frac{1}{1+\me^{-\lambda \bm{\hat{W_i} \hat{A}}}} - \frac{1}{1+\me^{-\lambda \bm{\hat{W_i} \hat{A'}}}} \right)^2 + \right.\\
    & \left. \left(\left(\frac{1}{1+\me^{-\lambda \bm{\hat{W_i} \hat{A}}}}\right)^\intercal\bm{(W_iA)^\circ}-\left(\frac{1}{1+\me^{-\lambda \bm{\hat{W_i} \hat{A'}}}}\right)^\intercal\bm{(W_iA')^\circ}\right)^2\right) \\
    =& \sum_{i=1}^n \left(\frac{1}{1+\me^{-\lambda \bm{\hat{W_i} \hat{A}}}} - \frac{1}{1+\me^{-\lambda \bm{\hat{W_i} \hat{A'}}}} \right)^2 + GRS\\
    \leqslant &  \sum_{i=1}^n \left(\frac{\lambda}{4}\right)^2 \left(\bm{\hat{W_i} \hat{A}} - \bm{\hat{W_i} \hat{A'}} \right)^2 + GRS\\
    = &\left(\frac{\lambda}{4}\right)^2  \sum_{i=1}^n \left(\sum_{j=1}^n \hat{w}_{ij}\left(\hat{A}_j - \hat{A'}_j\right)\right)^2 + GRS\\
    \leqslant & \left(\frac{\lambda}{4}\right)^2 \sum_{i=1}^n \left(\sum_{j=1}^n \hat{w}_{ij}^2 \right) \left(\sum_{j=1}^n \left(\hat{A}_j - \hat{A'}_j\right)^2 \right)+ GRS\\
  \end{split}
  \end{equation}

where 
\begin{equation*}
  \begin{split}
    GRS= \sum_{i=1}^n \left(\left(\frac{1}{1+\me^{-\lambda \bm{\hat{W_i} \hat{A}}}}\right)^\intercal\bm{(W_iA)^\circ}-\left(\frac{1}{1+\me^{-\lambda \bm{\hat{W_i} \hat{A'}}}}\right)^\intercal\bm{(W_iA')^\circ}\right)^2
  \end{split}
\end{equation*}
Observing the kernel part, it can deduce that if
\begin{equation*}
  \begin{split}
    \left(\frac{\lambda}{4}\right)^2 \sum_{i=1}^n \left(\sum_{j=1}^n \hat{w}_{ij}^2 \right) \left(\sum_{j=1}^n \left(\hat{A}_j - \hat{A'}_j\right)^2 \right) <\sum_{j=1}^n \left(\hat{A}_j - \hat{A'}_j\right)^2
  \end{split}
\end{equation*}
holds, which is equivalent to the condition
\begin{equation}
  \label{sigkernelconvergenceinproof}
  \left(\sum_{i=1}^n\sum_{j=1}^n \hat{w}_{ij}^2 \right)^\frac{1}{2} < \frac{4}{\lambda}
\end{equation}
being satisfied, then the kernel of FGGCM must converge to a unique fixed point.

When the sigmoid function is used as the activation function, the iteration method of the greyness can be described as:
\begin{equation}
  \label{sig_grey_iter}
  \begin{split}
    {A^\circ_j}^{t + 1}=&f({A_1^\circ}^t,{A_2^\circ}^t,\cdots,{A_n^\circ}^t,\hat{A}_1^t,\hat{A}_2^t\cdots \hat{A}_n^t)\\
     = &\frac{1}{\left(1+\me^{-\lambda \sum_{j=1}^{n}\hat{w}_{ij}\hat{A}_j^t}\right)}\sum_{j=1}^{n}\frac{\max (w_{ij}^\circ,{A_j^\circ}^t)\left|\hat{w}_{ij}\hat{A}_j^t\right|}{\sum_{j=1}^{n}\left|\hat{w}_{ij}\hat{A}_j^t\right|}\\
     = &\hat{A}_i^{t+1}\sum_{j=1}^{n}\frac{ \max (w_{ij}^\circ,{A_j^\circ}^t)\left|\hat{w}_{ij}\hat{A}_j^t\right|}{\sum_{j=1}^{n}\left|\hat{w}_{ij}\hat{A}_j^t\right|}.\\
  \end{split}
\end{equation}

According to the mean value theorem of multivariate functions, Eq. \eqref{multivariate} must hold. Then, calculate the partial derivative with respect to the greyness:
\begin{equation}
  \label{dsg}
  \begin{split}
    \frac{\partial f}{\partial C_j^\circ}= &\frac{\left|{\hat{C}_{j} w_{ij}}\right| \theta\left(C_j^\circ - \hat{w}_{ij}^\circ\right)}{\left(1+\me ^{\sum_{j=1}^{n}\hat{w}_{ij}\hat{C}_j} \right)\sum_{j=1}^{n}\left|\hat{w}_{ij}\hat{C}_j\right|}\\
    =&\frac{\hat{C}_i^{t+1} \left|{\hat{C}_{j} w_{ij}}\right| \theta\left(C_j^\circ - \hat{w}_{ij}^\circ\right)}{\sum_{j=1}^{n}\left|\hat{w}_{ij}\hat{C}_j\right|}\\
  \end{split}
\end{equation}

Suppose the kernel of FGGCM has converged to a fixed point, then the influence of the kernel on the greyness can be neglected. The calculation of the greyness part in Eq. \eqref{dSAA} is Eq. \eqref{sig_grey_cal}:

  \begin{equation}
    \label{sig_grey_cal}
    \small
    \begin{split} 
      &\sum_{i=1}^n \left(\sum_{j=1}^{n}\frac{\max(w_{ij}^\circ,A_j^\circ)\left|\hat{w}_{ij}\hat{A}_j\right|}{\left(1+\me^{-\lambda \sum_{j=1}^{n}\hat{w}_{ij}\hat{A}_j}\right)\sum_{j=1}^{n}\left|\hat{w}_{ij}\hat{A}_j\right|}- \sum_{j=1}^{n}\frac{\max(w_{ij}^\circ,{A'}_j^\circ)\left|\hat{w}_{ij}\hat{A'}_j\right|}{\left(1+\me^{-\lambda \sum_{j=1}^{n}\hat{w}_{ij}\hat{A'}_j}\right)\sum_{j=1}^{n}\left|\hat{w}_{ij}\hat{A'}_j\right|}\right)^2\\ 
      =&\sum_{i=1}^{n}\left(f({A_1^\circ},{A_2^\circ},\cdots,{A_n^\circ},\hat{A}_1,\hat{A}_2\cdots \hat{A}_n) -  f({{A'}_1^\circ},{{A'}_2^\circ},\cdots,{{A'}_n^\circ},\hat{{A'}}_1,\hat{{A'}}_2\cdots \hat{{A'}}_n)\right)^2\\
      =& \sum_{i=1}^{n} \left(\sum_{j=1}^{n}\frac{\partial f}{\partial C_j^\circ} ({A_j^\circ} - {{A'}_j^\circ})+ \sum_{j=1}^{n}\frac{\partial f}{\partial \hat{C}_j} ({\hat{A}_j} - {\hat{A'}_j})\right)^2\\
      =& \sum_{i=1}^{n} \left(\sum_{j=1}^{n}\frac{\left|{\hat{C}_{j} w_{ij}}\right| \theta\left(C_j^\circ - \hat{w}_{ij}^\circ\right)}{\left(1+\me ^{\sum_{j=1}^{n}\hat{w}_{ij}\hat{C}_j} \right)\sum_{j=1}^{n}\left|\hat{w}_{ij}\hat{C}_j\right|} ({A_j^\circ} - {{A'}_j^\circ})\right)^2\\
      \leqslant & \sum_{i=1}^{n} \sum_{j=1}^{n}\left(\frac{\left|{\hat{C}_{j} w_{ij}}\right| \theta\left(C_j^\circ - \hat{w}_{ij}^\circ\right)}{\left(1+\me ^{\sum_{j=1}^{n}\hat{w}_{ij}\hat{C}_j} \right)\sum_{j=1}^{n}\left|\hat{w}_{ij}\hat{C}_j\right|}\right)^2\sum_{j=1}^{n} \left({A_j^\circ} - {{A'}_j^\circ}\right)^2 \\
    \end{split}
  \end{equation}
 
  Since $\hat{A}_j$ is equivalent to $\hat{{A'}}_j$, meaning $\hat{C}_j = \hat{A}_j + t(\hat{{A'}}_j - \hat{A}_j) = \hat{A}_j$, if
\begin{equation*}
  \sum_{i=1}^{n} \sum_{j=1}^{n}\left(\frac{\left|{\hat{A}_{j} \hat{w}_{ij}}\right|\theta\left(C_j^\circ - \hat{w}_{ij}^\circ\right) }{\left(1+\me ^{-\lambda \sum_{j=1}^{n}\hat{w}_{ij}\hat{A}_j} \right)\sum_{j=1}^{n}\left|\hat{w}_{ij}\hat{A}_j\right|}\right)^2 < 1
\end{equation*}
or
\begin{equation*}
  \left(\sum_{i=1}^{n} \sum_{j=1}^{n}\left(\frac{\hat{A}_{i}\left|{\hat{A}_{j} \hat{w}_{ij}}\right|\theta\left(C_j^\circ - \hat{w}_{ij}^\circ\right) }{\sum_{j=1}^{n}\left|\hat{w}_{ij}\hat{A}_j\right|}\right)^2\right)^\frac{1}{2} < 1,
\end{equation*}
The greyness of the sigmoid FGGCM must converge to a unique fixed point.

Similarly, with the aid of Lemma \ref{Browder}, it can be concluded that when both Eq. \eqref{sigeqkernel} and \eqref{sigeqgreyness} are satisfied simultaneously, FGGCM has at least one fixed point.

\end{pf}

 The following corollary can also be derived easily from the above proof process.

\begin{corollary}
  If Eq. \eqref{sigkernelconvergence} holds, then the kernel of this FGGCM node must converge to a unique fixed point; if Eq. \eqref{sigeqkernel} holds, then the kernel of FGGCM has at least one fixed point.
\end{corollary}

\begin{corollary}
  If the kernel of the FGGCM converges and inequalities \eqref{siggreynessconvergence} holds, then the greyness of the FGGCM node must converge to a unique fixed point. On the other hand, if the kernel of the FGGCM converges and Eq. \eqref{sigeqgreyness} holds, then the greyness of the FGGCM node must have at least one fixed point.
\end{corollary}

After an examination of the greyness iteration within the FGGCM, Corollary \ref{sig_grey_convergence} can be established. Since Eq. \eqref{sig_grey_convergence_eq} is really just a specific instance of Eq. \eqref{siggreynessconvergence} as outlined in Theorem \ref{thmFGGCMsig}, this revelation reinforces the criteria for greyness convergence in Eq. \eqref{siggreynessconvergence} from Theorem \ref{thmFGGCMsig}.

\begin{corollary}
  \label{sig_grey_convergence}
  If the kernel of an FGGCM with the sigmoid as the activation function converges, and for any iteration step, it holds that $\max (w_{ij}^\circ, A_j^\circ) = A_j^\circ$, and \begin{equation} \label{sig_grey_convergence_eq}
    \sum_{i=1}^{n} \sum_{j=1}^{n}\left(\frac{\hat{A}_{i}\left|\hat{A}_{j} \hat{w}_{ij}\right|}{\sum_{j=1}^{n}\left|\hat{w}_{ij}\hat{A}_j\right|}\right)^2 = {\left\lVert \mathbf{M}\right\rVert }^2_F <1
  \end{equation}
   is satisfied, then the greyness of the FGGCM will necessarily converge to a unique fixed point. This fixed point is a characteristic vector corresponding to the eigenvalue $1$ of the $\mathbf{M}$ or zero vector, where
   \begin{equation}
    \label{sig_M}
    \mathbf{M} = 
    \begin{pmatrix}
      \frac{\hat{A}_1^{t+1}\left|\hat{w}_{11}\hat{A}_1^t\right|}{\sum_{j=1}^{n}\left|\hat{w}_{1j}\hat{A}_j^t\right|} & \frac{\hat{A}_1^{t+1}\left|\hat{w}_{12}\hat{A}_2^t\right|}{\sum_{j=1}^{n}\left|\hat{w}_{1j}\hat{A}_j^t\right|} & \cdots &  \frac{\hat{A}_1^{t+1}\left|\hat{w}_{1n}\hat{A}_n^t\right|}{\sum_{j=1}^{n}\left|\hat{w}_{1j}\hat{A}_j^t\right|}\\
      \frac{\hat{A}_2^{t+1}\left|\hat{w}_{21}\hat{A}_1^t\right|}{\sum_{j=1}^{n}\left|\hat{w}_{2j}\hat{A}_j^t\right|} & \frac{\hat{A}_2^{t+1}\left|\hat{w}_{22}\hat{A}_2^t\right|}{\sum_{j=1}^{n}\left|\hat{w}_{2j}\hat{A}_j^t\right|} & \cdots &  \frac{\hat{A}_2^{t+1}\left|\hat{w}_{2n}\hat{A}_n^t\right|}{\sum_{j=1}^{n}\left|\hat{w}_{2j}\hat{A}_j^t\right|}\\
      \vdots &\vdots & \ddots & \vdots \\
      \frac{\hat{A}_n^{t+1}\left|\hat{w}_{n1}\hat{A}_1^t\right|}{\sum_{j=1}^{n}\left|\hat{w}_{nj}\hat{A}_j^t\right|} & \frac{\hat{A}_n^{t+1}\left|\hat{w}_{n2}\hat{A}_2^t\right|}{\sum_{j=1}^{n}\left|\hat{w}_{nj}\hat{A}_j^t\right|} & \cdots &  \frac{\hat{A}_n^{t+1}\left|\hat{w}_{nn}\hat{A}_n^t\right|}{\sum_{j=1}^{n}\left|\hat{w}_{nj}\hat{A}_j^t\right|}\\
    \end{pmatrix}.
  \end{equation}
   If  ${\left\lVert \mathbf{M}\right\rVert }^2_F = 1$, the greyness of the FGGCM will have at least one fixed point,
\end{corollary}

\begin{pf}
Eq. \eqref{sig_grey_iter} can also be written in the matrix form as Eq. \eqref{matrixeq}. The $\mathbf{M}$ matrix becomes Eq. \eqref{sig_M}.

If for any $i$ and $j$, there is $\max (w_{ij}^\circ, {A_j^\circ}) = {A_j^\circ}$, then the Eq. \eqref{matrixeq} can be reduced to Eq. \eqref{simplified_eq}.

Observe the matrix $\mathbf{M}$, it can be found that the sum of each row of the matrix $\mathbf{M}$ is $\hat{{A}}_i$, not $1$. Different from the case of tanh activation function, the eigenvalue of the matrix $\mathbf{M}$ does not necessarily include $1$ at this time. Even then, it cannot be guaranteed that there exists $\bm x \in \mathbb{G}^n$ such that $\mathbf{M} \bm x = \bm x$ holds true.

When $\max (\bm{w_{i}^\circ,{A^\circ}}) ={\bm{A}^\circ}$ holds, there is
\begin{equation*}
  \begin{split}
    \sum_{i=1}^{n} \sum_{j=1}^{n}\left(\frac{\hat{A}_{i}\left|{\hat{A}_{j} \hat{w}_{ij}}\right|\theta\left(C_j^\circ - \hat{w}_{ij}^\circ\right) }{\sum_{j=1}^{n}\left|\hat{w}_{ij}\hat{A}_j\right|}\right)^2 
    =  \sum_{i=1}^{n} \sum_{j=1}^{n}\left(\frac{\hat{A}_{i}\left|{\hat{A}_{j} \hat{w}_{ij}}\right| }{\sum_{j=1}^{n}\left|\hat{w}_{ij}\hat{A}_j\right|}\right)^2 = {\left\lVert \mathbf{M}\right\rVert }^2_F
  \end{split}
\end{equation*}

Therefore, when the kernel of FGGCM converges to a certain fixed point and a sigmoid function is used as the activation function, if the condition $\max (\bm{w_{i}^\circ,{A^\circ}}) ={\bm{A}^\circ}$ is satisfied, it can be inferred that if the ${\left\lVert \mathbf{M}\right\rVert }_F<1$, the greyness level of FGGCM will converge to a unique fixed point. This fixed point is a certain eigenvector corresponding to the eigenvalue of $1$ of $\mathbf{M}$ or zero vector. Moreover, according to Lemma \ref{Browder}, if ${\left\lVert \mathbf{M}\right\rVert }_F = 1$, the greyness of FGGCM will converge to a certain fixed point, which is also a certain eigenvector  corresponding to the eigenvalue of $1$ of the matrix $\mathbf{M}$ or zero vector, but it is not guaranteed that this fixed point is unique.
\end{pf}

\section{Experiments Design}\label{sec:case_study}

This section first presents a case study of a FCM. Based on this, in conjunction with the case study of web experience provided in the introduction, the compatibility of the theorems proposed in this paper with existing literature is verified.

\subsection{An FCM Applied in Civil Engineering}

The concepts utilized in the Civil Engineering FCM, as depicted in Fig. \ref{civil_fcm_graph}, which investigates the effects of urban population growth and modernization on public health, are as Table \ref{nodes_civil}. For the convenience, this FCM will be called the Civil FCM in the following text.
\begin{figure}[htbp]
  \centering
  \begin{tikzpicture}[every node/.style={circle,draw,inner sep=5pt,font=\footnotesize}]
    \node[fill=blue!20] (c1)at (-2,2){$C_{1}$};
    \node[fill=blue!20] (c2)at (2,2){$C_{2}$};
    \node[fill=blue!20] (c3)at (0,0){$C_{3}$};
    \node[fill=blue!20] (c4)at (-2,-3){$C_{4}$};
    \node[fill=blue!20] (c5)at (2,0){$C_{5}$};
    \node[fill=blue!20] (c6)at (0,-2){$C_{6}$};
    \node[fill=blue!20] (c7)at (2,-3){$C_{7}$};
    \draw[-latex] (c2) -- (c1) node[draw = none,midway, sloped, above, rectangle] {0.1};
    \draw[-latex] (c1) -- (c3) node[draw = none,midway, sloped, above, rectangle] {0.6};
    \draw[-latex] (c3) -- (c2) node[draw = none,midway, sloped, above, rectangle] {0.7};
    \draw[-latex] (c1) -- (c4) node[draw = none,midway, sloped, above, rectangle] {0.9};
    \draw[-latex] (c6) -- (c1) node[draw = none,midway, sloped, above, rectangle] {-0.3};
    \draw[-latex] (c3) -- (c5) node[draw = none,midway, sloped, above, rectangle] {0.9};
    \draw[-latex] (c5) -- (c6) node[draw = none,midway, sloped, above, rectangle] {-0.9};
    \draw[-latex] (c4) -- (c7) node[draw = none,midway, sloped, above, rectangle] {0.9};
    \draw[-latex] (c7) -- (c6) node[draw = none,midway, sloped, above, rectangle] {0.8};
    \draw[-latex] (c5) -- (c7) node[draw = none,midway, sloped, above, rectangle] {-0.9};
  \end{tikzpicture}
  \caption{The Civil Engineering FCM graph}
  \label{civil_fcm_graph}
\end{figure}
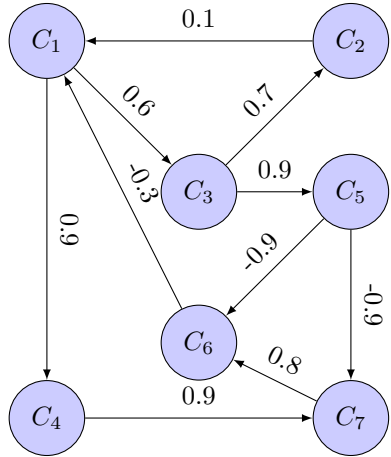
\begin{table}[htbp]
  \centering
  \caption{The nodes' meanings of Civil Engineering FCM.}\label{nodes_civil}
  \begin{tabular*}{.63\linewidth}{cc}
    \toprule    
    Nodes & Meanings \\ 
    \midrule 
    $C_1 $& Urban Population  \\ 
    $C_2 $& Influx of Migrants   \\ 
    $C_3 $& Degree of Modernization  \\ 
    $C_4 $& Waste per Unit Area  \\ 
    $C_5 $& Sanitary Infrastructure  \\ 
    $C_6 $& Incidence of Diseases per Thousand Inhabitants  \\ 
    $C_7 $& Bacterial Concentration per Unit Area  \\ 
    \bottomrule    
    \end{tabular*}
  \end{table}

The weight matrix is 
\begin{equation}
  \label{w_civil}
  \mathbf{W}_{civil}=
\begin{pmatrix}
  0 & 0.1 & 0 & 0 & 0 & -0.3 & 0 \\
  0 & 0 & 0.7 & 0 & 0 & 0 & 0 \\
  0.6 & 0 & 0 & 0 & 0 & 0 & 0 \\
  0.9 & 0 & 0 & 0 & 0 & 0 & 0 \\
  0 & 0 & 0.9 & 0 & 0 & 0 & 0 \\
  0 & 0 & 0 & 0 & -0.9 & 0 & 0.8 \\
  0 & 0 & 0 & 0.9 & -0.9 & 0 & 0 \\
  \end{pmatrix}
\end{equation}

\begin{figure}[h]
  \centering
    \includegraphics[width=0.75\linewidth]{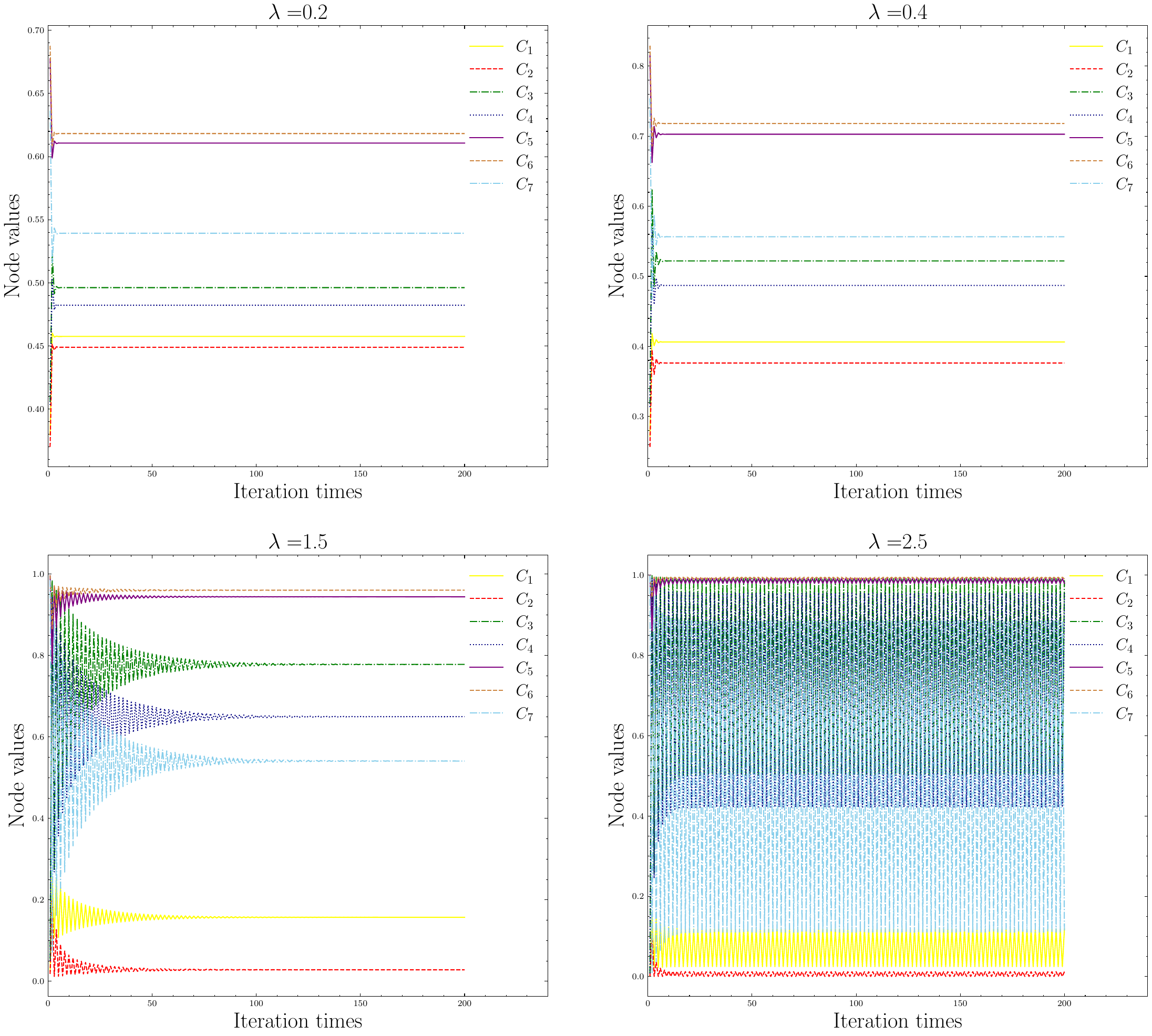}
  \caption{The output of the Civil FCM under different $\lambda$s.}
  \label{figfcmc}
\end{figure}

The initial vector of the Civil FCM is set as $\begin{pmatrix}
  0.8 & 0.5 & 0.3 & 0 & 0 & 0 & 0
  \end{pmatrix},$ and the reasoning process of the FCM under different $\lambda$s is shown in Fig. \ref{figfcmc}.
 
The results show that the FCM converges to a fixed point when the $\lambda$ value is $0.2$, $0.4$, and $1.5$. However, when the $\lambda$ value is $2.5$, the FCM forms a limit cycle. It is noteworthy that, despite $\left\lVert W\right\rVert _F  = 2.375 > \frac{1}{2.5}$, this phenomenon does not violate Theorem \ref{thmfcm}. The reason is that Theorem \ref{thmfcm} only provides sufficient conditions for the convergence of the FCM to a fixed point, and the Civil FCM has multiple fixed points \cite{Tsadiras2008}.

\subsection{Correctness Verification}

To initiate the verification process, we will explore the integration of greyness into both the Civil FCM and the Web Experience FCM, leading to transformation into the FGCM and FGGCM, respectively. These models correspond to the tanh and sigmoid activation functions. By maintaining consistent levels of greyness across these models, we will conduct a comparative analysis of their convergence properties. This involves a concurrent examination of Theorems \ref{thmfcm}, \ref{thmfgcm}, \ref{thmFGGCMtanh}, and \ref{thmFGGCMsig}. The objective of this comparison is to elucidate that Theorems \ref{thmfcm} and \ref{thmfgcm} are indeed specific cases of the more general Theorems \ref{thmFGGCMtanh} and \ref{thmFGGCMsig}.

Following this, a detailed comparative analysis will be conducted between Theorem \ref{thmfgcm} and the proposed Theorems \ref{thmFGGCMtanh} and \ref{thmFGGCMsig}. A key limitation to be noted in Theorem \ref{thmfgcm} is its inability to address scenarios where the weights are restricted strictly to values greater than $0$ or less than $0$. This limitation makes it unsuitable for cases where the weight matrix includes elements ranging from negative to positive values, such as $[-a, +b]$ (where $a, b > 0$, and $a, b \in \mathbb{R}$). In contrast, the proposed Theorems \ref{thmFGGCMtanh} and \ref{thmFGGCMsig} in this research extend their applicability precisely to these types of scenarios. When the weight matrix contains elements like $[-a, +b]$, these new theorems are capable of determining the convergence of FGGCM. Under the purview of Theorem \ref{thmfgcm}, the presence of such elements in the weight matrix hinders the calculation of $w^*$, thus preventing the evaluation of convergence for FGCM. However, the application of the Theorems \ref{thmFGGCMtanh} and \ref{thmFGGCMsig} presented here enables an effective determination of convergence. To demonstrate the expanded scope of these new theorems, a portion of the weight matrix will be altered to the form of $[-a, +b]$, showcasing their capability to address situations beyond the scope of Theorem \ref{thmfgcm}.

In addition to considering the case where the weights of FGGCM contain the form $[-a, +b]$, it should also test FGGCM weights that include more GGN forms, such as $w^\pm_{{ij}} = [-0.95, -0.89] \cup -0.83 \cup [-0.8, -0.75]$. This capability, which involves reasoning with weights that include the above-mentioned data forms, is the most essential characteristic and advantage of FGGCM over FGCM. Theorems \ref{thmFGGCMtanh} and \ref{thmFGGCMsig} should have the ability to judge the convergence of FGGCM under these general cases, in order to fully demonstrate the flexibility and applicability of FGGCM in handling more complex weight structures.

\section{Experiments Results} \label{sec:result}
According to the above experimental design, this section presents the corresponding experimental results. 
\subsection{Simulations for Web Experience FGGCM }
Firstly, calculate that the $\left\lVert \mathbf{W}\right\rVert _F$ of the Web Experience FCM is $6.1359$, and its convergence situation under different $\lambda$ is shown as Fig. \ref{figfcmweb} in the introduction. Use Eqs. \eqref{addgreyl} and \eqref{addgreyu} to add greyness to Matrix \eqref{wweb}.
\begin{equation}
  \label{addgreyl}
  \underline{w_{ij}} = \left\{\begin{array} {c c} w_{ij} - g^\circ & \mathrm{if} \,\, w_{ij} - g > -1 \\ 
    -1 & \mathrm{if} \,\, w_{ij} - g^\circ \leqslant -1   \\ \end{array} \right.
\end{equation}

\begin{equation}
  \label{addgreyu}
  \overline{w_{ij}} = \left\{\begin{array} {c c} w_{ij} + g^\circ & \mathrm{if} \,\, w_{ij} + g \leqslant 1 \\ 
    1 & \mathrm{if} \,\, w_{ij} + g^\circ > 1   \\ \end{array} \right.
\end{equation}
In Eqs. \eqref{addgreyl} and \eqref{addgreyu}, $g^\circ>0, g \in \mathbb{R}$, which is used to control the magnitude of the greyness.

In order to ensure that Theorem \ref{thmfgcm} can also be successfully applied, in Matrix \eqref{wweb}, when $\left\lvert w_{ij}\right\rvert < g$, do not use \eqref{addgreyl} and \eqref{addgreyu} to add greyness to it. At this time, it can be ensured that $\underline{w_{ij}}\leqslant \overline{w_{ij}} \leqslant 0$ and $0 \leqslant \underline{w_{ij}}\leqslant \overline{w_{ij}}$ can always hold. Set $g = 0.01$, and the weight matrix in IGN form is Eq. \eqref{wwebgreyav}.

  \begin{equation}
    \footnotesize
    \setlength{\arraycolsep}{1pt}
    \begin{split}
      &\otimes \mathbf{W}_{web} =\\
    &\begin{pmatrix}  
      [0,0]  &[-0.91,-0.89]  &[-0.89,-0.87]  &[0.99,1.00]  &[-0.86,-0.84]  &[-0.84,-0.82]  &[0.99,1.00]  \\ 
    [0.99,1.00]  &[0,0]  &[-0.94,-0.92]  &[-0.90,-0.88]  &[-0.91,-0.89]  &[-0.95,-0.93]  &[0.99,1.00]  \\ 
    [-0.99,-0.97]  &[-0.94,-0.92]  &[-1.00,-0.99]  &[-1.00,-0.99]  &[0.99,1.00]  &[0.99,1.00]  &[0.99,1.00]  \\ 
    [-1.00,-0.98]  &[-0.90,-0.88]  &[-1.00,-0.99]  &[-0.40,-0.38]  &[0.72,0.74]  &[0.57,0.59]  &[0.69,0.71]  \\ 
    [0.99,1.00]  &[0.99,1.00]  &[0.99,1.00]  &[0.99,1.00]  &[-0.81,-0.79]  &[0.50,0.52]  &[0.99,1.00]  \\ 
    [0.99,1.00]  &[0.99,1.00]  &[0.82,0.84]  &[0.99,1.00]  &[0.50,0.52]  &[-0.40,-0.38]  &[0.99,1.00]  \\ 
    [0.99,1.00]  &[0.99,1.00]  &[0.99,1.00]  &[0.99,1.00]  &[-0.72,-0.70]  &[-0.50,-0.48]  &[-0.68,-0.66]  \\ 
    \end{pmatrix}
    \end{split}
    \label{wwebgreyav}
  \end{equation}
  
The Eq. \eqref{wwebgreyav} satisfies the condition for calculating $\mathbf{W}^{*}$ in Theorem \ref{thmfgcm}. According to Eq. \eqref{wstar}, $\mathbf{W}^{*}$ can be calculated as Eq. \eqref{wstarweb}, and it can be calculated that $\left\lVert \mathbf{W}_{web}^{*}\right\rVert _F= 6.1657$.
\begin{equation}
  \label{wstarweb}
  \mathbf{W}_{web}^{*}= \begin{pmatrix}  
    0  &0.91  &0.89  &1.00  &0.86  &0.84  &1.00  \\ 
  1.00  &0  &0.94  &0.90  &0.91  &0.95  &1.00  \\ 
  0.99  &0.94  &1.00  &1.00  &1.00  &1.00  &1.00  \\ 
  1.00  &0.90  &1.00  &0.40  &0.74  &0.59  &0.71  \\ 
  1.00  &1.00  &1.00  &1.00  &0.81  &0.52  &1.00  \\ 
  1.00  &1.00  &0.84  &1.00  &0.52  &0.40  &1.00  \\ 
  1.00  &1.00  &1.00  &1.00  &0.72  &0.50  &0.68  \\ 
  \end{pmatrix}
\end{equation}

Convert Eq. \eqref{wwebgreyav} to the GGN form, as shown in Eq. \eqref{wwebggn}, and it can be calculated that $\left\lVert \mathbf{\hat{W}}_{web}\right\rVert _F= 6.1172$.
  \begin{equation}
    \label{wwebggn}
     \begin{split}
      &\mathbf{W}_{web}^\pm = \\
     &\begin{pmatrix}  
       0.000_{0.000} &-0.900_{0.010} &-0.880_{0.010} &0.995_{0.005} &-0.850_{0.010} &-0.830_{0.010} &0.995_{0.005} \\ 
     0.995_{0.005} &0.000_{0.000} &-0.930_{0.010} &-0.890_{0.010} &-0.900_{0.010} &-0.940_{0.010} &0.995_{0.005} \\ 
     -0.980_{0.010} &-0.930_{0.010} &-0.995_{0.005} &-0.995_{0.005} &0.995_{0.005} &0.995_{0.005} &0.995_{0.005} \\ 
     -0.990_{0.010} &-0.890_{0.010} &-0.995_{0.005} &-0.390_{0.010} &0.730_{0.010} &0.580_{0.010} &0.700_{0.010} \\ 
     0.995_{0.005} &0.995_{0.005} &0.995_{0.005} &0.995_{0.005} &-0.800_{0.010} &0.510_{0.010} &0.995_{0.005} \\ 
     0.995_{0.005} &0.995_{0.005} &0.830_{0.010} &0.995_{0.005} &0.510_{0.010} &-0.390_{0.010} &0.995_{0.005} \\ 
     0.995_{0.005} &0.995_{0.005} &0.995_{0.005} &0.995_{0.005} &-0.710_{0.010} &-0.490_{0.010} &-0.670_{0.010} \\ 
     \end{pmatrix}
     \end{split}
  \end{equation}
  
\begin{figure}[htbp]
  \centering
    \includegraphics[width=0.75\linewidth]{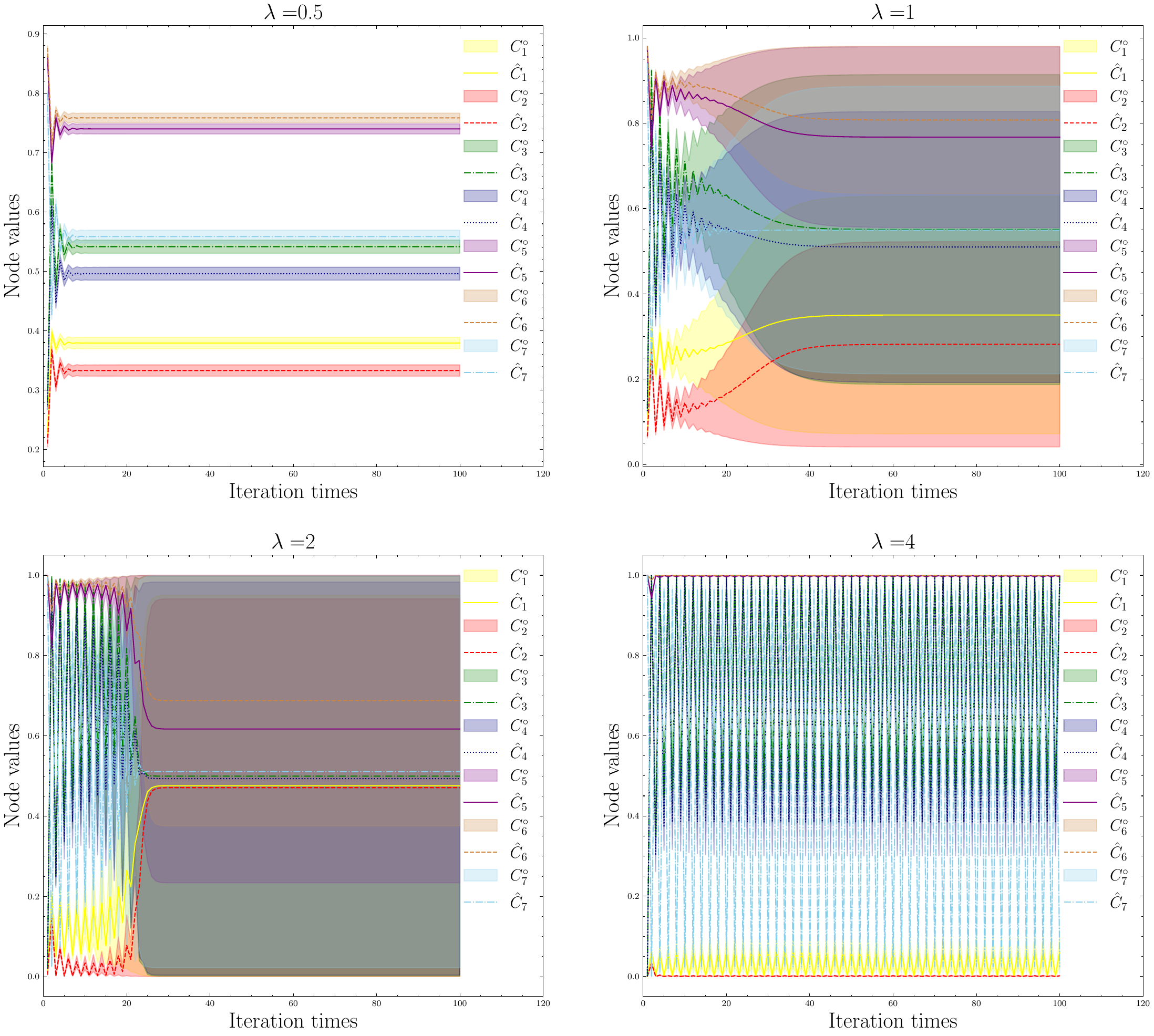}
  \caption{The output of the Web Experience FGCM under different $\lambda$s.}
  \label{figFGCM_web}
\end{figure}

\begin{figure}[htbp]
  \centering
    \includegraphics[width=0.75\linewidth]{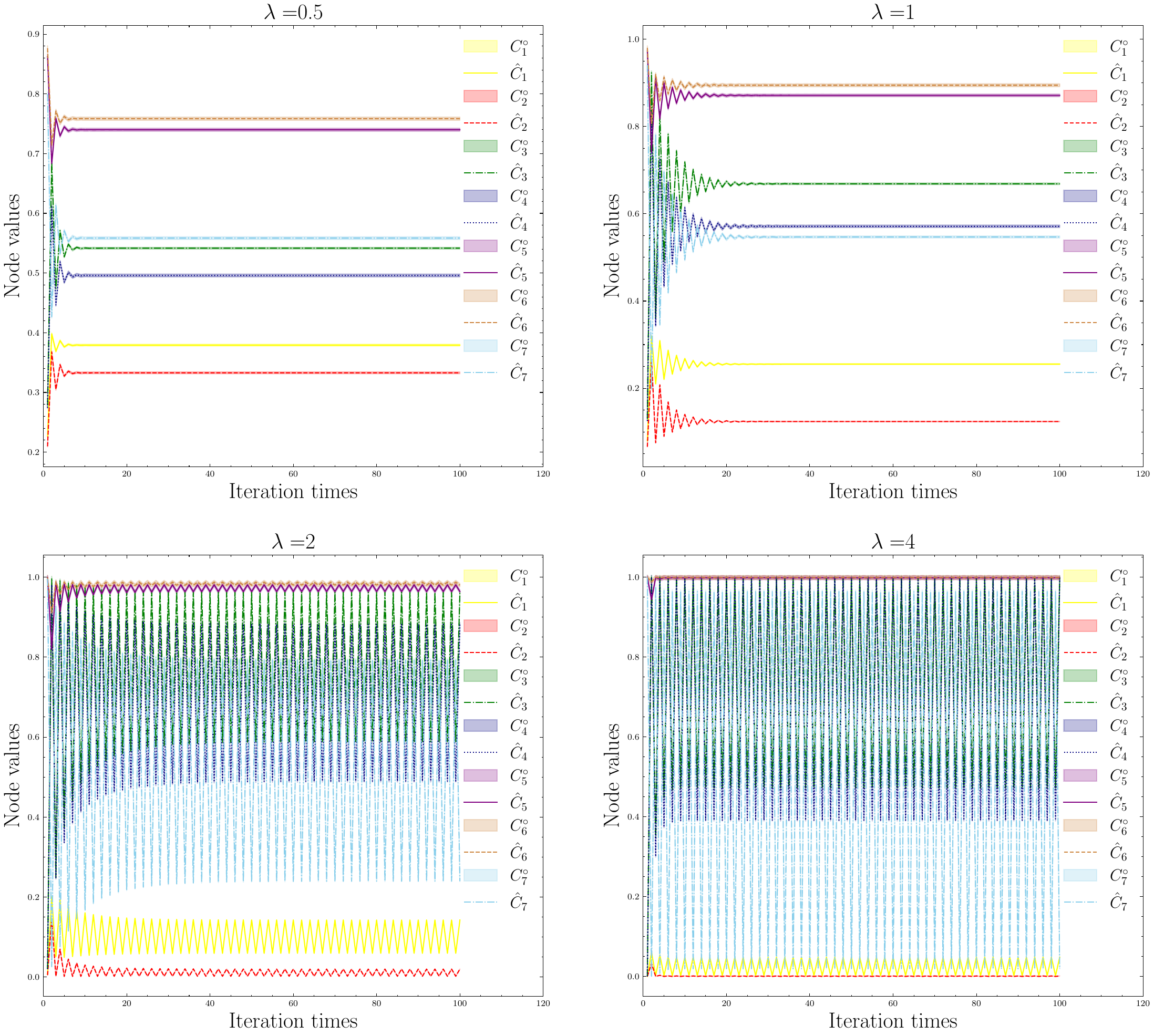}
  \caption{The output of the Web Experience FGGCM under different $\lambda$s.}
  \label{figFGGCM_web}
\end{figure}

The FGCM input vector is 
\begin{equation*}
  \begin{split}
    \left([0.99,1.00],[0.99,1.00],[0.99,1.00],[0.99,1.00],[0.99,1.00],[0.99,1.00],[0.00,0.00] \right)
  \end{split}
\end{equation*} 
The corresponding simplified general grey number form is 
\begin{equation}
  \label{webinputggn}
\begin{split}
    \left(0.995_{(0.010)},0.995_{(0.010)},0.995_{(0.010)},0.995_{(0.010)},0.995_{(0.010)},0.995_{(0.010)},0.000_{(0.000)}\right).
\end{split}
\end{equation}
According to the matrix \eqref{wwebgreyav} and \eqref{wwebggn}, the reason results of the corresponding FGCM and FGGCM are shown as Fig. \ref{figFGCM_web} and \ref{figFGGCM_web}.


The following explains several special cases. These are the cases where Theorem \ref{thmFGGCMsig} proposed in this paper can judge whether it converges, while the previous Theorem \ref{thmfcm} and \ref{thmfgcm} cannot make a judgment.

\begin{itemize}
  \item  Case 1: where $\otimes \mathbf{W}$ contains $[-a,b]$, $a,b>1, a,b \in \mathbb{R}$. Let $\otimes w_{web_{11}} = [-0.1,0.1]$.
  In this case, ${\mathbf{W}^*}$ cannot be calculated, but ${\hat{\mathbf{W}}}$ can be obtained. According to Eqs. \eqref{uniong}, \eqref{unionp}, and \eqref{union}, $\otimes w_{web_{11}}  = 0_{0.1}$. The convergence of FGCM and the FGGCM is shown as Fig. \ref{figFGCM_web_cw} and \ref{figFGGCM_web_cw}.
  
  \begin{figure}[htbp]
    \centering
      \includegraphics[width=0.75\linewidth]{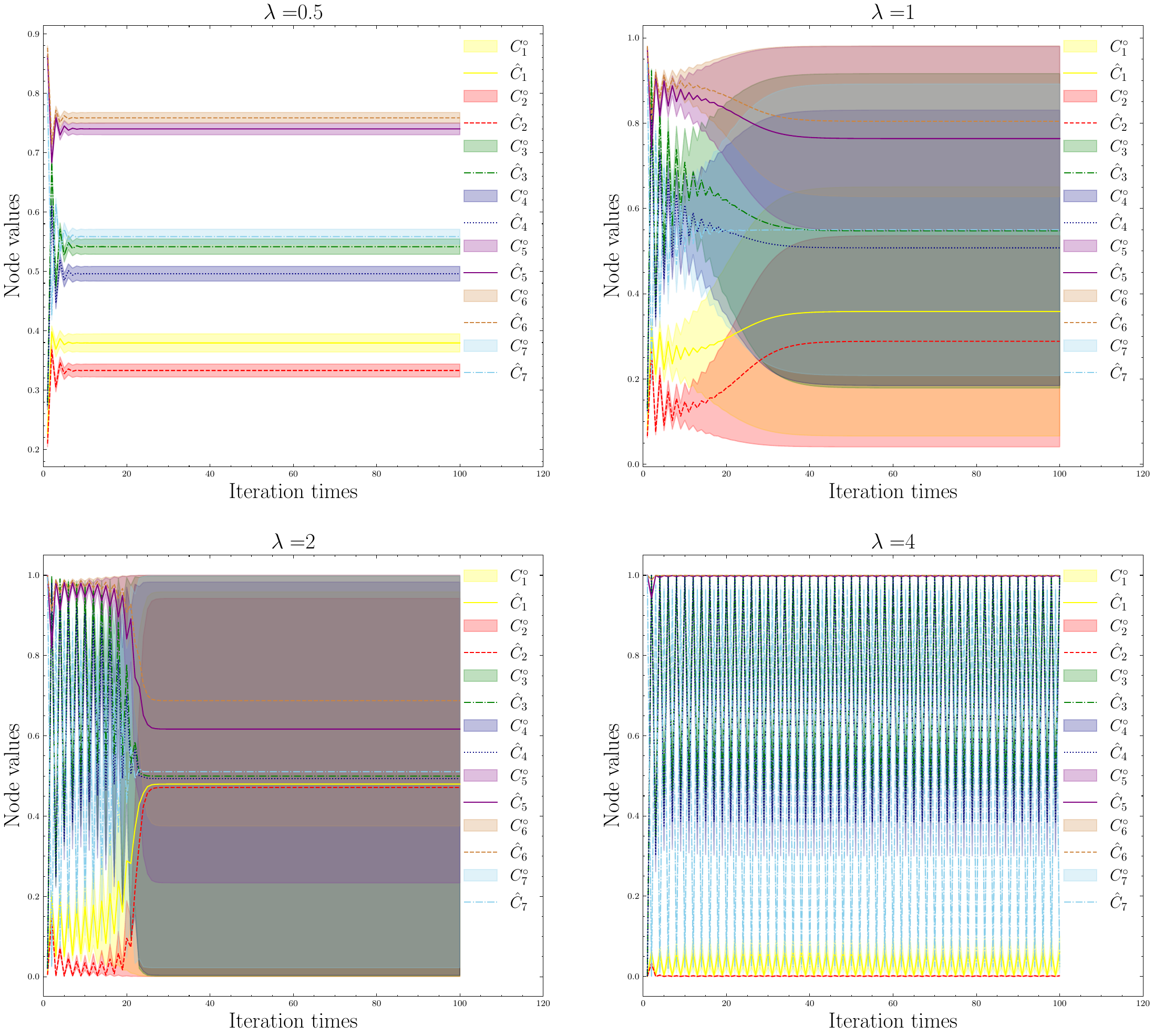}
    \caption{The output of the Web Experience FGCM under different $\lambda$s, with $\otimes w_{web_{11}}= [-0.1,0.1]$}
    \label{figFGCM_web_cw}
  \end{figure}
  \begin{figure}[htbp]
    \centering
      \includegraphics[width=0.75\linewidth]{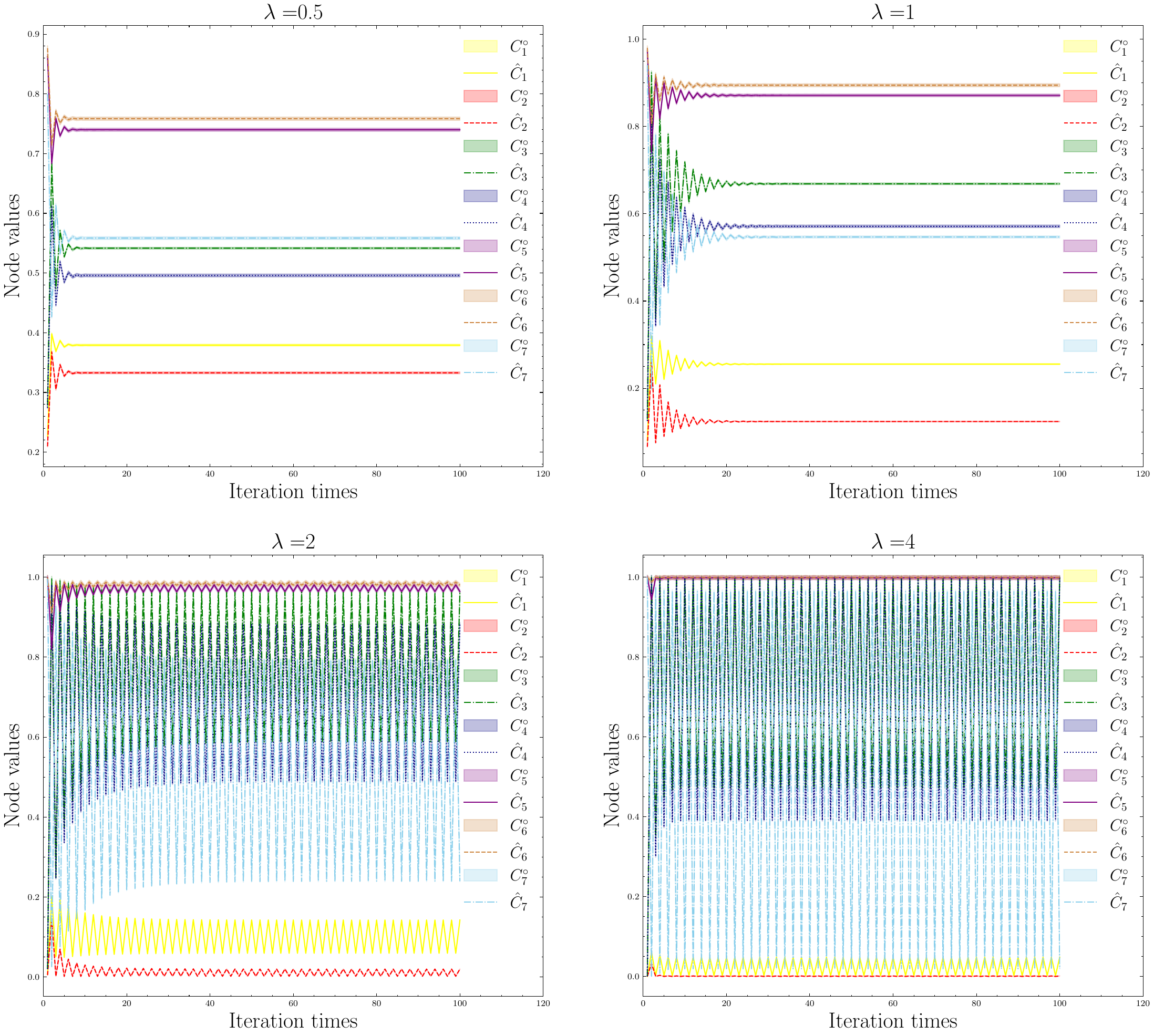}
    \caption{The output of the Web Experience FGGCM under different $\lambda$s, with $ w_{web_{11}}^\pm= 0_{0.1}$}
    \label{figFGGCM_web_cw}
  \end{figure}
  \item Case 2: where the $\mathbf{W}^\pm$ involves multiple IGNs or fuzzy numbers. In this context, conventional FCM and FGCM are  incapable of executing reasoning operations due to the constraints imposed by data format requirements. Consequently, Theorem \ref{thmfgcm} is unable to assess the convergence of the system. In response to this challenge, the adoption of FGGCM becomes imperative for operational purposes. Additionally, Theorem \ref{thmFGGCMsig} can be employed to evaluate the convergence of the system under these complex conditions. This approach ensures that the system's convergence can be effectively determined even when dealing with the intricate data structures.
  Given the weights composed of the following GGNs: $ w_{web_{11}}^\pm = [-0.9, -0.75] \cup [0.4, 0.9]$, $ w_{web_{12}}^\pm = [-0.95, -0.89] \cup -0.83 \cup [-0.8, -0.75]$, $ w_{web_{33}}^\pm = [-1, -0.95] \cup [-0.94, -0.90] \cup [-0.89, 0.88]$, $ w_{web_{15}}^\pm = [0.99, 1] \cup [0.95, 0.98] \cup [-0.90, 0.93]$.
  
  In this case, only FGGCM can be used for reasoning, and its output is shown in Fig. \ref{figFGGCM_web_complex}.
  
  \begin{figure}[htbp]
    \centering
      \includegraphics[width=0.75\linewidth]{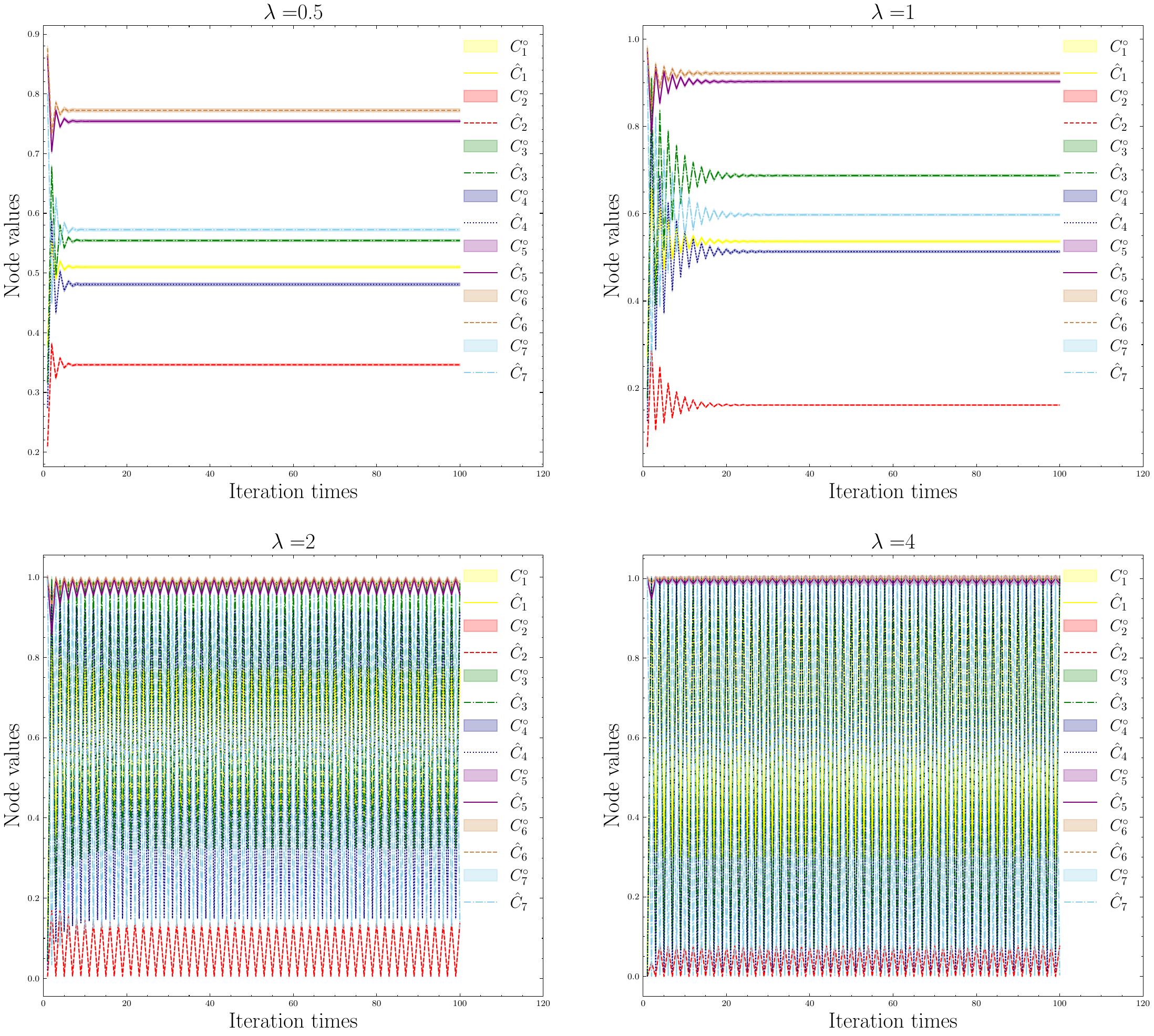}
    \caption{The output of the Web Experience FGGCM under different $\lambda$s, with more complex weights}
    \label{figFGGCM_web_complex}
  \end{figure}
\end{itemize}

\subsection{Simulations for Civil FGGCM }
Similar to the processing process of Web Experience FCM, use Eqs. \eqref{addgreyl} and \eqref{addgreyu} to add greyness to Matrix \eqref{w_civil}. Similarly, when $\left\lvert w_{ij}\right\rvert < g$, do not use \eqref{addgreyl} and \eqref{addgreyu} to add greyness to it to ensure that $\underline{w_{ij}}\leqslant \overline{w_{ij}} \leqslant 0$ and $0 \leqslant \underline{w_{ij}}\leqslant \overline{w_{ij}}$ can always hold. Set $g = 0.01$, and the resulting weight matrix is Eq. \eqref{wcivilgreyav}.

  \begin{equation}
    \setlength{\arraycolsep}{2pt}
    \begin{split}
      &\otimes\mathbf{W}_{civil} =\\
      &\begin{pmatrix}  
        [0,0]  &[0.09,0.11]  &[0,0]  &[0,0]  &[0,0]  &[-0.31,-0.29]  &[0,0]  \\ 
      [0,0]  &[0,0]  &[0.69,0.71]  &[0,0]  &[0,0]  &[0,0]  &[0,0]  \\ 
      [0.59,0.61]  &[0,0]  &[0,0]  &[0,0]  &[0,0]  &[0,0]  &[0,0]  \\ 
      [0.89,0.91]  &[0,0]  &[0,0]  &[0,0]  &[0,0]  &[0,0]  &[0,0]  \\ 
      [0,0]  &[0,0]  &[0.89,0.91]  &[0,0]  &[0,0]  &[0,0]  &[0,0]  \\ 
      [0,0]  &[0,0]  &[0,0]  &[0,0]  &[-0.91,-0.89]  &[0,0]  &[0.79,0.81]  \\ 
      [0,0]  &[0,0]  &[0,0]  &[0.89,0.91]  &[-0.91,-0.89]  &[0,0]  &[0,0]  \\ 
      \end{pmatrix}
    \end{split}
    \label{wcivilgreyav}
  \end{equation}

The matrix \eqref{wcivilgreyav} meets the requirements for determining $\mathbf{W}^{*}$ as stated in Theorem \ref{thmfgcm}. Referring to Eq. \eqref{wstar}, $\mathbf{W}^{*}$ can be computed as Eq. \eqref{wstarcivil}.

\begin{equation}
  \label{wstarcivil}
  \mathbf{W}_{civil}^{*}= \begin{pmatrix}  
    0  &0.11  &0  &0  &0  &0.31  &0  \\ 
  0  &0  &0.71  &0  &0  &0  &0  \\ 
  0.61  &0  &0  &0  &0  &0  &0  \\ 
  0.91  &0  &0  &0  &0  &0  &0  \\ 
  0  &0  &0.91  &0  &0  &0  &0  \\ 
  0  &0  &0  &0  &0.91  &0  &0.81  \\ 
  0  &0  &0  &0.91  &0.91  &0  &0  \\  
  \end{pmatrix}
\end{equation}

Convert Matrix \eqref{wcivilgreyav} to the GGN form, 
% \clearpage
  \begin{equation}
    \label{wcivilggn}
    \setlength{\arraycolsep}{-1pt}
    \begin{pmatrix}  
   0_{0} &0.10_{0.01} &0_{0} &0_{0} &0_{0} &-0.30_{0.01} &0_{0} \\ 
   0_{0} &0_{0} &0.70_{0.01} &0_{0} &0_{0} &0_{0} &0_{0} \\ 
   0.60_{0.01} &0_{0} &0_{0} &0_{0} &0_{0} &0_{0} &0_{0} \\ 
   0.90_{0.01} &0_{0} &0_{0} &0_{0} &0_{0} &0_{0} &0_{0} \\ 
   0_{0} &0_{0} &0.90_{0.01} &0_{0} &0_{0} &0_{0} &0_{0} \\ 
   0_{0} &0_{0} &0_{0} &0_{0} &-0.90_{0.01} &0_{0} &0.80_{0.01} \\ 
   0_{0} &0_{0} &0_{0} &0.90_{0.01} &-0.90_{0.01} &0_{0} &0_{0} \\ 
   \end{pmatrix}
  \end{equation}

\begin{figure}[htbp]
    \centering
      \includegraphics[width=0.75\linewidth]{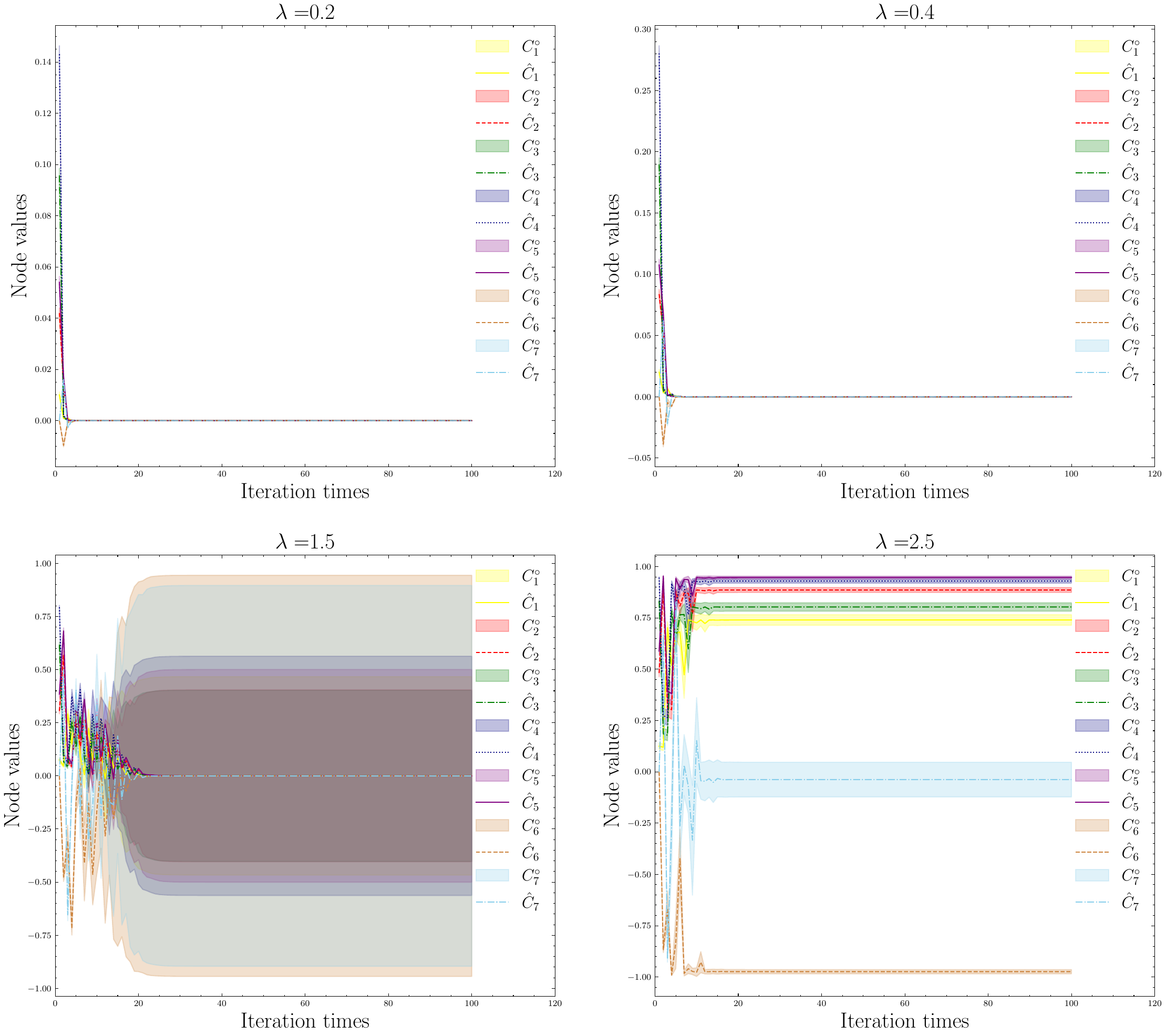}
    \caption{The output of the Civil FGCM under different $\lambda$s.}
    \label{figFGCM_c}
  \end{figure}

\begin{figure}[htbp]
    \centering
      \includegraphics[width=0.75\linewidth]{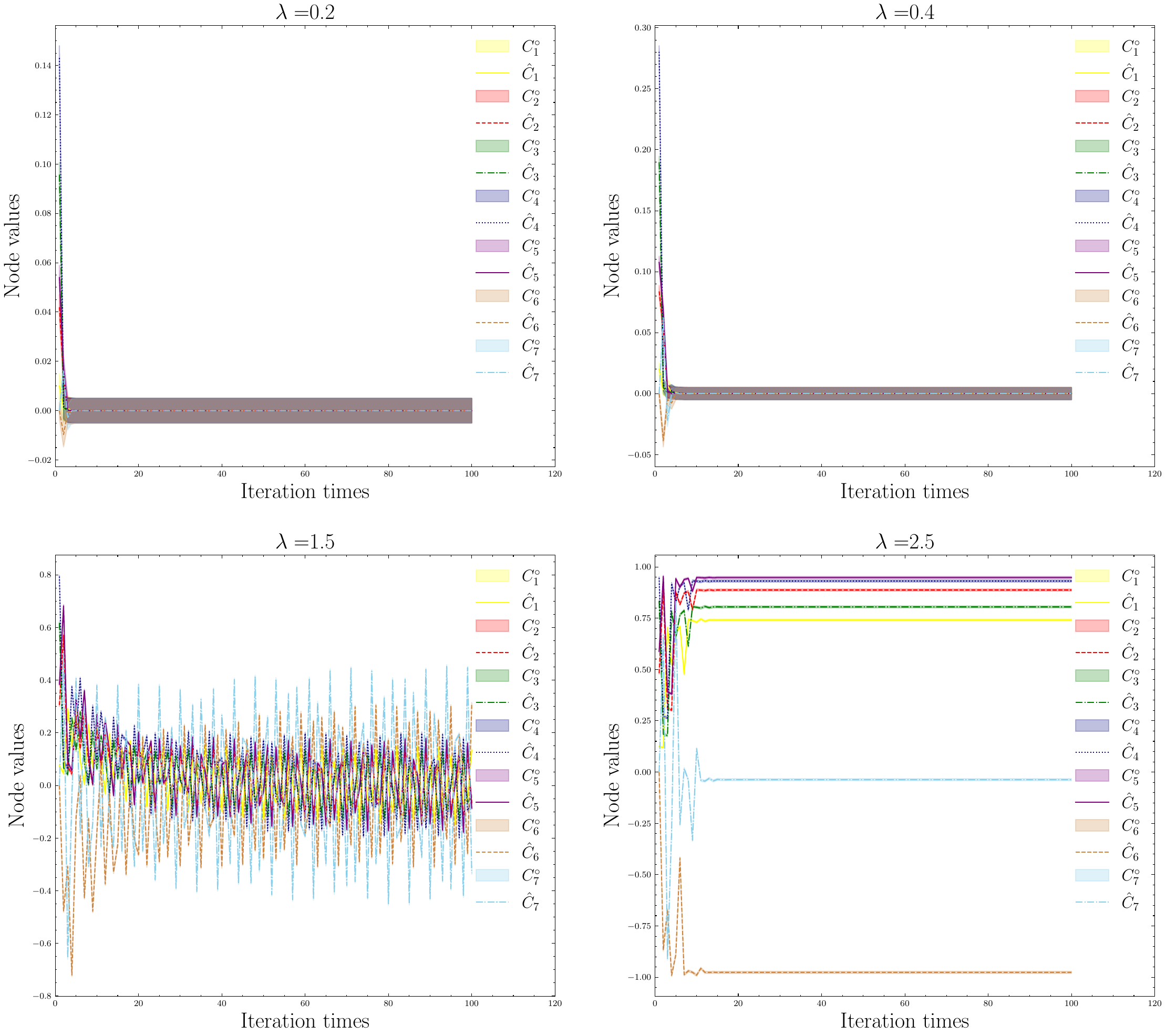}
    \caption{The output of the Civil FGGCM under different $\lambda$s.}
    \label{figFGGCM_c}
  \end{figure}

   Consider the input vector:
   \begin{equation*}
    \begin{split}
      &\left([0.79,0.81],[0.49,0.51],[0.29,0.31],[0.00,0.00],[0.00,0.00],[0.00,0.00],[0.00,0.00] \right).
    \end{split}
   \end{equation*}
   with the corresponding simplified GGN being:
   \begin{equation}
    \label{civilggnv}
    \begin{split}
      &\left(0.80_{(0.01)},0.50_{(0.01)},0.30_{(0.01)},0.00_{(0.00)},0.00_{(0.00)},0.00_{(0.00)},0.00_{(0.00)} \right).
    \end{split}
   \end{equation}
   Utilizing matrices \eqref{wcivilgreyav} and \eqref{wcivilggn}, the outcomes of the FGCM and FGGCM are depicted in Fig. \ref{figFGCM_c} and \ref{figFGGCM_c}.

  Similar to the experiments with web FCM, the subsequent experiments outline two specific scenarios. These are instances where Theorem \ref{thmFGGCMsig} introduced in this paper can determine convergence, whereas the earlier Theorem \ref{thmfcm} and \ref{thmfgcm} are unable to provide such a determination.

   \begin{itemize}
    \item Case 1: When $\otimes \mathbf{W}$ encompasses the interval $[-a,b]$. Suppose ${\otimes w_{civil}}_{11} = [-0.1,0.1]$.
    Under these circumstances, ${\mathbf{W}^*}$ cannot be computed, whereas ${\hat{\mathbf{W}}}$ is obtainable. Referring to Eqs. \eqref{uniong}, \eqref{unionp}, and \eqref{union}, ${ w^\pm_{civil}}_{11} = 0_{0.1}$. The convergence of FGCM and FGGCM is depicted in Fig. \ref{figFGCM_c_change_w} and \ref{figFGGCM_c_change_w}, respectively.
    \begin{figure}[htbp]
      \centering
        \includegraphics[width=0.75\linewidth]{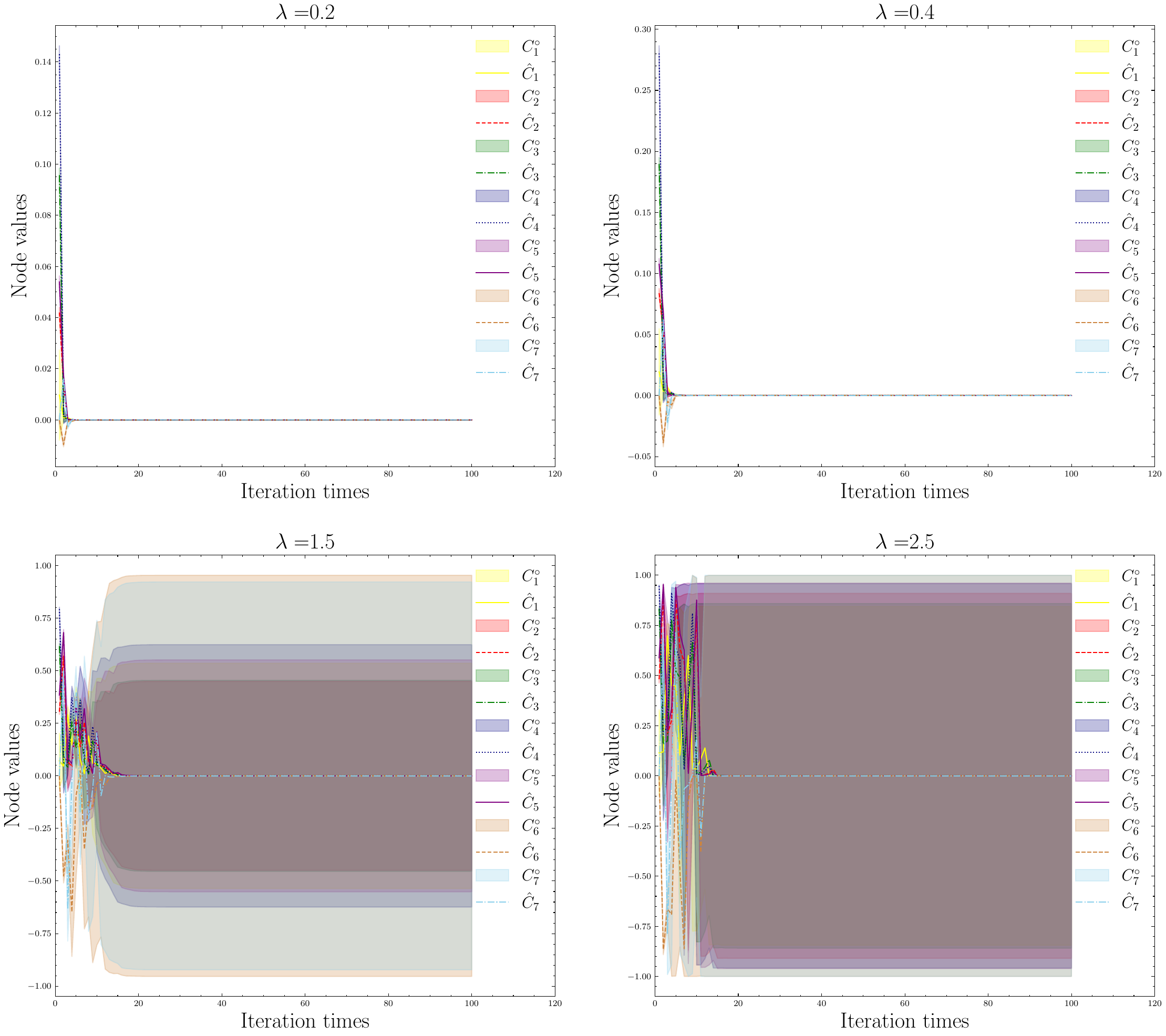}
      \caption{The output of the Civil FGCM under different $\lambda$s,with $\otimes w_{{civil}_{11}} = [-0.1,0.1]$}
      \label{figFGCM_c_change_w}
    \end{figure}
    
    \begin{figure}[htbp]
      \centering
        \includegraphics[width=0.75\linewidth]{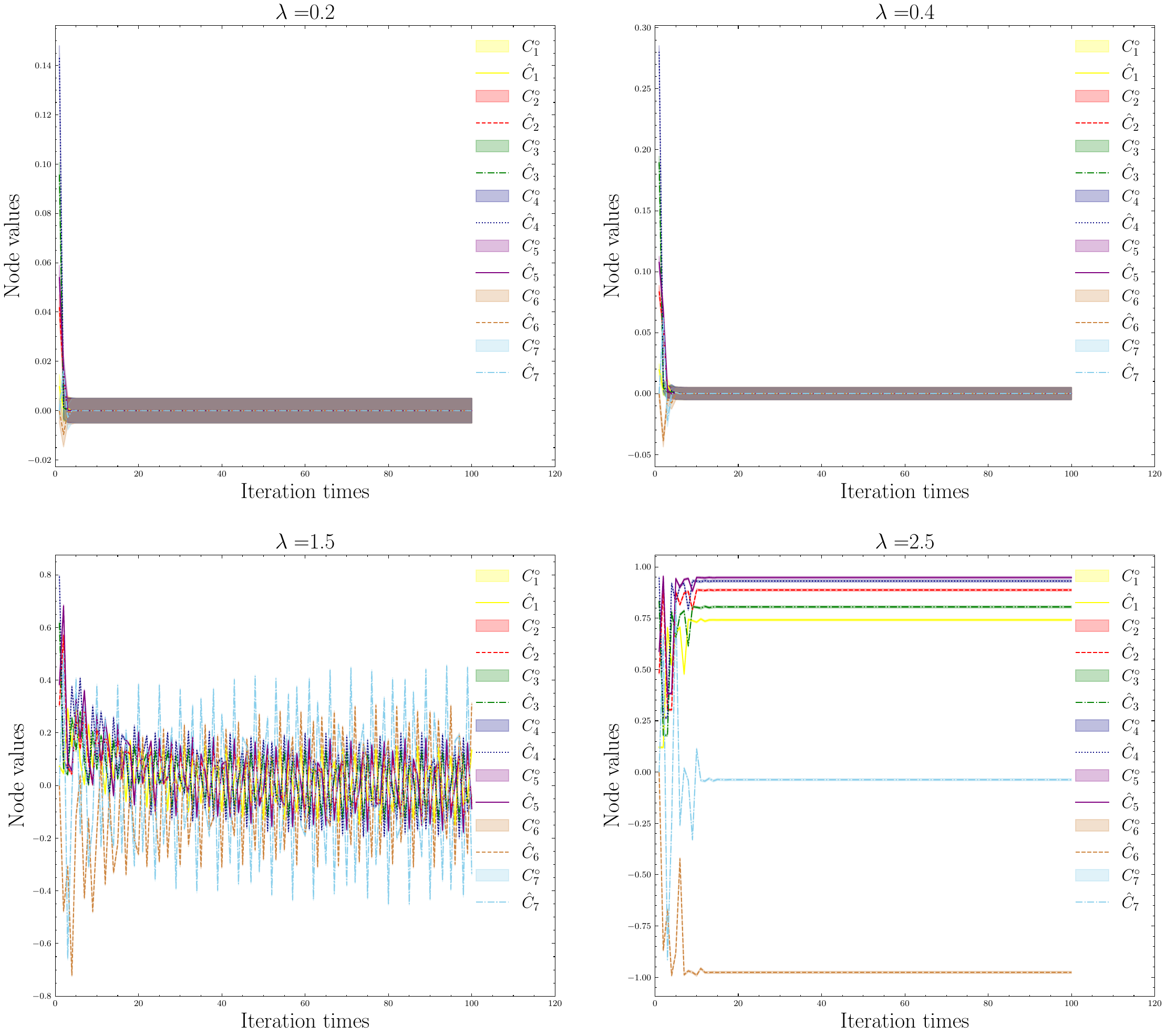}
      \caption{The output of the Civil FGGCM under different $\lambda$s, with $ w^\pm_{{civil}_{11}} = 0_{0.1}$}
      \label{figFGGCM_c_change_w}
    \end{figure}
    \item Case 2: where $ \otimes w_{ij} $ is a blend of multiple IGN or fuzzy numbers. In such cases, data format issues preclude the use of FCM and FGCM for inference operations, and even the determination of their convergence status becomes infeasible. Under these circumstances, the operation can only be conducted using FGGCM, with its convergence assessed utilizing Theorem \ref{thmFGGCMtanh}. Thus, set some elements in Matrix \eqref{wcivilgreyav}: $ w^\pm_{civil_{11}} = [-0.1, 0.1] $, $ w^\pm_{civil_{12}} = [0.07, 0.08] \cup [0.09, 0.11] \cup  [0.13, 0.15] $, $ w^\pm_{civil_{23}} = [0.65, 0.68] \cup [0.685, 0.715] \cup 0.72 \cup  [0.725, 0.73] $, $w^\pm_{civil_{65}} = [-0.97, -0.93] \cup [-0.92, -0.88] \cup [-0.85, -0.8]$. In this case, only FGGCM can be used for reasoning, and the results are shown in Fig. \ref{figFGGCM_c_complex}.

\begin{figure}[htbp]
  \centering
    \includegraphics[width=0.75\linewidth]{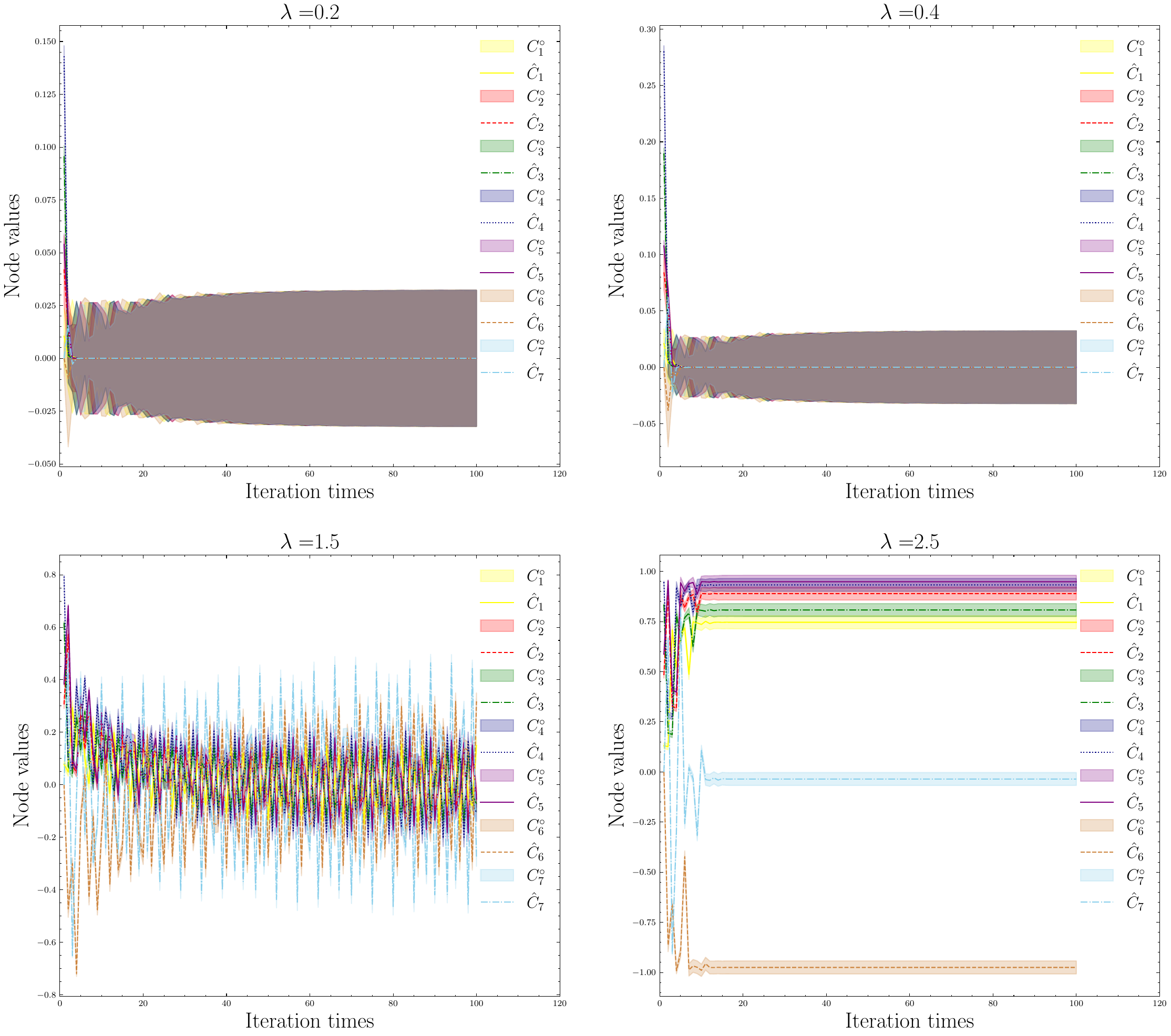}
  \caption{The output of the Civil FGGCM under different $\lambda$s, with more complex weight}
  \label{figFGGCM_c_complex}
\end{figure}
    
\end{itemize}

\subsection{The Convergence of Greyness}
In the current literature, there is a notable absence of conditions specifically detailing the greyness convergence of FGGCM. Thus, this paper addresses this gap by introducing a condition that elucidates the convergence of the greyness in FGGCM. This contribution is significant as it provides a critical understanding of the convergence behavior in FGGCM, which is essential for the analysis and application of these models in various domains.

\begin{figure}[htbp]
  \centering
    \includegraphics[width=0.75\linewidth]{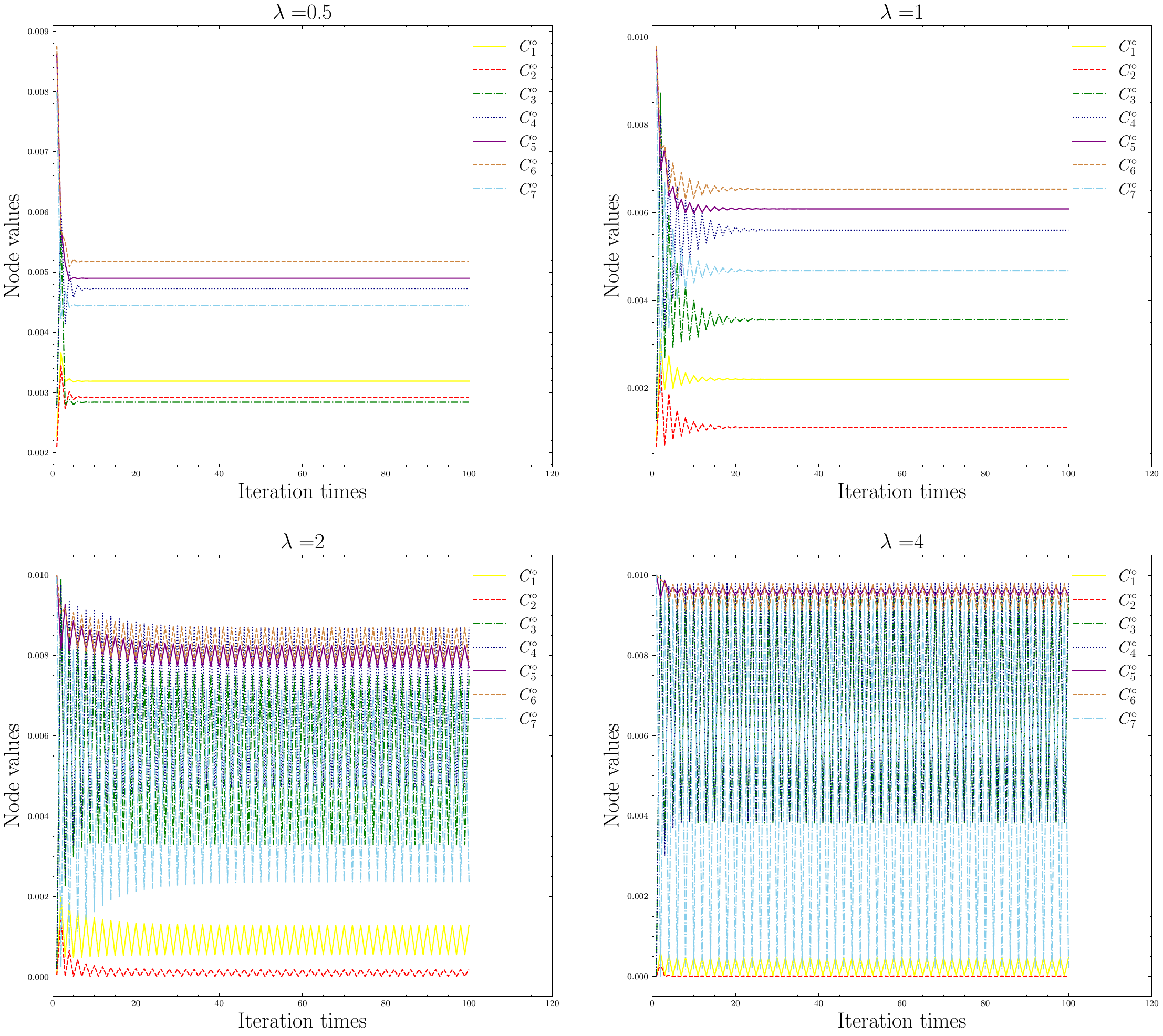}
  \caption{The greyness of the Web Experience FGGCM under different $\lambda$s}
  \label{figFGGCM_web_g}
\end{figure}

Utilizing Web Experience FGGCM's weight \eqref{wwebggn} and an input vector \eqref{webinputggn}, the variation of the greyness over the iterations can be observed, as illustrated in Fig. \ref{figFGGCM_web_g}. 

It can be found that when $\lambda = 0.5$ and $1$, the greyness of FGGCM converges. However, when $\lambda = 2$ or $4$, the greyness  of FGGCM does not converge but oscillates periodically with the oscillation of the kernels.

\begin{figure}[htbp]
  \centering
    \includegraphics[width=0.75\linewidth]{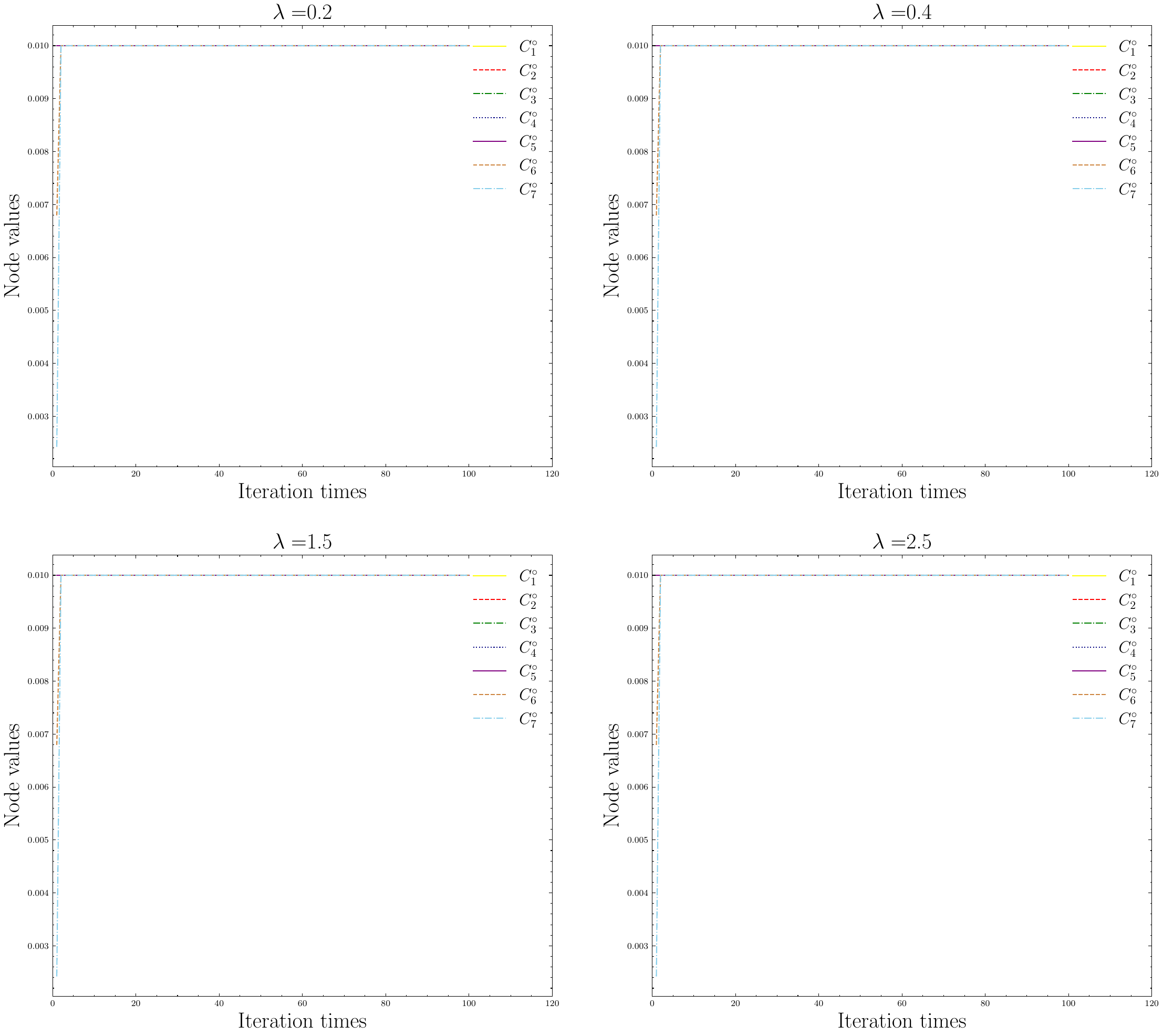}
  \caption{The greyness of the Civil FGGCM under different $\lambda$s}
  \label{figFGGCM_c_g}
\end{figure}

By employing the Civil FGGCM's weight \eqref{wcivilggn} and an input vector \eqref{civilggnv}, the progression of the greyness as it changes with the number of iterations can be determined. This process is depicted in Fig. \ref{figFGGCM_c_g}, offering a visual representation of how the greyness evolves over time.

It can be observed that under different values of $\lambda$, the greyness of all Civil FGGCMs converge to $0.01$.

\section{Discussion}   \label{sec:discussion}
This section is dedicated to the discussion of the simulated results, which serve to validate the theorems introduced in this paper. The aim is to provide empirical evidence that supports the theoretical propositions made, thereby reinforcing the validity and applicability of these new theorems in the context of FGGCM.

\subsection{The Convergence of Kernels}
This part focuses on analyzing the convergence of the kernels of FGGCM, first focusing on the convergence of FGGCM with the sigmoid function as the activation function, and then on the convergence of FGGCM with the tanh function.

\subsubsection{The Kernels' Convergence under Sigmoid Context}
The convergence of Web Experience FCM, FGCM, and FGGCM with the sigmoid function as the activation function under different $\lambda$s is investigated. To evaluate the convergence of these models, the $\left\lVert \mathbf{W}\right\rVert _F \times \lambda$ values are calculated according to Theorems \ref{thmfcm}, \ref{thmfgcm} and \ref{thmFGGCMsig} respectively, and listed in Table \ref{webcompare}. If $\left\lVert \mathbf{W}\right\rVert _F \times \lambda < 4$, it can judge that the kernels of corresponding FCM, FGCM, and FGGCM will converge to a unique fixed point. 

\begin{table}[htbp]
  \caption{In the context of Web engineering, the values of $\mathbf{W}_F \times \lambda$ under different $\lambda$s.
  }\label{webcompare}
  \centering
  \begin{tabular*}{.6\linewidth}{ccccc}
    \toprule    
     & $\lambda = 0.5$ &$\lambda = 1$& $\lambda = 2$ & $\lambda = 4$ \\ 
    \midrule 
    $\left\lVert \mathbf{W}_{web}\right\rVert _F \times \lambda$& \textbf{3.0680}  &6.1359  &12.2719  &24.5437  \\ 
    $\left\lVert \mathbf{W}_{web} ^*\right\rVert _F \times \lambda$&\textbf{3.0829}  &6.1657  &12.3315  &24.6630  \\ 
    $\left\lVert \hat{\mathbf{W}}_{web}\right\rVert _F \times \lambda$&\textbf{3.0586} &6.1172  &12.2344  &24.4688  \\ 
    $\left\lVert \hat{\mathbf{W}}_{web_1}\right\rVert _F \times \lambda$&\textbf{3.0586} &6.1172  &12.2344  &24.4688  \\ 
    $\left\lVert \hat{\mathbf{W}}_{web_{mc}}\right\rVert _F \times \lambda$&\textbf{3.0186}   &6.0372  &12.0745  &24.1489  \\ 
    \bottomrule    
    \end{tabular*}
  \end{table}

  In the Table \ref{webcompare}, $\mathbf{W}_{web_1}$ represents replacing the element $\otimes w_{11}$ in the weight matrix \eqref{wwebgreyav} with the interval $[-0.1,0.1]$, while $\mathbf{W}_{web_{mc}}$ indicates that the weight matrix contains multiple different IGNs or fuzzy numbers. In these two special cases, the convergence can only be judged by the theorems proposed in this paper. The bold numbers in the table indicate that in these cases, the kernels of FGGCM are certain to converge to a unique fixed point. This conclusion is supported by the convergence of each model shown in Fig. \ref{figfcmweb}, \ref{figFGCM_web}, \ref{figFGGCM_web}, \ref{figFGCM_web_cw}, \ref{figFGGCM_web_cw}, and \ref{figFGGCM_web_complex}. It is worth noting that even in some cases where $\left\lVert \mathbf{W}\right\rVert _F \times \lambda > 4$, the FCM, FGCM, and FGGCM still show a trend of convergence. This is because Theorems \ref{thmFGGCMtanh} and \ref{thmFGGCMsig} proposed in this paper and Theorems \ref{thmfcm} and \ref{thmfgcm}  are not necessary and sufficient conditions for judging the convergence of these models, but rather sufficient conditions. Therefore, in practical applications, it is necessary to comprehensively judge the convergence by combining specific situations and theorem conditions.

  By observing Fig. \ref{figfcmweb}, \ref{figFGCM_web}, and \ref{figFGGCM_web}, it can be seen that the reasoning process of the Web Experience FCM, FGCM, and FGGCM, which is in agreement with the relevant descriptions in \cite{Chen2020}. Particularly, the FGGCM not only shows good compatibility with the reasonsing results of FGCM and FCM but also exhibits convergence characteristics similar to FCM. The main reason for this phenomenon is that FGGCM is highly similar to FCM in the operation process of its kernels, and it can minimize the amplification of greyness while calculating the kernel, which becomes a significant advantage of FGGCM. For FGCM, due to the inevitable amplification of greyness during the iteration process, when FGGCM and FCM enter a limit cycle, FGCM will further enter a fixed point (when $\lambda=1$ and $2$) due to the amplification of greyness in its calculation process. This situation does not violate the judgment of Theorem \ref{thmfgcm}, because when $\left\lVert \otimes \mathbf{W}\right\rVert _F \times \lambda > 4$, FGGM does not necessarily guarantee convergence.

  Among the experiments, special attention is paid to the case where the weight matrix of FGGCM contains the interval $[-a,+b]$, with $\otimes w_{11} =[-0.1,0.1]$. In this situation, since the weight does not satisfy the conditions $\underline{w_{ij}}\leqslant \overline{w_{ij}} \leqslant 0$ or $0 \leqslant \underline{w_{ij}}\leqslant \overline{w_{ij}}$ in Theorem \ref{thmfgcm}, it is not possible to directly use Theorem \ref{thmfgcm} to judge the convergence of FGCM. However, when the IGN in the form of $[-a,+b]$ is converted into simplified GGN, FGGCM can be used for reasoning, and Theorem \ref{thmFGGCMsig} can be used to judge the convergence of FGGCM. Its reasoning results are shown in Fig. \ref{figFGGCM_web_cw}, and its convergence situation is consistent with the judgment of Theorem \ref{thmFGGCMsig}.

  Further comparison between Fig. \ref{figFGCM_web_cw} and \ref{figFGGCM_web_cw} reveals that their reasoning results are similar to those in Fig. \ref{figFGCM_web} and \ref{figFGGCM_web}. When $\lambda = 1$ and $2$, FGCM, due to the inevitable amplification of greyness during its iteration process, the increase in the greyness of $\otimes w_{11}$ directly leads to the increase in the greyness of $C_1$, which can be inferred from the comparison between Fig. \ref{figFGCM_web} and  \ref{figFGCM_web_cw}.
  Next, examining Fig. \ref{figFGGCM_web_complex}, when the weights have a more complex structure of GGNs, its convergence can only be judged using Theorem \ref{thmFGGCMsig}. By observing Table \ref{webcompare} and Fig. \ref{figFGGCM_web_complex}, it can be find that the experimental results support the judgment of Theorem \ref{thmFGGCMsig}.

\subsubsection{The Kernels' Convergence under tanh Context}
 The convergence of the FGGCM with the tanh function as the activation function under different $\lambda$s is examined in this part. To evaluate the convergence, the calculated results of $\left\lVert \mathbf{W}\right\rVert _F \times \lambda$ under different $\lambda$s are listed in Table \ref{civilcompare}. According to Theorems \ref{thmfcm}, \ref{thmfgcm}, and \ref{thmFGGCMtanh}, if $\left\lVert \mathbf{W}\right\rVert _F \times \lambda<1$, it can be inferred that the kernels of the FGGCM must converge to a unique fixed point.

 \begin{table}[htbp]
  \caption{In the context of Civil FCM, the values of $\mathbf{W}_F \times \lambda$ under different $\lambda$s.}\label{civilcompare}
  \centering
  \begin{tabular*}{0.62\linewidth}{ccccc}
    \toprule    
     & $\lambda = 0.2$ &$\lambda = 0.4$& $\lambda = 1.5$ & $\lambda = 2.5$ \\ 
    \midrule 
    $\left\lVert \mathbf{W}_{civil}\right\rVert _F \times \lambda$& \textbf{0.4750}  &\textbf{0.9499}  &3.5623  &5.9372  \\ 
  $\left\lVert \mathbf{W}_{civil} ^*\right\rVert _F \times \lambda$&\textbf{0.4809}  &\textbf{0.9617}  &3.6066  &6.0109  \\ 
  $\left\lVert \hat{\mathbf{W}}_{civil}\right\rVert _F \times \lambda$&\textbf{0.4750}  &\textbf{0.9499} &3.5623  &5.9372   \\ 
  $\left\lVert \hat{\mathbf{W}}_{civil_1}\right\rVert _F \times \lambda$&\textbf{0.4750}  &\textbf{0.9499} &3.5623  &5.9372   \\ 
  $\left\lVert \hat{\mathbf{W}}_{civil_{mc}}\right\rVert _F \times \lambda$&\textbf{0.4746 } &\textbf{0.9491}  &3.5593  &5.9322   \\ 
    \bottomrule    
    \end{tabular*}
  \end{table}

  Similar to the Table \ref{webcompare},  In Table \ref{civilcompare}, $\mathbf{W}_{civil_1}$ also represents replacing the element $\otimes w_{11}$ in the weight matrix \eqref{wcivilgreyav} with the interval $[-0.1, 0.1]$, while $\mathbf{W}_{civil_{mc}}$ indicates that the weight matrix contains multiple different IGNs or fuzzy numbers. In these two special cases, it is only possible to rely on the theorems proposed in this paper (Theorem \ref{thmFGGCMtanh}) to judge the convergence of the models. The bold numbers indicate that the kernels of FGGCM must converge to a unique fixed point. This conclusion is supported by the convergence of each model shown in Fig. \ref{figfcmc}, \ref{figFGCM_c}, \ref{figFGGCM_c}, \ref{figFGCM_c_change_w}, \ref{figFGGCM_c_change_w}, \ref{figFGGCM_c_complex}. It can be observed that when $\lambda = 0.2$ and $0.4$, the kernels of FGGCM converge to $\bm 0$, which also supports the judgment of Corollary \ref{corollary_tanh3}. In some cases, $\left\lVert \mathbf{W}\right\rVert _F \times \lambda > 1$, FCM, FGCM, and FGGCM still show a trend of convergence. This is because the Theorems \ref{thmFGGCMtanh} proposed in this paper, as well as the theorems \ref{thmfcm} and \ref{thmfgcm} proposed in previous literature, are not necessary and sufficient conditions for judging the convergence of these models, but merely sufficient conditions. This means that even if the conditions of the theorems are not met, FCM, FGCM, and FGGCM may still converge.

  When $\otimes w_{11} = [-0.1, 0.1]$, Theorem \ref{thmfgcm} can not judge the convergence of FGCM. However, after transfer it into a GGN form, using FGGCM for reasoning, it can use Theorem \ref{thmFGGCMtanh} to judge whether if converge, as shown in Fig. \ref{figFGGCM_c_change_w} and Table \ref{civilcompare}. When the weight matrix has a complex structure, the convergence can only be judged using Theorem \ref{thmFGGCMtanh}, as shown in Fig. \ref{figFGGCM_c_complex}. Combining Table \ref{civilcompare} and Fig.\ref{figFGCM_c_change_w}, \ref{figFGGCM_c_change_w}, \ref{figFGGCM_c_complex}, it can be found that the judgment of Theorem \ref{thmFGGCMtanh} and \ref{thm_tanh_grey_convergence} are correct: when $\left\lVert \mathbf{W}_{civil}\right\rVert _F \times \lambda <1$, the kernels of FGGCM converge to a unique fixed point $\bm 0$.

  Fig. \ref{figfcmc}, \ref{figFGCM_c}, and \ref{figFGGCM_c} respectively represent the operation process of Civil FCM, FGCM, and FGGCM. Except for $\lambda = 1.5$, the models show similar convergence. When $\lambda = 1.5$, FGCM exhibits different behavior due to the amplification of greyness during iteration: when FGGCM and FCM enter a limit cycle, the greyness of FGCM amplifies and eventually enters a fixed point. This phenomenon is due to the fuzziness in the modeling of FGCM, which needs to perform multiplication on two IGN during reasoning, and the result of each multiplication is always the product of their respective boundaries, leading FGCM to amplify the input uncertainty compared to FGGCM, making it difficult to represent the periodic changes of the system.
  Further comparison between Fig. \ref{figFGCM_c} and \ref{figFGCM_c_change_w}, and Fig.\ref{figFGGCM_c}, \ref{figFGGCM_c_change_w} and \ref{figFGGCM_c_complex}, reveals that as the greyness of the weight matrix increases, the greyness of the output result also increases synchronously. This observation underscores the impact of the greyness of the weight matrix on the model's output, especially when dealing with highly uncertain data.

\subsection{The Convergence of Greyness}
This part first analyzes and discusses the greyness convergence of FGGCM with tanh as the activation function. Subsequently, the convergence of FGGCM with sigmoid as the activation function is also discussed. Finally, this paper will use Theorems \ref{thmFGGCMtanh}, \ref{thm_tanh_grey_convergence} and \ref{thmFGGCMsig} to analyze the greyness convergence of various FGGCMs. 

\subsubsection{The Context of Sigmoid}
Fig. \ref{figFGGCM_web_g} and \ref{figFGGCM_c_g} respectively show the changes in greyness for the Web Experience FGGCM with a sigmoid activation function and the Civil FGGCM with a tanh activation function.

Upon examining Fig. \ref{figFGGCM_web_g}, it is noticed that when the value of $\lambda$ is $0.5$ and $1$, the greyness of FGGCM shows a trend of convergence. This indicates that under these specific parameter settings, the greyness of the model stably tends towards a fixed value. However, when the value of $\lambda$ increases to $2$ or $4$, a significant change occurs. In this case, the kernel of FGGCM no longer exhibits convergence but instead enters a state of periodic oscillation, which causes that the greyness of FGGCM is not stably tending towards a fixed point, but rather periodically varying within a certain range as the kernel oscillates.

The convergence of the greyness for the web engineering FGGCM can be judged according to Eq. \eqref{siggreynessconvergence} in Theorem \ref{thmFGGCMsig}. Let
$$
\frac{\hat{A}_{i}\left|{\hat{A}_{j} \hat{w}_{ij}}\right|\theta\left(C_j^\circ - \hat{w}_{ij}^\circ\right) }{\sum_{j=1}^{n}\left|\hat{w}_{ij}\hat{A}_j\right|} = \widetilde{m}_{ij}
$$
where $\widetilde{m}_{ij}$ is an element of $\mathbf{\widetilde{M}}$. The values of $\mathbf{\widetilde{M}}_F$ under different $\lambda$s are listed in Table \ref{mfweb}.

\begin{table}[htbp]
  \caption{The $\mathbf{\widetilde{M}}_F$ values of the Web Experience FGGCM under different $\lambda$s.
  }\label{mfweb}
  \centering
  \begin{tabular*}{.48\linewidth}{ccccc}
    \toprule    
    & $\lambda = 0.5$ &$\lambda = 1$& $\lambda = 2$ & $\lambda = 4$ \\ 
    \midrule 
    $\left\lVert \widetilde{M }\right\rVert _F  $ & 0.1984 &0.3466  &0.5217  &0.6076  \\ 
    \bottomrule    
    \end{tabular*}
  \end{table}

  It is noted that when $\lambda = 0.5$ or $1$, $\left\lVert \widetilde{M }\right\rVert _F  <1$ holds true, and at this time, the kernel of FGGCM converges to a fixed point. Therefore, it can be concluded that at this point, the greyness of FGGCM converges to a unique fixed point, and consequently, FGGCM converges to a unique fixed point according to Theorem \ref{thmFGGCMsig}. However, when $\lambda = 2$ or $4$, although the calculated $\left\lVert \widetilde{M }\right\rVert _F  <1$, the kernel of the FGGCMs do not converge, which does not satisfy the conditions of Corollary \ref{sig_grey_convergence}. Hence, the theorems proposed in this paper cannot judge the convergence of the greyness. However, from Eq. \eqref{sig_grey_iter}, it can be observed that when the sigmoid function is used as the activation function, the greyness of FGGCM is influenced by the kernels in two adjacent iterative steps. This means that the greyness of FGGCM depends not only on the state of the kernel in the current iteration step but also on the state of the kernel in the previous iteration step. If the kernels of FGGCM form a limit cycle, it will be difficult for the greyness of FGGCM to converge to a fixed point. This is because the cyclic state of the kernel will cause the greyness values to also exhibit periodic changes during the iteration process, rather than trending towards stability. This phenomenon reveals the dynamic complexity of FGGCM under certain conditions, especially when using the sigmoid function.

  \subsubsection{The Context of tanh}
  When observing the greyness change graph in Fig. \ref{figFGGCM_c_g} for the Civil FGGCM, it can be found that all greynesss show a trend of convergence. To judge whether the greyness of the civil FGGCM converges, an evaluation can be made based on Eq. \eqref{tanhgreynessconvergence} in Theorem \ref{thmFGGCMtanh}. Let
  $$
  \frac{\left|{\hat{A}_{j} \hat{w}_{ij}}\right|\theta\left(C_j^\circ - \hat{w}_{ij}^\circ\right) }{\sum_{j=1}^{n}\left|\hat{w}_{ij}\hat{A}_j\right|} = \widetilde{m}_{ij}
  $$
  where $\widetilde{m}_{ij}$ represents an element in the matrix $\mathbf{\widetilde{M}}$. The values of $\mathbf{\widetilde{M}}_F$ under different $\lambda$s are listed in Table \ref{mfcivil} for further analysis.

  \begin{table}[htbp]
    \caption{The $\mathbf{\widetilde{M}}_F$ values of the Civil FGGCM under different $\lambda$s. }\label{mfcivil}
    \centering
    \begin{tabular*}{.5\linewidth}{ccccc}
      \toprule    
      & $\lambda = 0.2$ &$\lambda = 0.4$& $\lambda = 1.5$ & $\lambda = 2.5$ \\ 
      \midrule 
      $\left\lVert \widetilde{M }\right\rVert _F  $ & 1.1634 &1.1198  &0.8699  &1.0877  \\ 
      \bottomrule    
      \end{tabular*}
    \end{table}

    It can be observed that when the value of $\lambda$ is $0.2$, $0.4$, and $2.5$, the $\left\lVert \widetilde{M }\right\rVert _F  $ is greater than $1$. According to Theorem \ref{thm_tanh_grey_convergence}, it can be inferred that the greyness of FGGCM has at least one fixed point. However, when $\lambda$ is $1.5$, the kernel of FGGCM enters a state of limit cycle. In this case, Theorem \ref{thmFGGCMtanh} cannot be used to judge whether the greyness of FGGCM has a unique fixed point. Nevertheless, according to Theorem \ref{thm_tanh_grey_convergence}, it can conclude that the greyness of FGGCM has at least one fixed point. This indicates that even when the behavior of the kernel exhibits periodic changes, the greyness of the model with tanh as the activation function can still tend to stabilize, although it may not be the unique stable state. These findings reveal the dynamic behavior and stability characteristics of FGGCM under different parameter settings, providing an important theoretical basis for understanding and predicting the behavior of the model under various conditions.

    By analyzing the data in Tables \ref{webcompare} and \ref{mfweb}, combined with Theorems \ref{thmFGGCMsig}, it can be found that when $ \lambda = 0.5 $, FGGCMs using ${\mathbf{W}}^\pm_{web} $, ${\mathbf{W}}^\pm_{web_{1}} $, ${\mathbf{W}}^\pm_{web_{mc}} $ as weights will converge to a unique fixed point. However, when $ \lambda = 1 $, $2$, or $4$ the above FGGCMs may exhibit multiple fixed points, form limit cycles, or enter chaotic states.Furthermore, by combining the data in Tables \ref{civilcompare} and \ref{mfcivil}, along with Theorems \ref{thmFGGCMtanh}, \ref{thm_tanh_grey_convergence} and Corollary \ref{corollary_tanh1}, \ref{corollary_tanh2},\ref{corollary_tanh3}, it can be inferred that when $ \lambda = 0.2 $ or $0.4$, the kernel of the FGGCM using ${\mathbf{W}}^\pm_{civil} $, ${\mathbf{W}}^\pm_{civil_{1}} $, ${\mathbf{W}}^\pm_{civil_{mc}} $ as weights will converge to a unique fixed point, and this fixed point is $\bm 0$, with at least one fixed point in the greyness. When $ \lambda = 1.5 $, the convergence of the kernel of the above FGGCMs are unknown, but its greyness has a unique fixed point. When $ \lambda = 2.5 $, although the convergence of the kernel of the above FGGCM is also unknown, its greyness has at least one fixed point. These findings provide important references for understanding and predicting the dynamic behavior of FGGCM under different parameter settings.

    In fact, when the greyness of FGGCM is set to $0$, it can be found that Eqs. \eqref{tanhkernelconvergence} and \eqref{sigkernelconvergence}  degenerates into Eq. \eqref{eqfcmtanh}  and Eq. \eqref{eqfcmsig}. This means that Theorems \ref{thmfcm} and \ref{thmfgcm} are actually special cases of Theorems \ref{thmFGGCMtanh} and \ref{thmFGGCMsig}. This phenomenon is consistent with FCM and FGCM as special cases of FGGCM, thus unifying these models theoretically and revealing their intrinsic connections.

\section{Conclusions}   \label{sec:conclusion}

This paper mainly studies the convergence conditions of Fuzzy General Grey Cognitive Maps (FGGCM). Firstly, the metrics for the GGN space and its vector counterpart are presented and confirmed using the Minkowski inequality. By utilizing the characteristic that Cauchy sequences are sequences that converge, the completeness of these two spaces is established. Based on this, utilizing the Banach fixed point theorem and the Browder-Gohde-Kirk fixed point theorem, alongside Lagrange's mean value theorem and Cauchy's inequality, this article establishes the sufficient conditions for FGGCM to converge to a unique fixed point when using tanh and sigmoid functions as activation functions. Furthermore, the conditions required for the kernels and greyness of FGGCM to converge to a unique fixed point are also outlined separately. Finally, the correctness of the proposed theory is verified through case studies. These cases include FGGCM with tanh and sigmoid activation functions, and their convergence under different parameter settings. The research results show that the convergence conditions proposed in this paper can effectively judge the convergence behavior of FGGCM. Moreover, it is demonstrated that the convergence theorems of FCM are particular cases of the theorems introduced here.

The future work will include the following aspects:

\begin{itemize}
  \item Verification in Complex Space: Currently, the conclusions are mainly based on real number space, and the convergence of FGGCM in complex space has not been verified. In the future, further exploration of the behavior of FGGCM in complex space is needed to verify whether the above conclusions hold in complex space, in order to expand the application scope of FGGCM.
  \item Criteria for Limit Cycles or Chaotic States: The paper has not yet studied the limit cycles or chaotic states of FGGCM in depth. In the future, it is necessary to find relevant criteria to fully understand the behavior patterns of FGGCM under different conditions. This will help to better predict and control the running state of FGGCM, avoiding unstable situations.
  \item Research on the Convergence Speed of FGGCM: The convergence speed of FGGCM is of great importance to its practical application. In the future, it is necessary to study the factors affecting the convergence speed of FGGCM and explore how to improve its convergence speed to improve the efficiency and performance of FGGCM in practical applications.
  \item Extension to Other Activation Functions: Extending convergence analysis to other activation functions, such as radial basis functions, to explore whether new conditions are needed.
  \item Expansion of Application Scenarios: Applying FGGCM to more practical scenarios, such as intelligent control, decision support systems, etc., to verify its effectiveness and reliability in practical applications. At the same time, combined with specific application scenarios, further optimization and improvement of FGGCM can be carried out to better meet actual needs.
  \item Learning Algorithms: By utilizing the convergence characteristics of FGGCM under different conditions, designing corresponding FGGCM learning algorithms for different application scenarios can optimize the convergence speed of FGGCM or enhance the accuracy of FGGCM learning algorithms.
  \item Computational Efficiency: Investigating methods to improve the computational efficiency of FGGCMs, especially when dealing with large-scale systems, which may involve parallel processing techniques or more efficient algorithms to handle GGNs.
  \item Integration with Other Models: Exploring the integration of FGGCMs with other models, such as Bayesian networks or agent-based models, to create more powerful hybrid systems for complex system analysis.
\end{itemize}

In summary, this paper provides convergence determination schemes for FGGCM when the activation function is tanh and sigmoid. These schemes lay a theoretical foundation for the FGGCM application, stability design, and development of learning algorithms of FGGCM. Through future work, including further research and development of FGGCM, a deeper understanding and application of FGGCM can be achieved, thereby promoting its practical application and development in related fields.

\section*{Declaration of Competing Interest}

The authors declare that they have no known competing financial interests or personal relationships that could have appeared to influence the work reported in this paper.

\section*{Acknowledgments}

This work was supported in part by the 
National Natural Science Foundation of China under Grant 61305133, 61876187, 52372398.

\bibliography{convergence}

\begin{thebibliography}{10}

\bibitem{Napoles2023}
G.~Nápoles, N.~Ranković, and Y.~Salgueiro, ``On the interpretability of fuzzy
  cognitive maps,'' {\em Knowledge-Based Systems}, vol.~281, p.~111078, 2023.

\bibitem{Li2024}
N.~Li, W.~Zou, Y.~Zhu, B.~Wang, S.~Bu, and H.~Shen, ``A compact embedded flight
  parameter detection system for small soaring uavs,'' {\em IEEE/ASME
  Transactions on Mechatronics}, vol.~29, no.~1, p.~52 – 63, 2024.

\bibitem{Zhe2020}
Z.~ZHAO, Y.~NIU, and L.~SHEN, ``Adaptive level of autonomy for human-uavs
  collaborative surveillance using situated fuzzy cognitive maps,'' {\em
  Chinese Journal of Aeronautics}, vol.~33, no.~11, pp.~2835--2850, 2020.
\newblock SI: Emerging Technologies of Unmanned Aerial Vehicles.

\bibitem{Meghabghab2001}
G.~Meghabghab, ``Mining user's web searching skills through fuzzy cognitive
  state map,'' in {\em JOINT 9TH IFSA WORLD CONGRESS AND 20TH NAFIPS
  INTERNATIONAL CONFERENCE, PROCEEDINGS, VOLS. 1-5} (M.~Smith, W.~Gruver, and
  L.~Hall, eds.), (345 E 47TH ST, NEW YORK, NY 10017 USA), pp.~429--434, Int
  Fuzzy Syst Assoc; N Amer Fuzzy Informat Proc Soc; IEEE Syst, Man \& Cybernet
  Soc; IEEE, Neural Networks Council, IEEE, 2001.
\newblock 9th International-Fuzzy-Systems-Association World Congress/20th
  North-American-Fuzzy-Information-Processing-Society, International
  Conference, VANCOUVER, CANADA, JUL 25-28, 2001.

\bibitem{Meghabghab2003}
G.~Meghabghab, ``Fuzzy cognitive map and people's web behavior,'' in {\em
  PROCEEDINGS OF THE 7TH JOINT CONFERENCE ON INFORMATION SCIENCES} (S.~Chen,
  H.~Cheng, D.~Chiu, S.~Das, R.~Duro, E.~Kerre, H.~Leong, Q.~Li, M.~Lu,
  M.~Romay, D.~Ventura, and J.~Wu, eds.), (PO BOX 90291, DURHAM, NC 27708-0291
  USA), pp.~253--258, Assoc Intelligent Machinery; Duke Univ; Elsevier Publ
  Inc, Informat Sci Journal; Harbin Inst Technol; NIEHS, ASSOC INTELLIGENT
  MACHINERY, 2003.
\newblock 7th Joint Conference on Information Sciences (JCIS), RES TRIANGLE PK,
  NC, SEP 26-30, 2003.

\bibitem{Boutalis2009}
Y.~Boutalis, T.~L. Kottas, and M.~Christodoulou, ``Adaptive estimation of fuzzy
  cognitive maps with proven stability and parameter convergence,'' {\em IEEE
  TRANSACTIONS ON FUZZY SYSTEMS}, vol.~17, pp.~874--889, AUG 2009.

\bibitem{Harmati2018}
I.~A. Harmati and L.~T. Koczy, ``On the existence and uniqueness of fixed
  points of fuzzy set valued sigmoid fuzzy cognitive maps,'' in {\em 2018 IEEE
  INTERNATIONAL CONFERENCE ON FUZZY SYSTEMS (FUZZ-IEEE)}, IEEE International
  Conference on Fuzzy Systems, (345 E 47TH ST, NEW YORK, NY 10017 USA), IEEE,
  IEEE, 2018.
\newblock IEEE International Conference on Fuzzy Systems (FUZZ-IEEE), Rio de
  Janeiro, BRAZIL, JUL 08-13, 2018.

\bibitem{Harmati2020}
I.~A. Harmati and L.~T. Koczy, ``On the convergence of input-output fuzzy
  cognitive maps,'' in {\em ROUGH SETS, IJCRS 2020} (R.~Bello, D.~Miao,
  R.~Falcon, M.~Nakata, A.~Rosete, and D.~Ciucci, eds.), vol.~12179 of {\em
  Lecture Notes in Artificial Intelligence}, (GEWERBESTRASSE 11, CHAM, CH-6330,
  SWITZERLAND), pp.~449--461, SPRINGER INTERNATIONAL PUBLISHING AG, 2020.
\newblock International Joint Conference on Rough Sets (IJCRS), ELECTR NETWORK,
  JUN 29-JUL 03, 2020.

\bibitem{Tsadiras1999}
A.~K. Tsadiras and K.~G. Margaritis, ``An experimental study of the dynamics of
  the certainty neuron fuzzy cognitive maps,'' {\em Neurocomputing}, vol.~24,
  no.~1, pp.~95--116, 1999.

\bibitem{Taber2001}
R.~Taber, R.~Yager, and C.~Helgason, ``Small-sample quantization effects on the
  equilibrium behavior of combined fuzzy cognitive maps,'' in {\em 10TH IEEE
  INTERNATIONAL CONFERENCE ON FUZZY SYSTEMS, VOLS 1-3: MEETING THE GRAND
  CHALLENGE: MACHINES THAT SERVE PEOPLE}, (345 E 47TH ST, NEW YORK, NY 10017
  USA), pp.~1567--1572, IEEE, IEEE, 2001.
\newblock 10th IEEE International Conference on Fuzzy Systems, UNIV MELBOURNE,
  MELBOURNE, AUSTRALIA, DEC 02-05, 2001.

\bibitem{Taber2007}
R.~Taber, R.~R. Yager, and C.~M. Helgason, ``Quantization effects on the
  equilibrium behavior of combined fuzzy cognitive maps,'' {\em INTERNATIONAL
  JOURNAL OF INTELLIGENT SYSTEMS}, vol.~22, pp.~181--202, FEB 2007.

\bibitem{Karatzinis2018}
G.~Karatzinis, Y.~S. Boutalis, and T.~L. Kottas, ``System identification and
  indirect inverse control using fuzzy cognitive networks with functional
  weights,'' in {\em 2018 EUROPEAN CONTROL CONFERENCE (ECC)}, (345 E 47TH ST,
  NEW YORK, NY 10017 USA), pp.~2069--2074, IEEE, 2018.
\newblock European Control Conference (ECC), Limassol, CYPRUS, JUN 12-15, 2018.

\bibitem{Karatzinis2018a}
G.~Karatzinis, Y.~S. Boutalis, and Y.~L. Karnavas, ``Switching control of dc
  motor using multiple fuzzy cognitive network models,'' in {\em 2018 7th
  International Conference on Systems and Control (ICSC)}, pp.~384--390, Oct
  2018.

\bibitem{Karatzinis2018b}
G.~Karatzinis, Y.~S. Boutalis, and Y.~L. Karnavas, ``Motor fault detection and
  diagnosis using fuzzy cognitive networks with functional weights,'' in {\em
  2018 26th Mediterranean Conference on Control and Automation (MED)},
  pp.~709--714, 2018.

\bibitem{Karatzinis2021}
G.~D. Karatzinis and Y.~S. Boutalis, ``Fuzzy cognitive networks with functional
  weights for time series and pattern recognition applications,'' {\em Applied
  Soft Computing}, vol.~106, p.~107415, 2021.

\bibitem{Karatzinis2021a}
G.~D. Karatzinis, Y.~S. Boutalis, and Y.~L. Karnavas, ``An accurate multiple
  cognitive classifier system for incipient short-circuit fault detection in
  induction generators,'' {\em Electrical Engineering}, pp.~1--16, 2021.

\bibitem{Napoles2014}
G.~Napoles, R.~Bello, and K.~Vanhoof, ``How to improve the convergence on
  sigmoid fuzzy cognitive maps?,'' {\em INTELLIGENT DATA ANALYSIS}, vol.~18,
  no.~6, pp.~S77--S88, 2014.

\bibitem{Knight2014}
C.~J. Knight, D.~J. Lloyd, and A.~S. Penn, ``Linear and sigmoidal fuzzy
  cognitive maps: An analysis of fixed points,'' {\em Applied Soft Computing},
  vol.~15, pp.~193--202, 2014.

\bibitem{Hatwagner2017}
M.~F. Hatwágner, V.~A. Niskanen, and L.~T. Kóczy, ``Behavioral analysis of
  fuzzy cognitive map models by simulation,'' in {\em 2017 Joint 17th World
  Congress of International Fuzzy Systems Association and 9th International
  Conference on Soft Computing and Intelligent Systems (IFSA-SCIS)}, pp.~1--6,
  2017.

\bibitem{Harmati2020c}
I.~{\'A}. Harmati and L.~T. K{\'o}czy, {\em Notes on the Rescaled Algorithm
  for Fuzzy Cognitive Maps}, pp.~43--49.
\newblock Cham: Springer International Publishing, 2020.

\bibitem{Harmati2023a}
I.~{\'A}. Harmati, M.~F. Hatw{\'a}gner, and L.~T. K{\'o}czy, ``Global stability
  of fuzzy cognitive maps,'' {\em Neural Computing and Applications}, vol.~35,
  no.~10, pp.~7283--7295, 2023.

\bibitem{Koutsellis2022}
T.~Koutsellis, G.~Xexakis, K.~Koasidis, A.~Nikas, and H.~Doukas, ``Parameter
  analysis for sigmoid and hyperbolic transfer functions of fuzzy cognitive
  maps,'' {\em OPERATIONAL RESEARCH}, vol.~22, pp.~5733--5763, NOV 2022.

\bibitem{Maximov2023}
D.~Maximov, ``Multi-valued cognitive maps: Calculations with linguistic
  variables without using numbers,'' {\em FUZZY SETS AND SYSTEMS}, vol.~459,
  pp.~1--21, MAY 15 2023.

\bibitem{Concepcion2021}
L.~Concepcion, G.~Napoles, R.~Falcon, K.~Vanhoof, and R.~Bello, ``Unveiling the
  dynamic behavior of fuzzy cognitive maps,'' {\em IEEE TRANSACTIONS ON FUZZY
  SYSTEMS}, vol.~29, pp.~1252--1261, MAY 2021.

\bibitem{Napoles2022}
G.~Napoles, I.~Grau, L.~Concepcion, L.~K. Koumeri, and J.~P. Papa, ``Modeling
  implicit bias with fuzzy cognitive maps,'' {\em NEUROCOMPUTING}, vol.~481,
  pp.~33--45, APR 7 2022.

\bibitem{Xiaojie2022}
W.~Xiaojie, L.~Chao, and L.~Chen, ``The feedback stabilization of finite-state
  fuzzy cognitive maps,'' {\em TRANSACTIONS OF THE INSTITUTE OF MEASUREMENT AND
  CONTROL}, vol.~44, pp.~2485--2499, SEP 2022.

\bibitem{Harmati2019a}
I.~A. Harmati and L.~T. Koczy, ``On the convergence of sigmoidal fuzzy grey
  cognitivemaps,'' {\em INTERNATIONAL JOURNAL OF APPLIED MATHEMATICS AND
  COMPUTER SCIENCE}, vol.~29, pp.~453--466, SEP 2019.

\bibitem{Harmati2019}
I.~A. Harmati and L.~T. Koczy, ``Stability of fuzzy cognitive maps with
  interval weights,'' in {\em PROCEEDINGS OF THE 11TH CONFERENCE OF THE
  EUROPEAN SOCIETY FOR FUZZY LOGIC AND TECHNOLOGY (EUSFLAT 2019)} (V.~Novak,
  V.~Marik, M.~Stepnicka, M.~Navara, and P.~Hurtik, eds.), vol.~1 of {\em
  Atlantis Studies in Uncertainty Modelling}, (29 AVENUE LAVMIERE, PARIS,
  75019, FRANCE), pp.~756--763, European Soc Fuzzy Log \& Technol; Univ
  Ostrava, Inst Res \& Applicat Fuzzy Modeling; Czech Tech Univ Prague, Czech
  Inst Informat Robot \& Cybernet, ATLANTIS PRESS, 2019.
\newblock 11th Conference of the
  European-Society-for-Fuzzy-Logic-and-Technology (EUSFLAT), Czech Tech Univ,
  Prague, CZECH REPUBLIC, SEP 09-13, 2019.

\bibitem{Harmati2020a}
I.~A. Harmati and L.~T. Koczy, ``On the convergence of fuzzy grey cognitive
  maps,'' in {\em INFORMATION TECHNOLOGY, SYSTEMS RESEARCH, AND COMPUTATIONAL
  PHYSICS} (P.~Kulczycki, J.~Kacprzyk, L.~Koczy, R.~Mesiar, and R.~Wisniewski,
  eds.), vol.~945 of {\em Advances in Intelligent Systems and Computing},
  (GEWERBESTRASSE 11, CHAM, CH-6330, SWITZERLAND), pp.~74--84, SPRINGER
  INTERNATIONAL PUBLISHING AG, 2020.
\newblock 3rd Conference on Information Technology, Systems Research and
  Computational Physics (ITSRCP), Krakow, POLAND, JUL 02-05, 2018.

\bibitem{Concepcion2020}
L.~Concepcion, G.~Napoles, R.~Bello, and K.~Vanhoof, ``On the behavior of fuzzy
  grey cognitive maps,'' in {\em ROUGH SETS, IJCRS 2020} (R.~Bello, D.~Miao,
  R.~Falcon, M.~Nakata, A.~Rosete, and D.~Ciucci, eds.), vol.~12179 of {\em
  Lecture Notes in Artificial Intelligence}, (GEWERBESTRASSE 11, CHAM, CH-6330,
  SWITZERLAND), pp.~462--476, SPRINGER INTERNATIONAL PUBLISHING AG, 2020.
\newblock International Joint Conference on Rough Sets (IJCRS), ELECTR NETWORK,
  JUN 29-JUL 03, 2020.

\bibitem{Peng2016}
Z.~Peng, L.~Wu, and Z.~Chen, ``Research on steady states of fuzzy cognitive map
  and its application in three-rivers ecosystem,'' {\em Sustainability},
  vol.~8, no.~1, 2016.

\bibitem{Behrooz2019}
F.~Behrooz, R.~Yusof, N.~Mariun, U.~Khairuddin, and Z.~Hilmi~Ismail,
  ``Designing intelligent mimo nonlinear controller based on fuzzy cognitive
  map method for energy reduction of the buildings,'' {\em Energies}, vol.~12,
  no.~14, 2019.

\bibitem{Song2024}
X.~Song, ``Physical education teaching mode assisted by artificial intelligence
  assistant under the guidance of high-order complex network,'' {\em Scientific
  Reports}, vol.~14, no.~1, p.~4104, 2024.

\bibitem{Biloslavo2012}
R.~Biloslavo and A.~Grebenc, ``Integrating group delphi, analytic hierarchy
  process and dynamic fuzzy cognitive maps for a climate warning scenario,''
  {\em KYBERNETES}, vol.~41, no.~3-4, pp.~414--428, 2012.

\bibitem{Papageorgiou2005}
E.~I. Papageorgiou and P.~P. Groumpos, ``A weight adaptation method for fuzzy
  cognitive map learning,'' {\em Soft Computing}, vol.~9, pp.~846--857, 2005.

\bibitem{Ahmadi2014}
S.~Ahmadi, N.~Forouzideh, C.-H. Yeh, R.~Martin, and E.~Papageorgiou, ``A first
  study of fuzzy cognitive maps learning using cultural algorithm,'' in {\em
  2014 9th IEEE conference on industrial electronics and applications},
  pp.~2023--2028, IEEE, 2014.

\bibitem{Altundogan2018}
T.~G. Altundoğan and M.~Karaköse, ``An approach for online weight update
  using particle swarm optimization in dynamic fuzzy cognitive maps,'' in {\em
  2018 3rd International Conference on Computer Science and Engineering
  (UBMK)}, pp.~1--5, Sep. 2018.

\bibitem{Napoles2016}
G.~Napoles, E.~Papageorgiou, R.~Bello, and K.~Vanhoof, ``On the convergence of
  sigmoid fuzzy cognitive maps,'' {\em INFORMATION SCIENCES}, vol.~349,
  pp.~154--171, JUL 1 2016.

\bibitem{Napoles2017}
G.~Napoles, E.~Papageorgiou, R.~Bello, and K.~Vanhoof, ``Learning and
  convergence of fuzzy cognitive maps used in pattern recognition,'' {\em
  NEURAL PROCESSING LETTERS}, vol.~45, pp.~431--444, APR 2017.

\bibitem{Napoles2018}
G.~Napoles, L.~Concepcion, R.~Falcon, R.~Bello, and K.~Vanhoof, ``On the
  accuracy-convergence tradeoff in sigmoid fuzzy cognitive maps,'' {\em IEEE
  TRANSACTIONS ON FUZZY SYSTEMS}, vol.~26, pp.~2479--2484, AUG 2018.

\bibitem{Qiao2024}
Y.~Qiao, L.~Jian, and H.~Cai, ``A novel multi-attribute three-way decision
  model with three-parameter interval grey number decision-theoretic rough
  sets,'' {\em KYBERNETES}, 2024 JUN 12 2024.

\bibitem{Jiang2021}
S.~Q. Jiang, S.~Liu, and Z.~Liu, ``General grey number decision-making model
  and its application based on intuitionistic grey number set,'' {\em GREY
  SYSTEMS-THEORY AND APPLICATION}, vol.~11, pp.~556--570, OCT 19 2021.

\bibitem{Chen2020}
J.~Chen, X.~Gao, and J.~Rong, ``Enhance the uncertainty modeling ability of
  fuzzy grey cognitive maps by general grey number,'' {\em IEEE ACCESS},
  vol.~8, pp.~163844--163856, 2020.

\bibitem{Tsadiras2008}
A.~K. Tsadiras, ``Comparing the inference capabilities of binary, trivalent and
  sigmoid fuzzy cognitive maps,'' {\em INFORMATION SCIENCES}, vol.~178,
  pp.~3880--3894, OCT 15 2008.

\end{thebibliography}
\end{document}